**Manufacturing, processing, applications, and advancements of Fe-based shape memory alloys**

Anwar Algamal [a], Hossein Abedi [a], Umesh Gandhi [b], Othmane Benafan [c], Mohammad Elahinia [a], Ala Qattawi [a,]*

[a] Department of Mechanical, Industrial & Manufacturing Engineering, University of Toledo 2801 W. Bancroft, Toledo, OH 43606, USA

[b] Toyota Research Institute North America, Ann Arbor, MI 48105, USA

[c] NASA Glenn Research Center, Materials and Structures Division, Cleveland, OH 44135, USA

[*] Corresponding author: ala.qattawi@utoledo.edu

## Abstract

Fe-based shape memory alloys (Fe-SMAs) belong to smart metallic materials that can memorize or restore their preset shape after experiencing a substantial amount of deformation under heat, stress, or magnetic stimuli. Fe-SMAs have remarkable thermomechanical properties and have attracted significant interest because of their potential merits, such as cost-effective alloying elements, superior workability, weldability, a stable superelastic response, and low-temperature dependence of critical stress required for stress-induced martensitic transformation. Therefore, Fe-SMAs can be an intriguing and economical alternative to other SMAs. The recent advancements in fabrication methods of conventional metals and SMAs are helping the production of customized powder composition and then customized geometries by additive manufacturing (AM). The technology in these areas, i.e., fabrication techniques, experimental characterization, and theoretical formulations of Fe-SMAs for conventional and AM has been rapidly advancing and is lacking a comprehensive review. This paper provides a critical review of the recent developments in Fe-SMAs-related research. The conventional and AM-based methods of producing Fe-SMAs are discussed, and a detailed review of the current research trends on Fe-SMAs including 4-D printing of Fe-SMAs are comprehensively documented. The presented review provides a comprehensive review of experimental methods and processes used to determine the material characteristics and features of Fe-SMAs. In addition, the work provides a review of the reported computational modeling of Fe-SMAs to help design new Fe-SMA composition and geometry. Finally, different Fe-SMAs-based applications such as sensing and damping systems, tube coupling, and reinforced concrete are also discussed. This work will guide new research opportunities for working on Fe-SMAs and encourage new developments in the future.

## Graphical abstract

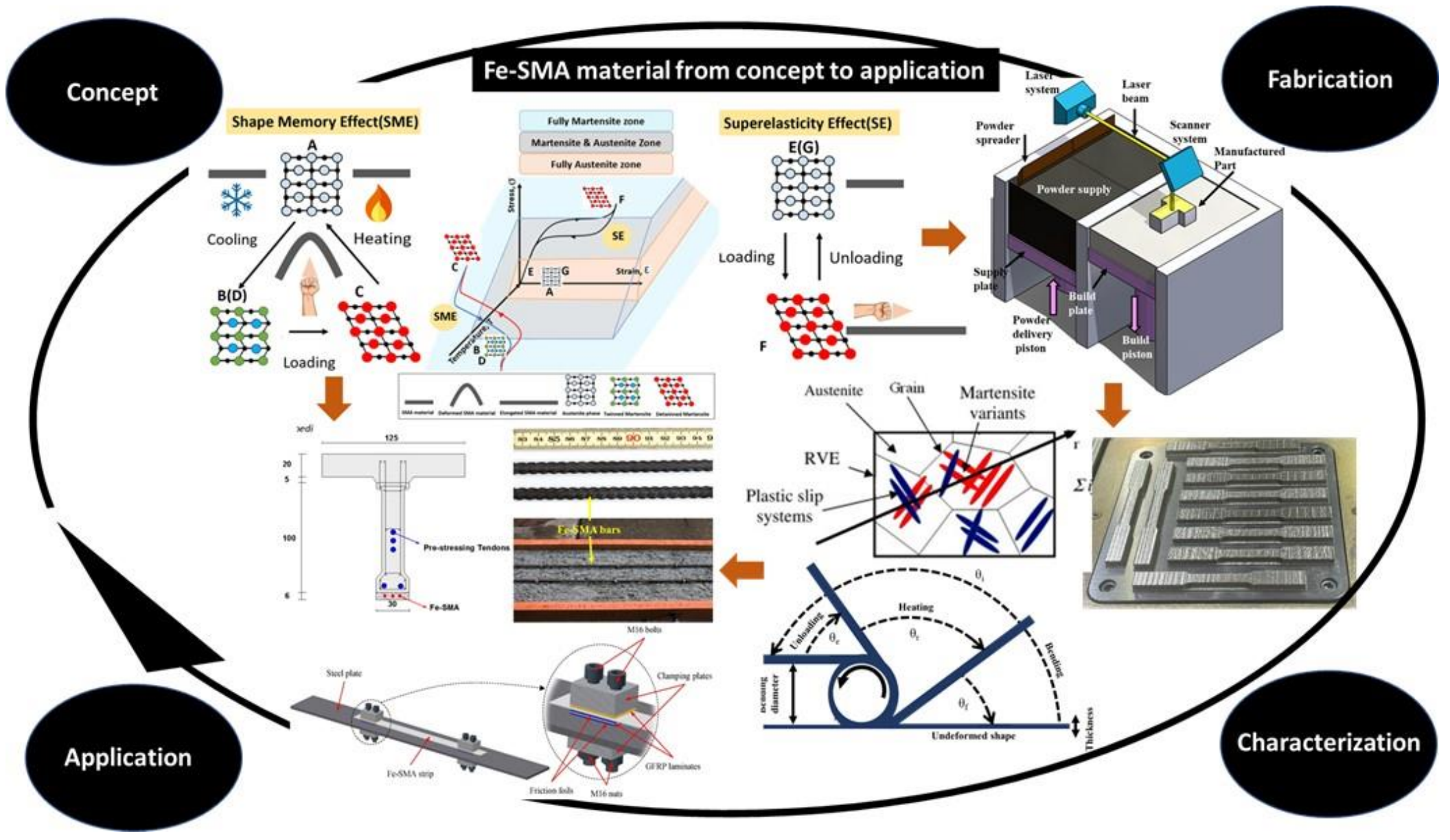

**Abbreviations**

**$A_f$**, austenite finish temperature; **$A_s$**, austenite start temperature; **AM**, additive manufacturing; **ANN**, artificial neural network; **BCC**, body-centered cubic; **BCT**, body-centered tetragonal; **CAD**, computer-aided design; **DMA**, dynamic mechanical analyzer; **DSC**, differential scanning calorimeter; **ECCI**, channeling contrast imaging; **EBSD**, electron backscatter diffraction; **FCC**, face-centered cubic; **HCP**, hexagonal- close-packed; **HTSMA**, high-temperature shape memory alloy; **ICP**, inductively coupled plasma; **LPBF**, laser powder bed fusion; **MA**, mechanical alloying; **$M_f$**, martensitic finish temperature; **$M_s$**, martensitic start temperature; **MTs**, martensitic transformations; **OM**, optical microscopy; **OWSME**, one-way shape memory effect; **PE** or **SE**, pseudo-elasticity or superelasticity; **SEM**, scanning electron microscopy; **SMAs**, shape-memory alloys; **SME**, shape memory effect; **SMMs**, shape-memory materials; **SR**, strain recovery; **TEM**, transmission electron microscopy; **TH**, thermal hysteresis; **TTs**, transformation temperatures; **TWSME**, two-way shape memory effect; **VAR**, vacuum arc melting; **VED**, volumetric energy density; **VIM**, vacuum induction melting; **XRD**, X-ray diffraction.

## 1. Introduction

Smart materials are defined as those that can perceive their environment and/or own states, make a judgment, and then modify their functions following a predetermined purpose [1]. Their smart behavior includes a change in the geometry size, shape, or color in reaction to a variety of externally performed stimuli including light, stress, temperature, pH, moisture, and electric or magnetic fields [2], [3].[4], [5]. Shape memory materials (SMMs) are smart materials that can restore their predefined shape after experiencing considerable deformation with exposure to a particular stimulus, such as heat, stress, or magnetic induction [6]. Numerous materials, including ceramics, polymers, metals, and composite materials, exhibit the shape memory phenomenon [7], [8], [9], [10], [11].

Shape memory alloys (SMAs) are one common category of SMMs that have excellent thermomechanical characteristics and can restore their undeformed shape after being subjected to a certain temperature or load [12], [13]. SMAs have been employed in various industrial and medical fields since their discovery in 1932, including robotics, biomedical, aerospace, and mechanical engineering [14]. Crystallographic phases are what enable SMA performance in which reversible transformations from one material phase to another take place via diffusion-less transformations known as reversible martensitic transformations (MTs). This phase change is related to the distinct properties including pseudo-elasticity (PE) or superelasticity (SE) and shape memory effect (SME) that set SMAs apart from other materials. Temperature-induced reversible transformations result in SME, whereas stress-induced transformations lead to PE [15], [16], [17], [18], [19], [20].

The shape memory performance was first investigated in Au-Cd alloys. After that, several SMAs including Ti-Ni, Ni-Co-Mn-In, Ni-Mn-Ga, and Cu-Al-Ni have been described. Due to PE, SMAs such as Cu-Zn-Al and Ni-Ti display a significant recoverable strain. The Ni-Ti–based alloy, particularly, is the only superelastic material currently produced due to its ductility and outstanding superelastic strain [21], [22]. SMAs are known for a variety of distinctive phenomena, including SME, SE, enormous damping capacity, and a two-way SME [23]. SMAs have high conductivity so electrical current can be employed for their electrical resistive heating [24].

Ni-Ti SMA has received interest since its discovery in 1959 due to its outstanding SE, high recovery strain, and corrosion resistance. Nevertheless, the high cost and time-consuming processing combined with the need to produce NiTi SMA with suitable properties minimize its applications on a broad scale [24], [25]. Despite all efforts exerted on NiTi SMA investigation, no viable binary NiTi SMA with a transition temperature greater than 150 °C has been discovered [26], [27]. In general, NiTi SMAs are still too costly for a variety of applications and there is a need to explore alternate options that have higher transition temperatures, are more stable, and have appropriate strength and ductility.

Iron-based SMAs or Fe-SMAs have outstanding characteristics including superior workability and weldability, low-temperature dependency of transformation critical stress, and stable superelastic behavior over a wide temperature range [28], [29], [30], [31], [32]. Fe-SMAs can exhibit lower thermal hysteresis and nearly identical martensitic phase transformation behavior to that of nitinol SMA such as Fe–Ni–Co, Fe–Pd, and Fe–Pt alloys. However, Fe–Pd or Fe–Pt Fe-based alloys do not typically display PE at ambient temperature. Fe-SMA with higher thermal hysteresis including Fe–Mn–Si and Fe–Ni–C can demonstrate significant SMEs [21], [33], [34], [35], [36].

Additive Manufacturing (AM) technologies advance metals and SMAs fabrication, enabling the processing of customized powder composition and customized geometries. However, a comprehensive review of the fabrication techniques, experimental characterization, and theoretical formulations of Fe-SMAs is still lacking. This work aims to fill this gap by providing an in-depth review of the development, properties, mechanisms, and classification of Fe-SMAs. The work examines the experimental methods and processes used to ascertain the essential characteristics and features of Fe-SMAs. Theoretical research on Fe-SMAs has been addressed to enable the development of models that would simplify the design and production of these functional materials. This research comprehensively highlights the various methods proposed for modifying Fe-SMAs and enhancing their performance—including alloying elements, the materials' textures, and structures, as well as the deformation temperature, grain (abnormal growth, size, orientation, and morphology), and using heat treatment techniques. The article highlights the most recent research on the Fe-SMAs and their future directions including AM-produced Fe-SMAs. Fe-SMAs-based applications such as reinforced concrete, tube coupling, and sensing and damping systems are comprehensively reviewed. Fig.1 illustrates the publication's survey of Fe-SMAs and other SMAs. The number of publications in 2000-2024 through Web of Science accessed on September 20, 2024, where the number of publications on Fe-SMAs is growing.

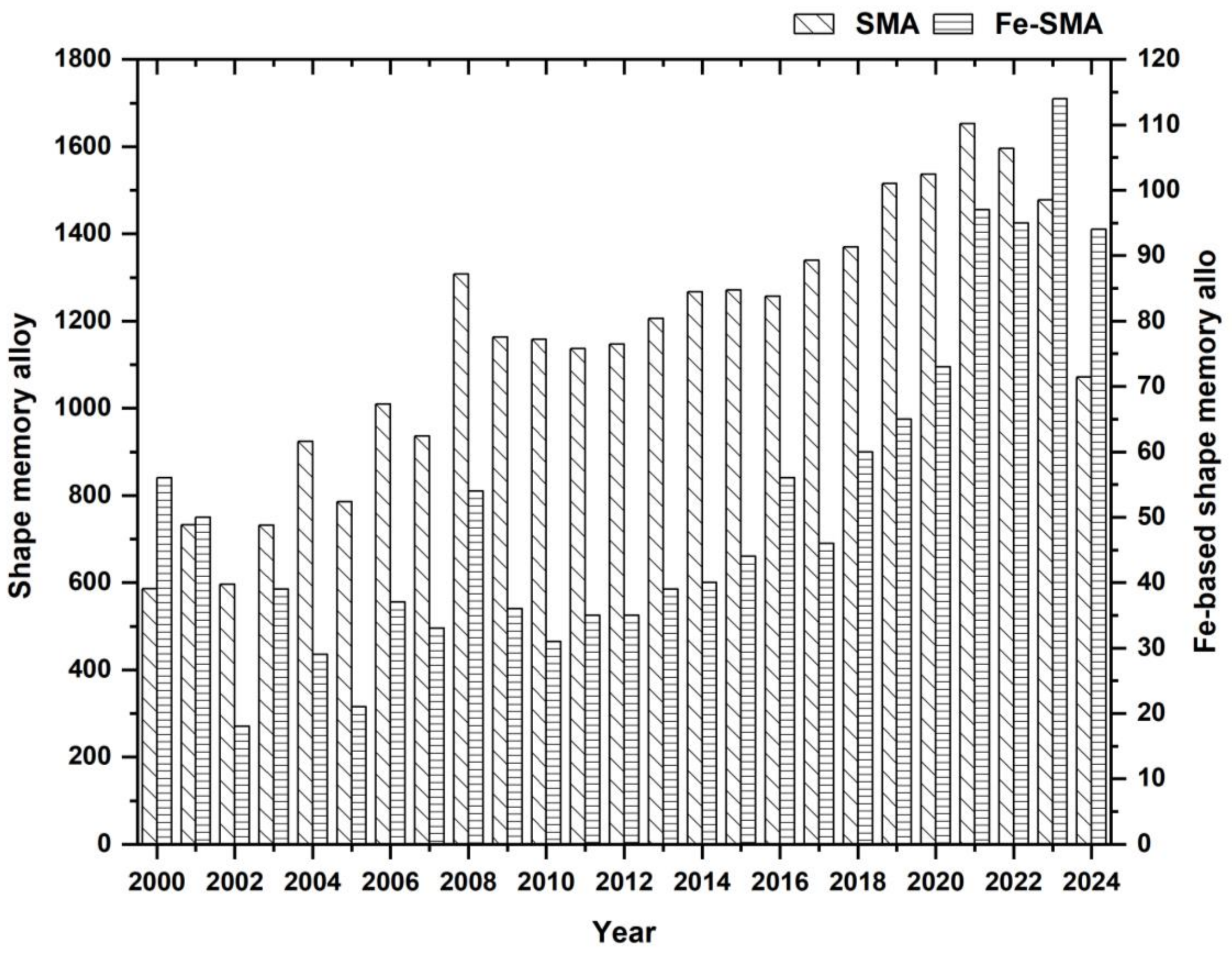

**Fig. 1. Publications survey of Fe-SMAs and other SMAs. The number of publications in 2000-2024 through Web of Science accessed on September 20, 2024.**

## 2. Evolution, Classification, and Mechanisms of Fe-SMAs

### 2.1. Ni- and Cu-based SMAs.

The SMAs are classified into copper-based, nickel-titanium-based, and iron-based SMAs as illustrated in Fig.2a. The NiTi-based SMAs are robust on several levels, namely large reversible strain, good actuation stress, and corrosion resistance. They are produced mostly using Ti and Ni elements and other alloying elements such as Hf, Ti, Cu, Fe, and Zr [37], [38], [39], [40]. They have notable features such as SE and SME, as well as better functional fatigue and biocompatibility. However, they are still considered expensive and difficult to process and produce with designed optimal features. In addition, they are quickly fractured when subjected to substantial deformations of more than 30% compared to those developed under intermediate annealing or cold rolling processing. They also exhibit insufficient ductility for practical applications and their maximal martensitic start temperature (Ms) of NiTi-SMAs cannot surpass 360K [41], [42]. The reduced production and increased manufacturing costs of NiTi SMAs due to their high sensitivity to compositional changes in their transformation temperatures (TTs), difficulties in cold forming due to their ordered intermetallic structure, and complexity in joining such materials, have slowed their commercialization and restricted their applicability to a broad range of industries [21], [36], [38], [39], [40], [43].

Fe-based and Cu-based SMAs, such as Fe-Mn-Si and Cu-Al-Ni, are other types of SMAs. They are characterized by their low costs, commercial availability, good workability/weldability, as well as large temperature range for SE [12], [44], [45]. They are also classified as alloys with high transformation temperatures (HTSMA) as their reverse transition begins only at temperatures above 120 °C (390 K) in a stress-free state [46], [47]. The wide range of TTs and flexibility of iron and copper-based SMAs allow for customizing the composition, processing, and optimizing their properties [48]. The common challenges in Cu-based SMAs include brittleness and thermal stability because of large grain sizes and γ1 (Cu9Al4) separation, especially in polycrystalline alloys. Several factors typically enabled the production of components with better strength and ductility such as the development of copper-based single-crystal SMAs, which can provide notable SE, and using AM methods to create complicated shape Cu-based SMAs with rapid cooling rates [49], [50].

SMAs can also be classified according to the type of SME, as illustrated in Fig.2b. The first type is the one-way SME (OWSME) with austenite arrangement recovered. This OWSME type is present in the majority of commonly used SMAs and is defined by the process of recovering the undeformed shape after heating the deformed specimen to a temperature higher than the finishing temperature of austenite. On the other hand, a two-way SME (TWSME) is defined when austenite shape recovers at high temperatures and martensite arrangement returns at low temperatures. Alloys with TWSME are less commonly utilized commercially because of the extra training demand. The TWSME generally yields less reversible strain compared to OWSME for the same alloy [51], [52], [53]. Furthermore, its strain quickly deteriorates, particularly at high temperatures [54]. Typically, OWSMA offers a more reliable and cost-effective solution. The TWSME was generated in Ni-29.7Ti-20Hf alloy using repeated thermal cycling over its martensitic transition under shear stress [55]. The TWSME was also apparent in the Ni-24.28Al-18.53Fe alloy with two phases of microstructure. The TWSME of martensite and γ -phase samples quenched from 11850 °C was as high as 34 % at a strain of 0.33 %, but the TWSME of γ-β phases specimens quenched from 1100 °C was only 8.8 % at a strain of 0.66 percent, rising to 22.5 % when the strain was increased to 1.38 % [56]. This indicates that the TWSME varies significantly with the phase composition and the quenching temperature, as well as the applied strain.

### 2.2. Evolution of Fe-SMAs

Fe-Mn-Si alloy is the first iron-based alloy with the SME to be reported in 1982 by Sato et al [57]. Since then, Fe-SMAs including Fe-Ni-C [58], Fe-Mn-Si [59], [57], Fe-Ni-Co-Ti [60], [61], [62], and Fe-Mn-Al [63] have gained significant interest due to their superior weldability, workability, and cheaper processing than the commonly produced NiTi SMAs. Those Fe-SMAs have a limitation in achieving SE at ambient temperature due to their MTs between different phases (γ to ε [face-centered cubic (fcc) / hexagonal- close-packed (hcp)] and γ to α′ [ body-centered cubic (bcc) / body-centered tetragonal (bct)] are generally none thermoelastic. Although the thermoelastic transition (fcc/face-centered tetragonal (fct)) of Fe-Pd [64] and Fe-Pt [65] alloys are well-known, no SE at ambient temperature has been reported since 1984 [21]. To enhance the performance of Fe-SMAs, extensive research has been carried out. These studies showed that the alloying additives had a significant impact on the SME of these alloys. Forming NbC precipitates in Fe–Mn–Si SMAs achieved higher SME [66], [67]. Precipitation and composition modifications were effective in inducing thermoelastic transformations with lower thermal hysteresis in certain Fe-SMAs [68]. Their recovered strains, on the other hand, were less than 3%, limiting their use to exceptional applications like big-diameter pipe couplings [21], [69]. The inclusion of VC in Fe-SMAs resulted in attractive and promising features for applications of reinforcement and concrete structures. VC-based Fe-SMAs showed good properties with no training after manufacturing them under conventional air melting and casting circumstances [24], [70], [71], [72], [73].

Fe-SMAs can be used in a variety of sectors that demand vast amounts of materials, such as building or construction. The main shortcomings of Fe-SMAs are their non-thermoelastic transformation, which results in extremely high-temperature hysteresis (i.e., greater than 300 °C) and a lack of SE [74]. Although Fe-SMAs have long been known, the development of FeNiCoAlTaB and FeMnAlNi [34] alloys marked a turning point in Fe-based SMA design. In their polycrystalline state, FeNiCoAlTaB alloys are reported to exhibit significantly higher levels of a superelastic strain than nickel-titanium alloys. These Fe-SMAs have a high superelastic strain and are significantly stronger, as a result, they have a better capacity for absorbing energy compared to Cu and NiTi SMAs. They produced a superelastic strain higher than 13% and an ultimate strength of more than 1 GPa, which is nearly double that obtained by Ni-Ti alloys [21]. These Fe-SMAs can be used in strain sensing and energy-damping applications due to their outstanding properties. To induce such properties, the system must be precipitate hardened, cold rolled to impart the [100] orientation, and embedded with the Boron element for grain refinement [21]. Fe-Mn-Al-Ni SMAs have a minor temperature dependency of the critical stress due to a low transformation entropy shift generated by the Gibbs energy effect. Across a temperature range of –196 to 240 °C, the superelastic stress for Fe-Mn-Al-Ni SMA was varying by 0.53 MPa/°C [34]. Later investigations were conducted on single-crystalline FeNiCoAlTa alloy and elaborated on the MTs and the orientation dependency of SE. Under tensile testing, FeNiCoAlTa SMAs showed a strain recovery of 6.8%, 130 MPa of stress hysteresis, and a wide temperature range of 130 °C in a single crystalline state of [001] orientation. On the other hand, high-stress hysteresis values of 350-400 MPa, low superelastic strain values of 2.0 %, and a superelastic temperature range of 55 °C were obtained in single crystals with [111] orientation [75], [76]. The shape memory characteristics were tested for FeNiCoAlNb SMA in single crystal form along [100] orientation using SE tests in tension and compression conditions. The highest SE of around 11.5% was obtained in compression testing. The results demonstrated that SE significantly relies on stress conditions [77]. The aging of the alloy resulted in γ′- and β-phase precipitates. The induced SE was investigated between the starting martensite temperature ($M_s$) and the austenite finishing temperature ($A_f$), as well as between $A_f$ and 323–373 K [78]. The transformation temperatures of FeNiCoAlTiNb SMA were confirmed using thermo-magnetization under superconducting quantum interference at 0.05 and 7 Tesla [79]. The summary of the development and classification of Fe-SMAs is shown in Fig.3 and Table 1.

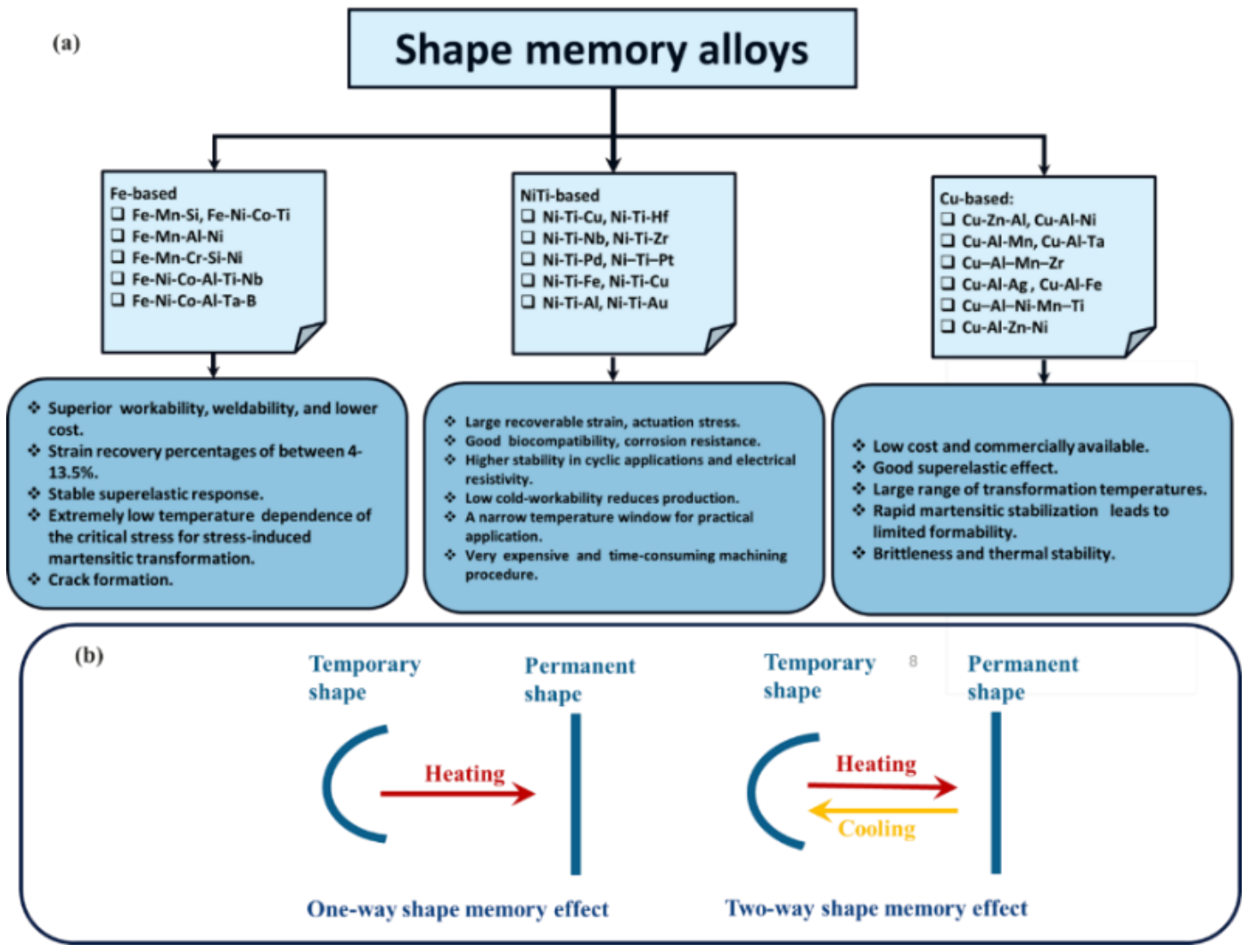

**Fig. 2. Classification and comparison of SMAs. (a) Classifications, advantages, and disadvantages of shape memory alloys and (b) A schematic of one-way SME and two-way SME behaviors of shape memory alloys.**

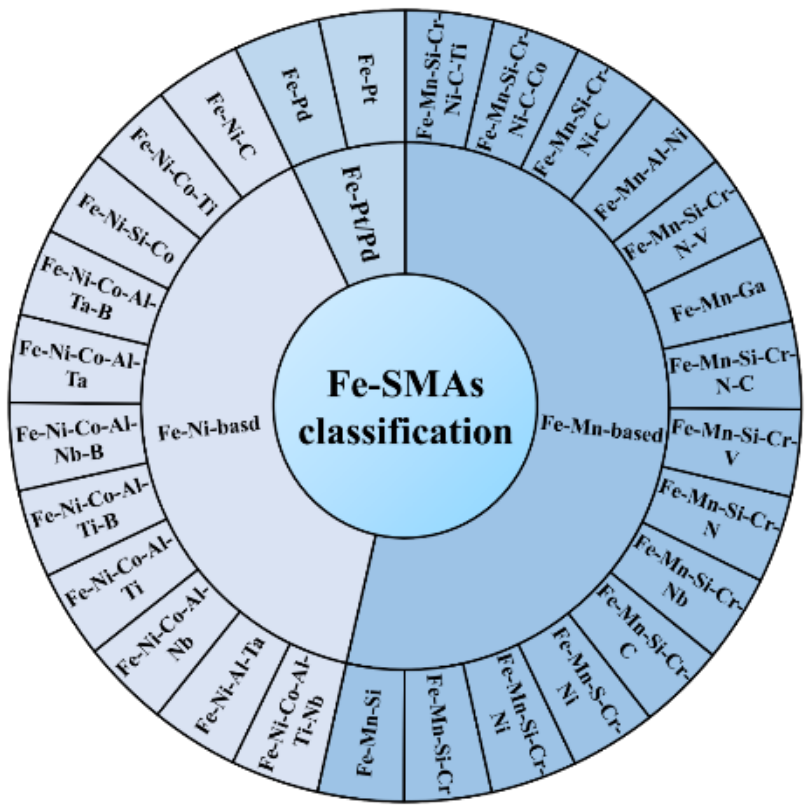

**Fig. 3. Classification of Fe-SMAs showing the concentration of research on Fe-Mn/Fe-Ni based SMAs.**

### 2.3. Mechanisms of Fe-SMAs

SE and SME are the distinctive functional properties that distinguish SMAs from other materials and allow for their usage in a variety of industrial applications [80], [81]. These metallic alloys' remarkable characteristics are due to the MT that is induced under either temperature or stress [82], [83], [84]. The contributing factors to this transformation are the austenite and martensite phases of SMAs. The austenite phase has a symmetric crystalline structure and is often stable at elevated temperatures and moderate stresses. The martensite phase, on the other hand, has a lattice structure with minimal symmetry and is usually stable under high stresses and low temperatures [85], [86]. The phase change occurs among them by rearranging the positions of particles within the solid's crystal structure as shown in Fig.4. Overall, beginning with the twinned martensite,(point D) in Fig.4, that is created after cooling the austenite phase (point G), external loading causes the twinned martensitic microstructure to deform due to the lattice's reorientation and detwinning resulting in so-called detwinned martensite (point F) as illustrated in Fig. 4. The ability of SMA material to accommodate strain in such a way as to produce a twinned martensite structure (point B) is its unique property. As the martensitic structure (detwinned, point C) experiences subsequent thermal flow (heating), the low-symmetry martensitic microstructure then undergoes a second transformation into the well-ordered, more stable, and high-symmetry austenite structure (point A). This microscopic distortion is the key to the recovery of the material's original shape [87], [88], [89], [90]. In the SE effect (Point G to F), external stress causes the material to transform from the high-symmetry, elevated-temperature, well-ordered austenite phase to the lower-symmetry martensite phase. This change takes place in a temperature interval where the martensitic phase is unstable without loading; hence, upon unloading, the phase reverts to the original austenitic form, reversing the deformation. Numerous iron alloys have been the subject of SE research, and various strategies have been developed to improve them [21], [91], [92], [93], [94], [95]. It can be concluded that the ability to reverse the austenite/ martensite interface is the most significant requirement for inducing the unique features of the Fe-SMA system.

**Table 1. Historical growth of Fe-SMAs with their specific compositions over the last decades**

| Year | Fe-shape memory alloy | Reference |
|---|---|---|
| 1974 | Fe- (25 or 27) Pt | [96] |
| 1980 | Fe-30Pd | [65], [88], [97] |
| 1982 | Fe-30Mn-1Si (single crystal) | [57] |
| 1984 | Fe-30Mn-6Si (single crystal) | [59] |
| 1985 | Fe-31Ni-0.4C<br>Fe-27Ni-0.8C | [98] |
| 1986 | Fe-32Mn-6Si | [99] |
| 1990 | Fe-28Mn-6Si-5Cr<br>Fe–20Mn–5Si–8Cr–5Ni<br>Fe-16Mn-5Si-12Cr-5Ni | [66] |
| 1992 | Fe-31.9Ni-9.8Co-4.1Ti<br>Fe-14Mn-6Si-9Cr-6Ni | [61], [100], [101] |
| 2001 | Fe-28Mn-6Si-5Cr-0.5(Nb, C)<br>Fe-28Ni-12Co-4.75Ti (single crystal) | [67], [68], [102] |
| 2002 | Fe-29Ni-18Co-4Ti | [69] |
| 2004 | Fe-28Mn-6Si-5Cr-1(V, N) | [103] |
| 2006 | Fe-28Ni-6Si-10Co | [104] |
| 2009 | Fe-17Mn-5Si-10Cr-4Ni-1(V, C)<br>Fe-28Mn-29Ga | [70], [72], [105] |
| 2010 | Fe-28Ni-17Co-11.5Al-2.5Ta-0.05B | [21] |
| 2011 | Fe-34Mn-15Al-7.5Ni | [34] |
| 2012 | Fe-28Ni-17Co-11.5Al-2.5Ta (single crystal) | [36] |
| 2013 | Fe-16Mn-5Si-10Cr-4Ni-1(V, N)<br>Fe-28Ni-17Co-10Al-2.5Nb-0.05B | [106], [107] |
| 2014 | Fe-28Ni-17Co-11.5Al-2.5Ti-0.05B<br>Fe-28Ni-17Co-11.5Al-2.5Ti (single crystal) | [108], [109] |
| 2015 | Fe-28Ni-17Co-11.5Al-2.5Nb (single crystal)<br>Fe-28Ni-17Co-11.5Al-2.5Nb-0.05B (single crystal)<br>Fe-28Ni-17Co-11.5Al-2.5Ti (single crystal)<br>Fe-34Mn-15Al-7.5Ni (single crystal) | [110], [111], [112] |
| 2016 | Fe-27.9Ni-17·2Co-9.9Al-2.4Nb (single crystal)<br>Fe-28Ni-17Co-11Al-2.5X (0.05% B) (X = Ti, Nb(B), (Ti + Nb) B) (single crystal) | [113], [114] |
| 2017 | Fe-13.51Mn-4.82Si-8.32Cr-3.49Ni-0.15C<br>Fe-34Mn-15Al-7.5Ni-1.5Ti<br>Fe-34Mn-16.5Al-7.5Ni | [115], [116] |
| 2018 | Fe-30Ni-15Co-10Al-2.5Ti (single crystal)<br>Fe-28Mn-(28+x) Ga (x = 1, 2, and 2.5)<br>Fe-13.51Mn-4.82Si-8.32Cr-3.49Ni-0.15C | [117], [118], [119] |
| 2019 | Fe-34Mn-7.5Ni-13.5Al<br>Fe-Mn-Al-Ni-X (X = Ti, Cr), (1.5 Ti, 3Cr)<br>Fe-34.8Mn-13.5Al-7.4Ni | [120], [121], [122] |
| 2020 | Fe-28Ni-11.5Al-xTa (x=0.5, 1, 1.5)<br>Fe-15Mn-5Si-8Cr-5Ni<br>Fe-15Mn-5Si-8Cr-5Ni-0.1C<br>Fe-15Mn-3Si-8Cr-5Ni-0.1C-2Co<br>Fe-34Mn-13.5Al-3Cr-7.5Ni<br>Fe-20.7Mn-9.4Cr-10.5Si-4.8Ni | [91], [123], [124], [125] |
| 2021 | Fe-17Mn-5Si-5Cr-4Ni-0.3C-1Ti<br>Fe-28Ni-17Co-11.5Al-1.25Ti-1.25Nb | [126], [127] |
| 2022 | Fe-34.72Mn-13.35Al-7.67Ni-1.54Cr<br>Fe-35.4Mn-14.8Al-7.9Ni-0.8Nb-0.07C | [128], [129], [130] |
| 2023 | Fe-32.9Mn-14.1Al-7.3Ni<br>Fe–28Mn–6Si–5Cr<br>Fe–20Mn–6Si–8Cr–5Ni-xTi (x = 0, 0.5, 1, 2)<br>Fe–17Mn–5Si–10Cr–4Ni-1(V, C) | [131], [132], [133], [134], [135], [136] |
| 2024 | Fe-33Mn-16Al-8Ni<br>Fe-34Mn-15Al-7.5Ni<br>Fe-18.80Mn-6.15Si-8.89Cr-4.81Ni<br>Fe-17Mn-5Si-4Ni-10Cr | [137], [138], [139], [140] |

The transformation can occur from fcc to hcp, as in the FeMnSi alloy, or from bcc to bct, as in the FeNiCoTi and FeNiCoAlX alloys [57], [74], [141]. The stress-induced transition of austenite to martensite (fcc to hcp) and its reversal (hcp to fcc) through heating is the basis of shape memory features in Fe-SMAs. The migration of Shockley partial dislocations via non-adjacent close-packed planes is investigated to be responsible for the creation of the martensite from the austenite [101], [142]. The typical phase transformation of Fe-SMAs in the stress-free state is fully described using four different temperatures. The starting ($M_s$) and finishing temperatures ($M_f$) of forward MT, as well as the starting ($A_s$) and finishing temperatures ($A_f$) of backward transformation. The martensitic forward transformation can be reversed by heating the material while it is in the martensitic region ($T < M_f$). The following is how the temperatures are organized: $M_f < M_s < A_s < A_f$. There are no phase shifts when the temperature is changed at the range (Ms < T < As), and both phases can exist at ($M_s < T < A_s$) [23], [72], [143], [144]. The shape memory characteristics can be fully represented using these four transition temperatures and the principle of self-accommodation. The MTs typically have thermal hysteresis (TH), which indicates that the forward and backward transformations do not proceed at the same temperature. TH in Fe-SMAs is often described as the difference between the $A_f$ and $M_s$. TH is caused by irreversible processes, such as dissipation of strain energy and resistance of friction to interfacial motion, during martensitic transformation. Large TH can produce substantial and steady recovery stresses, making Fe-SMAs with these characteristics, like Fe-Mn-Si SMAs, more suitable for structural applications [145]. Small TH, on the other hand, is favorable in other different applications, hence numerous experiments have been conducted to manufacture SMAs with thermal hysteresis of near-zero or as little as 0.4 °C [21], [146], [147], [148], [149], [150], [151]. Copper doping of Fe-Ni-Co-Ti SMA showed a significant decrease of the TH width to $\Delta T \approx 60$ K, maintaining the Curie temperature at about 300 °C [152].

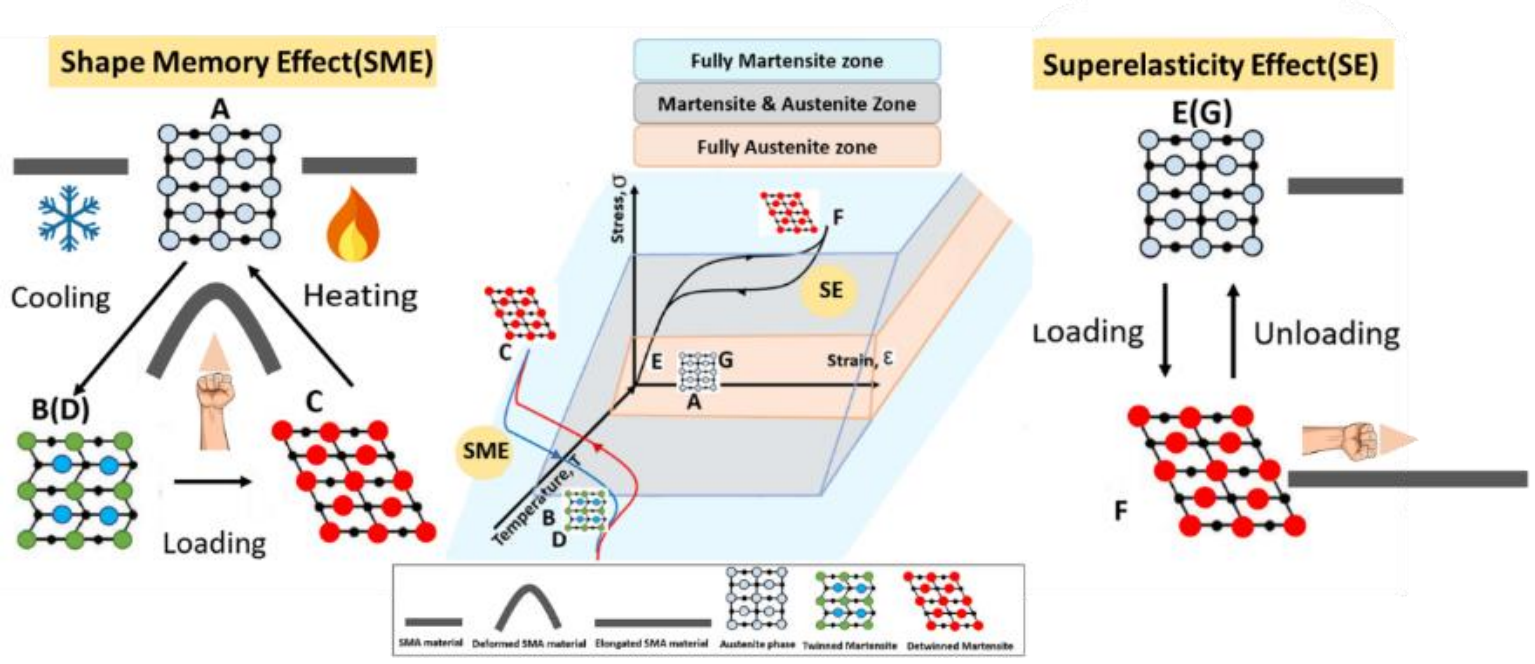

**Fig. 4. Diagram showing the entire cycle Fe-SMA mechanism. The phase transformation of austenite(γ) to martensite (ε) and its reversal is illustrated with a microscopic demonstration of SE and SME.**

## 3. Production, characterization, testing, and modeling of Fe-SMAs

### 3.1. Production routes of Fe-SMAs

Fe-SMAs are produced as semi-finished items, such as rods, wires, ribbons, and tubes, or as finished products, such as wire actuators and helical springs, in a variety of shapes and forms. They are manufactured in commercial quantities by several specialized companies all around the world [55], [153], [154], [155], [156], [157], [158]. Melting and casting (Fig.6 (a)) are often the first steps in the preparation of most Fe-SMAs. Casting is commonly applied using either vacuum arc melting (VAR) or induction melting (VIM). These are specialized processes used to ensure that the metals are evenly mixed, and impurities are minimized in the alloy. Other subsequent procedures such as heat treatment, hot rolling, and drawing are followed to produce parts with variable shapes and minimal defects. Heat treatment is mainly applied for the homogenization process, followed by hot working and further treatments to enhance the mechanical and thermomechanical properties of the Fe-SMAs [19], [72], [126]. After the preparation and treatment of the Fe-SMAs samples, thermomechanical, and other required properties can be determined. A summary of the different steps involved in Fe-SMAs conventional production is shown in Fig. 5.

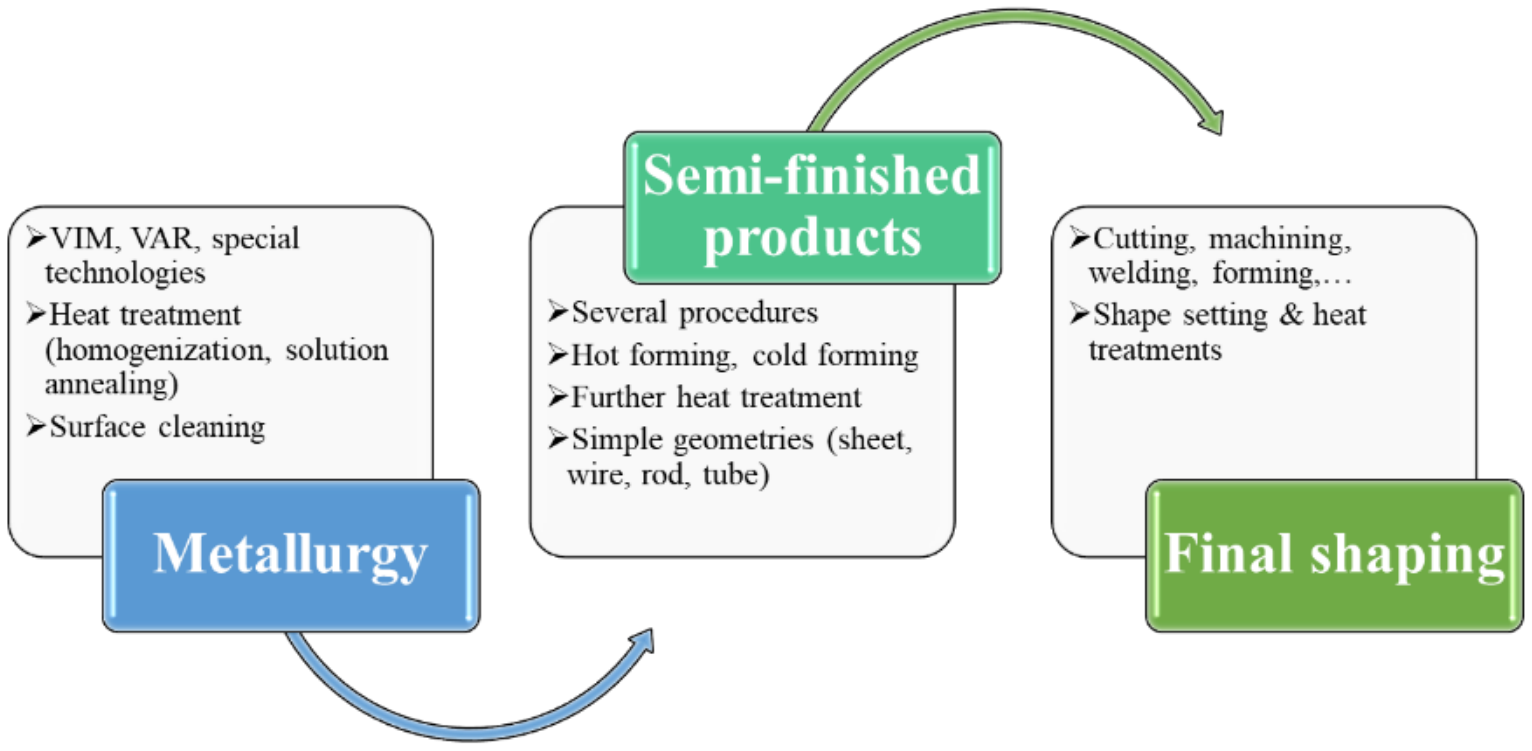

**Fig. 5. Conventional production procedures of Fe-based SMAs.**

Fe-17Mn-5Si-10Cr-4Ni SMA containing VC was fabricated using standard air-casting facilities followed by other heat treatments and characterizations to evaluate its properties [72]. Fe-17Mn-5Si-5Cr-4Ni-1Ti-0.3C SMA was produced using casting, followed by forging and hot-rolling to produce SMA with the intended shape [126]. A good summary of Fe-SMAs manufactured using conventional and modern methods is illustrated in Fig. 6 and a description of the fabrication processes used to synthesize Fe-SMA is listed in Table 1.

Alternatively, mechanical alloying (MA) has been used to produce Fe-SMA, Fig. 6(b) illustrates the steps of MA that involve solid-state reactions between powder particles brought on by high-energy interactions. MA is considered a powder metallurgy technology that has several benefits for the manufacture of Fe-SMA products, due to its ability to manufacture the components in almost net shape. This reduces the supplementary machining needed to shape the eventual product. The thermo-mechanical and mechanical properties of Fe-SMA produced by MA could be equivalent to that of conventional casting [143], [159]. On the other hand, the use of conventional methods in Fe-SMA production has some limitations. Even though the VIM method can achieve good chemical homogeneity using electromagnetic stirring, the presence of titanium in the processed alloy can react with the graphite crucibles which results in TiC formation. These particles will modify the alloy composition and, consequently, the TTs [160], [161], [162]. Chemical homogeneity of the ingots in the VAM technique needs multiple remelting processes due to insufficient stirring. The alloy absorbs oxygen and carbon during these remelting processes, which again has an impact on the TTs [163]. Complex structure fabrication using conventional techniques is time-consuming and so challenging. During manufacturing, excessive tool wear might be produced. In addition, it will be extremely difficult to create polycrystalline structures if the alloy is even slightly brittle [164], [165].

These issues and others can be overcome by using additive manufacturing (AM) techniques which eliminate the necessity for tooling and enable the manufacturing of SMA components of intricate geometries directly from CAD models. The manufacturing of Fe-SMAs has been improved with the new advancements in AM to create SMAs with improved characteristics [166], [167], [168], [169], [170]. Fig. 6(c) lists the steps followed to fabricate Fe-SMA using AM. AM technologies represent considerable potential for improving the manufacturing of Fe-SMAs. They have been recently used for several Fe-SMA production experiments [167], [171], [172], [173], [174], and it is anticipated that they will become widely used in the production of Fe-SMA. These studies demonstrated the viability of producing Fe-SMAs with improved characteristics for Fe-Mn-Si and Fe-Mn-Al-Ni SMAs. Recent AM technologies used for Fe-SMA are based on laser powder bed fusion (LPBF).

AM can produce parts with higher quality and performance than those produced using vacuum induction melting and other conventional techniques [175], [176], [177]. NiTi SMA produced by Electron beam melting (EBM) displayed higher purity than vacuum induction melting manufactured samples [178]. A separate section (6) is designed to analyze all reported studies about the AM of Fe-SMAs to clearly understand how the AM of Fe-SMAs is implemented, the equality of the produced parts, the types of Fe-SMAs that can be produced using AM techniques, and how these technologies can be applied further and generalized for a large number of Fe-SMAs with enhanced properties.

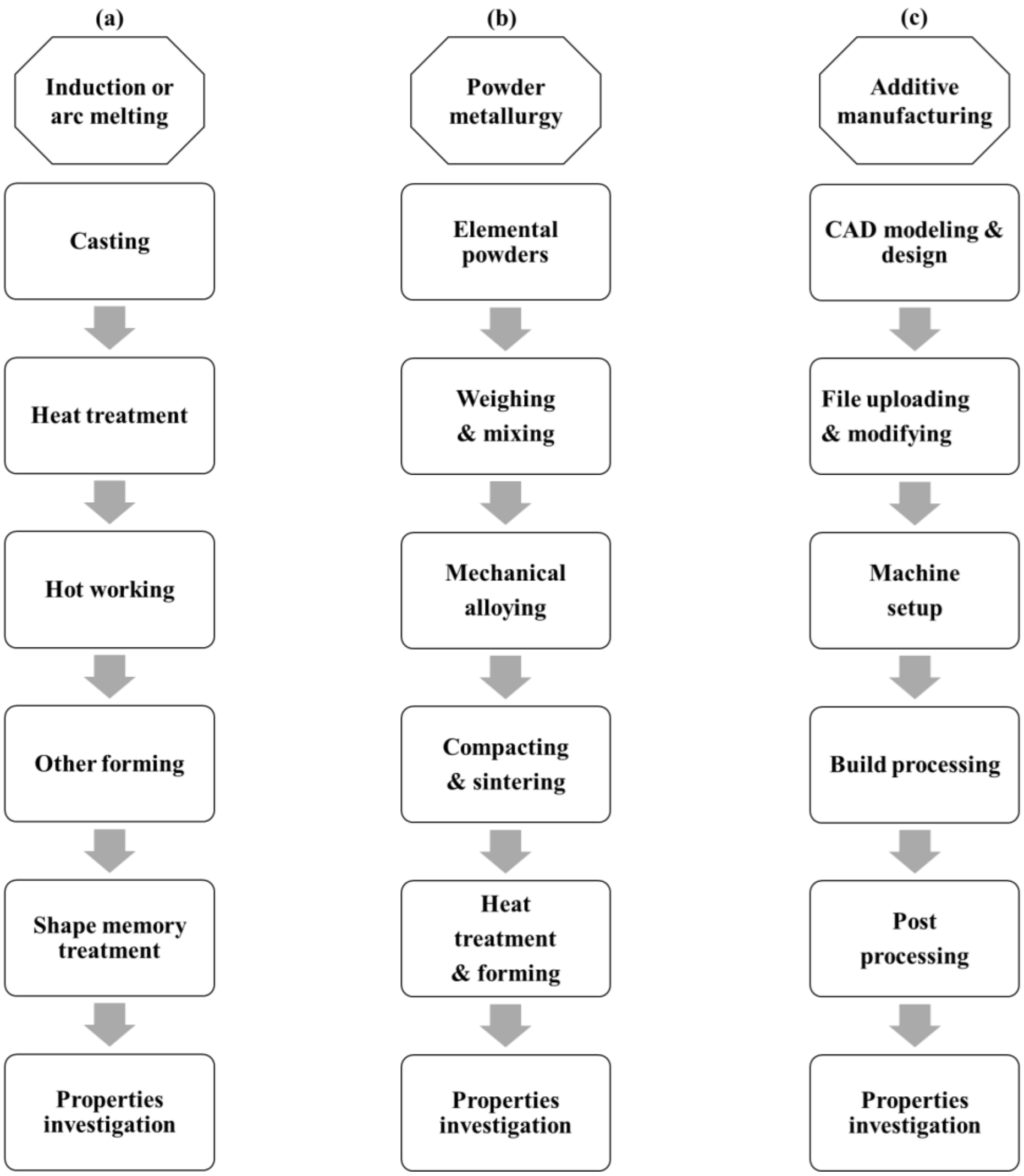

**Fig. 6. Production techniques. A flowchart of Fe-SMA production steps. (a) Casting processes. (b) Mechanical alloying. (c) Additive manufacturing.**

## 3.2. Characterization and testing of Fe-SMAs

### 3.2.1. Preparation of Fe-SMAs

Characterization is a crucial procedure in which several analytical techniques, methodologies, and tools are used to examine, measure, and establish Fe-SMA's chemical, microstructural, and physical properties. Fe-SMA is first produced using either modern processes like AM or more traditional methods including melting/casting and powder metallurgy [119], [179], [180]. Further heat treatments are carried out to homogenize Fe-SMA and enhance its properties. Hot forging, rolling, and other shaping methods are applied to the produced Fe-SMA to develop samples with the required geometry and dimensions [119]. The samples required for any characterization technique or test are cut from the as-built Fe-SMA sheets or blocks and machined into the required shape using electrical discharge machining [181]. Inductively coupled plasma optical emission spectroscopy is a typical technique to identify Fe-SMAs' elemental composition. It is an ionization source that completely breaks down a specimen into its component elements before transferring those elements into ions. It is usually made of argon gas; the plasma is created by "coupling" energy to it via an induction coil [182], [183], [184]. Energy-dispersive X-ray spectroscopy coupled with field-emission scanning electron microscopy (FESEM) can also be utilized to investigate the chemical composition of Fe-SMA. Before the microstructural evaluation and conducting the surface analysis, the samples are grounded using SiC sheets having grit sizes of 180-3000 or 400-5000, followed by polishing and etching. A Kalling 2 and Nital are common chemicals used as an etching solution for Fe-SMAs. Polishing is performed using a 50–100 nm colloidal silica suspension neutralized by H2O2 [185], [186], [187].

**Table 2. Manufacturing techniques of Fe-SMAs**

| Fe-SMA | Method description | Reference |
|---|---|---|
| Fe-17Mn-5Si-5Cr-4Ni-0.3C-1Ti | The SMA was initially melted under a vacuum in an induction melting pot before being cast into a graphite crucible with a cylindrical shape. After casting, the ingot was heat-treated for homogenization. After that, the alloy ingot underwent forging and hot rolling. | [126] |
| Fe-17Mn-5Si-10Cr-4Ni-1(V, C) | Under normal atmospheric settings, the alloy was induction melted and cast into a mold with a diameter and height of 90 and 300 mm, respectively. The ingot was cast using a feeder head and exothermic anti-piping powders to reduce cavities. | [72] |
| Fe-17Mn-5Si-10Cr-4Ni | LPBF was employed to manufacture this SMA. Gas atomization under an Argon environment was used to create the powder. Initial optimization focused on the LPBF processing parameters before producing the final SMA. | [171] |
| Fe-13.51Mn-4.82Si-8.32Cr-3.49Ni-0.15C | Induction melting in a vacuum was used. Homogenization was used before the ingot was hot-forged and subsequently hot-rolled. Solution annealing and water quenching were both used to process the rolled samples. | [115] |
| Fe-30Mn-6Si | MA was performed on elemental powders of Fe, Si, and Mn (99.9% purity) in a stainless-steel vial using numerous stainless-steel balls. A planetary ball mill was used to perform the MA. The alloyed powder was then compressed with a compression pressure of 20 MPa and sintered for 10 minutes at 900 °C under a vacuum (2 Pa). | [159] |
| Fe-33Mn-17Al-6Ni-0.15C | The alloy was produced by melting industrial raw materials in an induction furnace with argon gas and then casting it into ingots. These ingots underwent a 24-hour argon-filled homogenization process at 1000 °C. | [188] |
| Fe-19.4Mn-5.9Si-9.2Cr-5.1Ni | At the CEIT technological center, powders of the alloy were prepared using gas atomization and sieved before being inductively coupled plasma (ICP) analyzed. The powders were subsequently subjected to laser metal deposition treatment through a KukaKR30 with a ytterbium source using a scanning speed of 665.7 mm/min, laser power of 1150 W, a spot size of 2 mm, and powder flux of 9 g/min. | [189] |

### 3.2.2. Surface, phase, and microstructural analysis of Fe-SMAs

SEM and optical microscopy (OM) are performed to investigate the surface morphology and the microstructure of Fe-SMAs. SEM and ImageJ software can be used to analyze and measure the volume fraction of the precipitates that are formed in Fe-SMAs [190]. Electron backscatter diffraction (EBSD), a microstructural-crystallographic characterization method based on the SEM, can analyze the microstructure, texture, crystal orientation, grains, phase, or strain of the samples. A perchloric acid and ethyl alcohol electrolytic solution is used to electropolish the Fe-SMA samples before this evaluation. The EBSD data are analyzed using the OIM Analysis™ software. This program can calculate the overall length of a variety of boundaries, such as grain boundaries [187], [191], [192]. OM is used in metallographic observation to identify the defects in metals and determine the grain boundaries. Additionally, to differentiate the ε, α', and γ phases using an optical microscope, an optical color etching procedure is applied to the samples in a solution comprising 1.2 g $K_2S_2O_5$ and 0.5 g $NH_4HF_2$ in water. γ phase looks brown, α' is dark, and ε seems white except that thin ε plates show as black lines in color optical pictures [193], [194]. Using electron channeling contrast imaging (ECCI), the development of the martensite plate during strain is observed in situ [195]. The ECCI samples were initially ground with SiC sheets up to 4000 grit, polished for about 10 minutes with a diamond suspension (about 3 μm), and then polished precisely with a 50-100 nm colloidal silica liquid diluted by $H_2O_2$ [196]. For the identification of the deformation twin in the Fe-SMA, transmission electron microscopy (TEM) can be used. A twin jet polisher physically grinds and polishes the Fe-SMA specimens in a solution of sulfuric acid and methanol (1:4) to prepare foils for TEM studies [194]. To examine the microstructure of the stress-induced martensite (ε) plates, TEM characterization is also performed [196]. The X-ray diffraction (XRD) equipped with a diffractometer with Cu Kα radiation is performed on samples to examine the phase constituent and crystal structures of Fe-SMAs, where 10% HF and 90% H2O2 solution is used to etch the specimens after mechanically grinding them [187]. Fe-Mn-Si-Cr-Ni SMA samples are electro-polished using a 12.5% perchloric acid and 87.5% ethanol solution to eliminate surface stress before performing the XRD examination [124]. The volume fraction of the different phases in Fe-SMA such as γ-austenite, α′-martensite, and ε-martensite can be estimated using XRD, as well [124]. To obtain the TTs of the Fe-SMA sample, a differential scanning calorimeter (DSC) is used. The instrument is usually standardized for enthalpy and temperature, by considering some samples as standards such as zinc and indium samples [179], [191]. Superconducting quantum interference devices are used under fields of 0.05 and 7 Tesla to evaluate the TTs and the thermo-magnetization properties of FeNiCoAlTiNb SMA [197].

### 3.2.3. Shape memory testing of Fe-SMAs

To study the SME and damping behavior of Fe-SMAs, DSC and dynamic mechanical analyzer (DMA) can be applied [179]. The bending test for SME is carried out for Fe-SMA samples to evaluate the SME, as shown in Fig.7 (a), and (b). Before measuring the SME, the specimens should be cleaned in a diluted acidic solution to remove the oxidation layer [192]. The procedure of the bending test is first bending the sample to $\theta_i = 180°$, keeping the sample in this orientation for a specific time, and then releasing it. After unloading, the sample recovered with angles of $\theta_e$. Then, the specimen is heated at a specific temperature so the deformed shape will be partially recovered, and the recovery angle $\theta_r$ is measured when the samples are cooled to ambient temperature [187], [192]. The pre-strain is evaluated using this equation: $\varepsilon = t/(D+t)$, where D and t are the bending diameter, and thickness of the specimen, respectively. It is considered the greatest tensile strain at the outside edge of the specimen thickness. The residual strain $\varepsilon_r$ after recovery annealing is then recorded and utilized to determine the strain recovery ratio (SRR), which can be calculated as, $SRR = 100 \times (\varepsilon - \varepsilon_r)/\varepsilon$ [198].

Furthermore, pre-deformation was evaluated by $\varepsilon = (t/D) \times 100\ \%$, where D is the diameter of bending, t is the thickness of the sample, and ε is the pre-distortion [124]. To evaluate the SME using an extension test, a Vickers hardness tester is utilized to generate two indentations on each dog bone specimen's surface. The distance between them, $L_o$, is then precisely recorded using OM. The separation between the two points on each specimen, $L_1$, is then recorded accurately again after the indented specimens have undergone various degrees of tensile deformation at 293 K (i.e., ambient temperature). Finally, the distance between marked points, $L_2$, is calculated accurately after the distorted specimens have been recovered and annealed at 773 K for 15 min to regain their original shapes. The formula for calculating the SRR is $SRR = 100 \times (L_1 - L_2)/(L_1 - L_o)$ [198]. The shape recovery rate η) is determined using $\eta = (\theta_r / \theta_i - \theta_e) \times 100$, where $\theta_r$ is the strain recovery angle and $\theta_i - \theta_e$ is the remaining angle after relaxation [199].

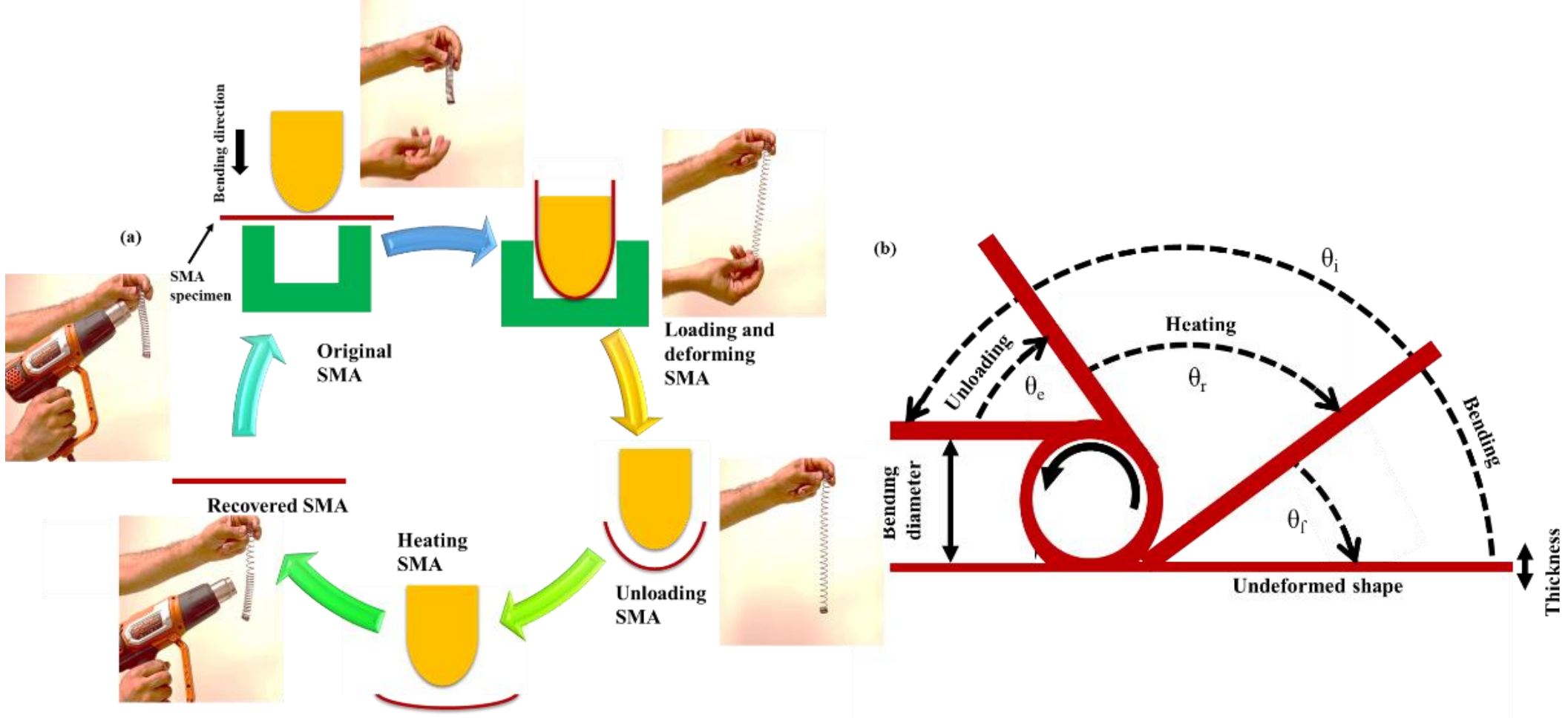

**Fig. 7. Testing procedure of Fe-SMA. (a) The procedure of testing and recovery of shape memory property. (b) Schematic diagram showing the evolution of radii during the bending test of the shape memory effect.**

**Table 3. Characterization and testing techniques summary of Fe-SMAs**

| Technique | Investigation | Reference |
|---|---|---|
| Bending test, and extension test | SME | [124], [187], [192], [198] |
| Inductively coupled plasma spectroscopy, energy-dispersive X-ray spectroscopy, spectrometer, X-ray fluorescence spectroscopy, optical emission spectrometry | Chemical composition | [183], [187], [191], [192] |
| Scanning electron microscopy<br>Electron backscatter diffraction<br>Optical microscope<br>Transmission electron microscope | Microstructural evolution | [124], [185], [187] |
| X-ray diffraction | Phases, microstructure, and precipitates | [124], [185] |
| Differential scanning calorimetry | Enthalpy and transformation temperature | [88], [120], [186] |
| Electrical discharge machining, Mo filament cutter | Sample cutting | [120], [198] |

### 3.3. Modeling and simulation of Fe-SMAs

Various studies focused on creating continuum formulations that can predict how SMAs will behave. The continuum state variable models mimic the stress-strain response by using yield functions related to the MT, the evolution of martensite volume percentage, and the associated flow rules [87], [200], [201], [202], [203], [204], [205]. More modeling work has been done to investigate the plastic strain buildup in SMAs. Multi-dimensional empirical models are provided to represent how the plastic and transformational strains evolve under cycling. The driving force of plastic strain agglomeration is the nucleation of slip dislocations at the interface of austenite/martensite when strains and stresses are relatively high, and the main contributing factor in this situation is the slip's critical resolved shear stress [206], [207], [208], [209], [210], [211].

Modeling techniques have also been recommended to visualize the recoverable strains caused by martensite reorientation. By including the dissipation potentials in their derivation, these models combine transformation and martensite reorientation in SMAs to accurately reflect the SME, SE, and deformation of martensite. The twinning stress controls the acquired rules and functions of these models for the martensite reorientation since it has a major impact on the twin interfaces' nucleation and motion, whereas the critical transformation stress governs the yield function of transformation [212], [213], [214]. Micromechanical models can also be based on the concept that critical transformation stress is what induces phase transition. To simulate the continuum-shape memory response, these models consider various variant-variant interactions as well as microstructurally informed parameters such as transformation shear magnitudes. These mesoscale formulations concluded that the polycrystalline response is mainly based on the synergy of transformation and slip [215], [216], [217], [218], [219].

The Modeling of Fe-SMAs starts by determining the TTs. An artificial neural network (ANN) model can be built on a gradient descent learning algorithm to investigate the start temperatures of austenite (As) and martensite (Ms) [86], as shown in Fig. 8. The model was created, trained, and verified using 85 Fe-SMAs reported in the literature. Ten input parameters were used in the model including the weight percentage of Fe, Mn, Ni, Si, Cr, Al, and Cu as well as hot rolling, quenching, and homogenizing temperature. ANN model can accurately predict the As and Ms temperatures in the range of considered input parameters and can show a strong agreement with the experimental data [86].

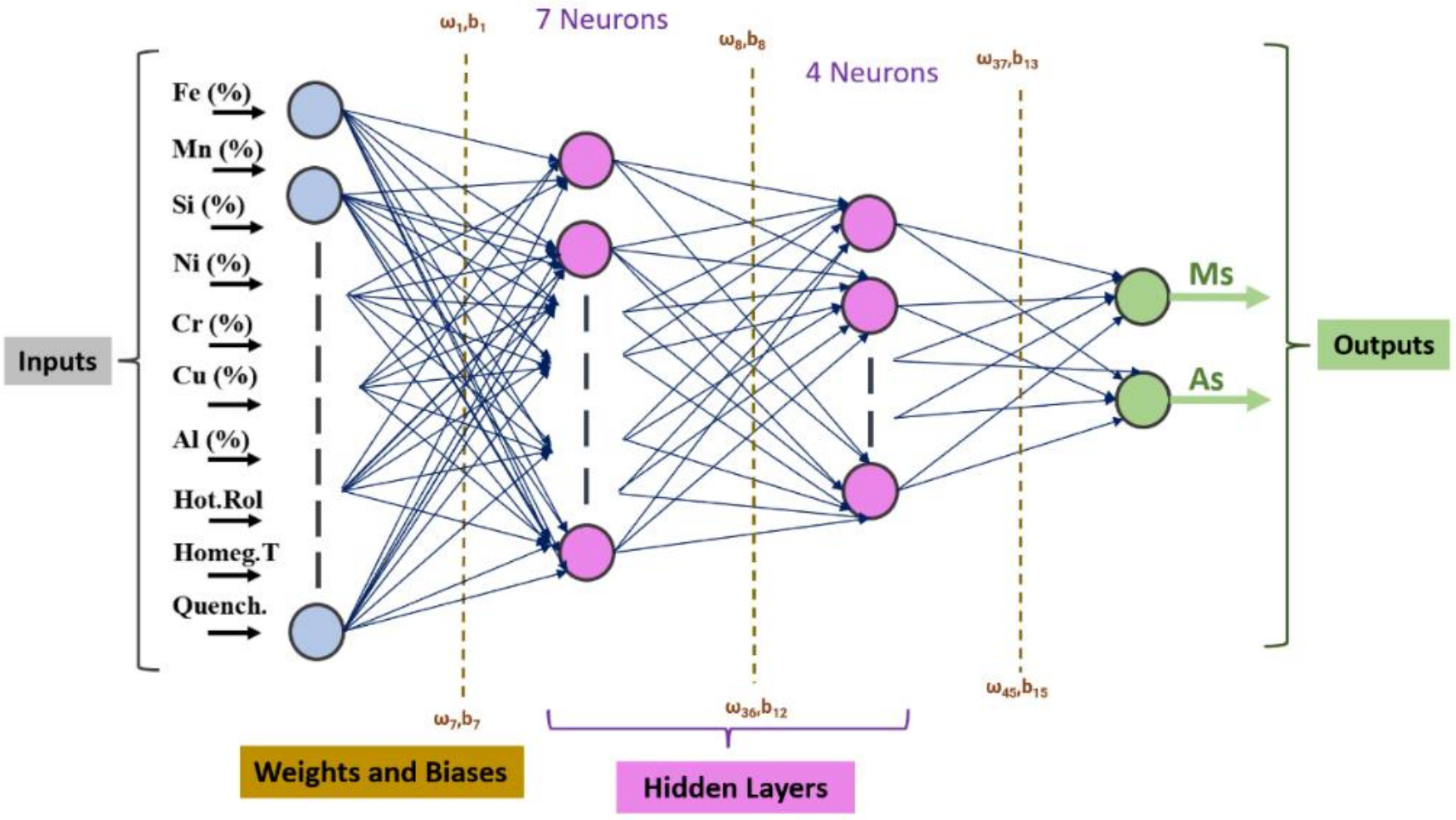

**Fig. 8 . Illustration of suggested ANN model of Ms and As temperatures.**

Additionally, a model was proposed to assess how plastic slip and MT affect the thermo-mechanical behavior of Fe-SMAs [220]. The adopted formulation was established on a condensed micromechanical description. The corresponding homogenous effect was considered on a representative volume element to determine the macroscopic behavior as shown in Fig. 9 (a). The model mainly considered the interactions between the plastic slip and MT mechanisms and their effects (Fig.8 (b)). The concepts of plastic gliding and the martensite volume fraction are introduced as two macroscopic internal variables. The experiments' validation at different steady temperatures can enable the identification of the material characteristics and the calibration of the suggested methodology for varied homogenous loadings. A good degree of agreement can be achieved between experimental and numerical data [220].

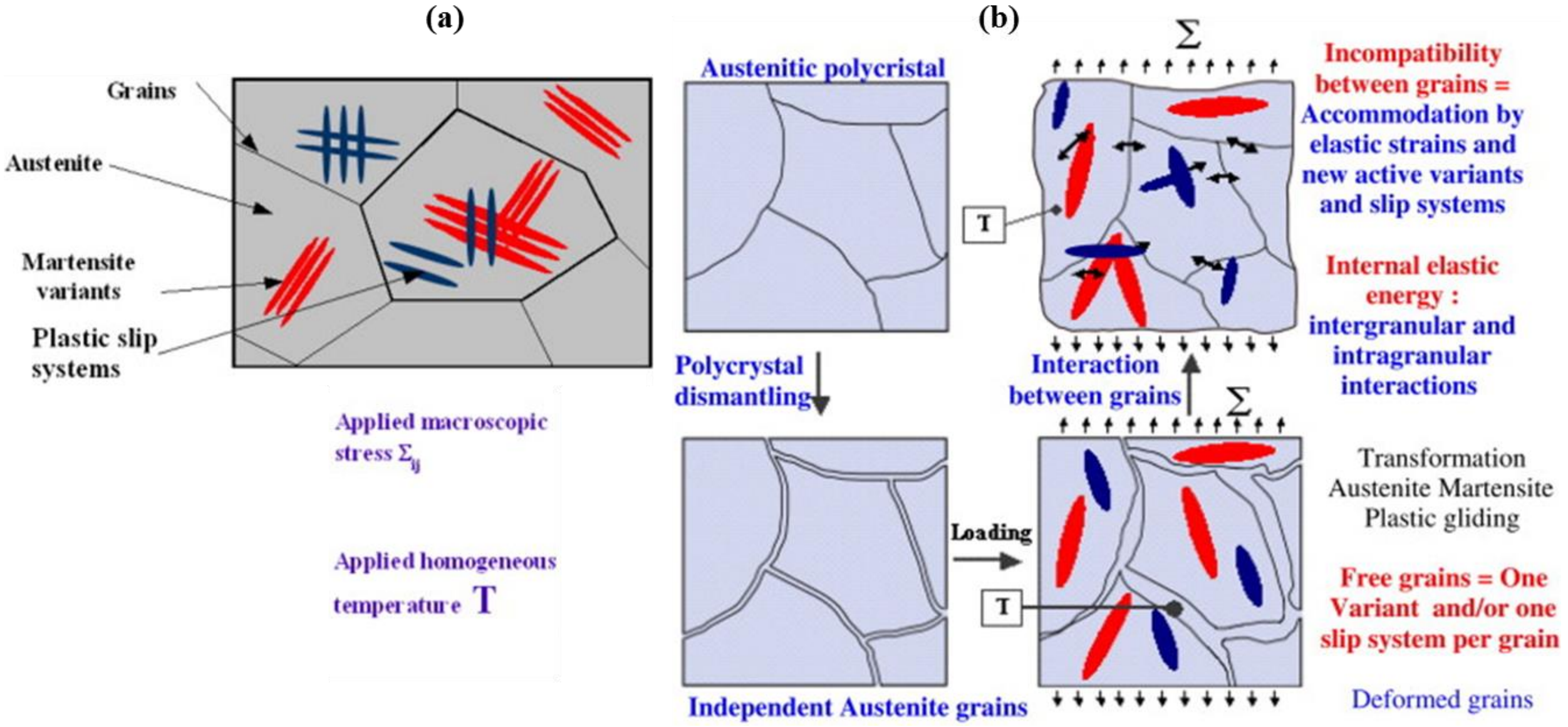

**Fig. 9. Representative of volume element and slip interaction. (a) An element of a polycrystalline Fe-SMA is used in the modeling. (b) Interactions between slip systems, martensite variations, and grains** [220]**.**

A finite element numerical tool was adapted for a Fe-SMA structural analysis [221], [222]. Mechanical and chemical variables as well as nonlinear interaction values related to inter- and intragranular incompatibilities were considered during the model derivation. The provided model successfully explained the complicated thermomechanical loading routes. The nonlinear stress-strain graph was precisely simulated and compared to the experiments during loading.

The findings display a strong correspondence with the conducted experiments as shown in Fig. 10. Small-strain thermomechanical models developed to address the link between phase transformation and plastic slip are not appropriate for higher loading. As a result, a finite-strain constitutive model for Fe-SMAs that includes such thermomechanical coupling can be developed [223]. The small-strain model was then expanded within a finite-strain thermodynamic framework to illustrate huge strains primarily brought on by plastic hardening in Fe-SMA. It was created on the concept of a total Lagrangian formulation with a local multiplicative split of the deformation gradient into elastic and inelastic components. The inelastic deformation gradient is also separated into plastic and transformational components [223].

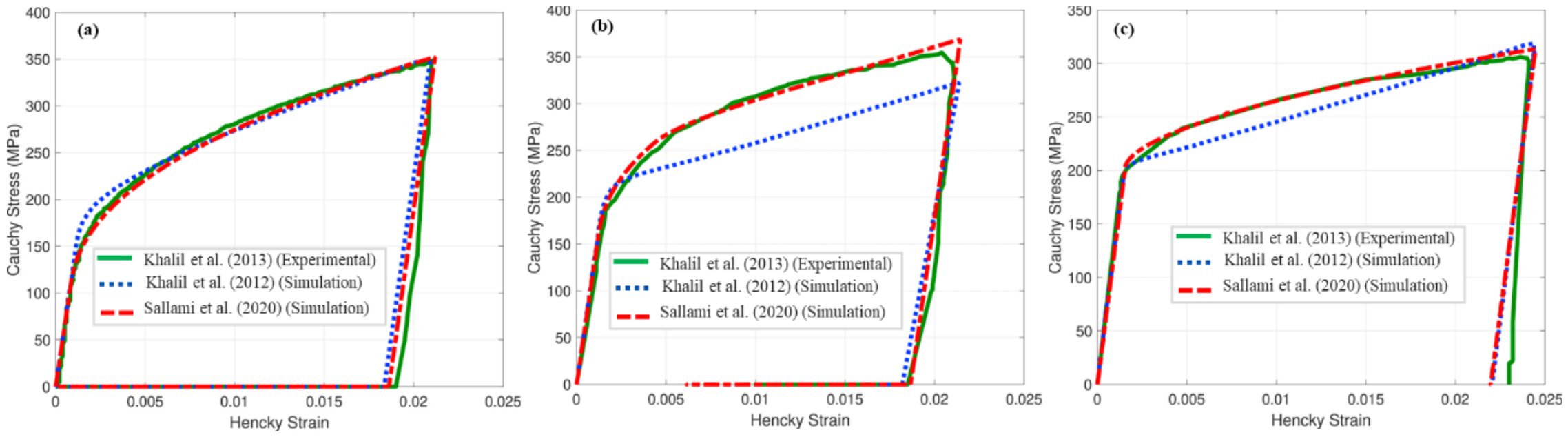

**Fig. 10. Simulation and experimental comparison. Simulation compared to experimental results of the stress-strain behavior under tension for Fe-Mn-Si-C SMA. (a) 20 °C. (b) 50 °C. (c) 130 °C [223].**

Furthermore, with the help of calculations for double shear in density functional theory, the bcc-fcc transformation in Fe-Mn-Al-Ni SMAs was investigated [87], as shown in Fig. 11. The dislocations' elastic interactions and transformation shear energy were included in energy expression that was developed to determine the fcc martensite formation stress. It was considered that the bcc to fcc transformation's double shear mechanism is achieved by the movement of two pairs of dislocations, (a0 /8) <110> and (a/6) <112>, on multiple planes in a bcc crystal, the connection of which generates the fcc crystal, shown in (Fig. 11). The stresses for the slip, twinning, and bcc-fcc transformation were reported to be 191 MPa, 201 MPa, and 335 MPa, respectively. These findings were remarkably consistent with the experimental data. The higher slip resistance combined with the low Clausius- Clapeyron slope (dσ/dT = 0.53 MPa/°C, [111]) of Fe-Mn-Al-Ni alloy was confirmed to be significant for the recoverability of the transformation because the higher slip resistance is required to reduce activation and the number of dislocations at transformation interface. The Bogers-Burgers double shear mechanism is preferred over the "traditional" Bain deformation since it was shown to continue with a substantially lower energy barrier [111].

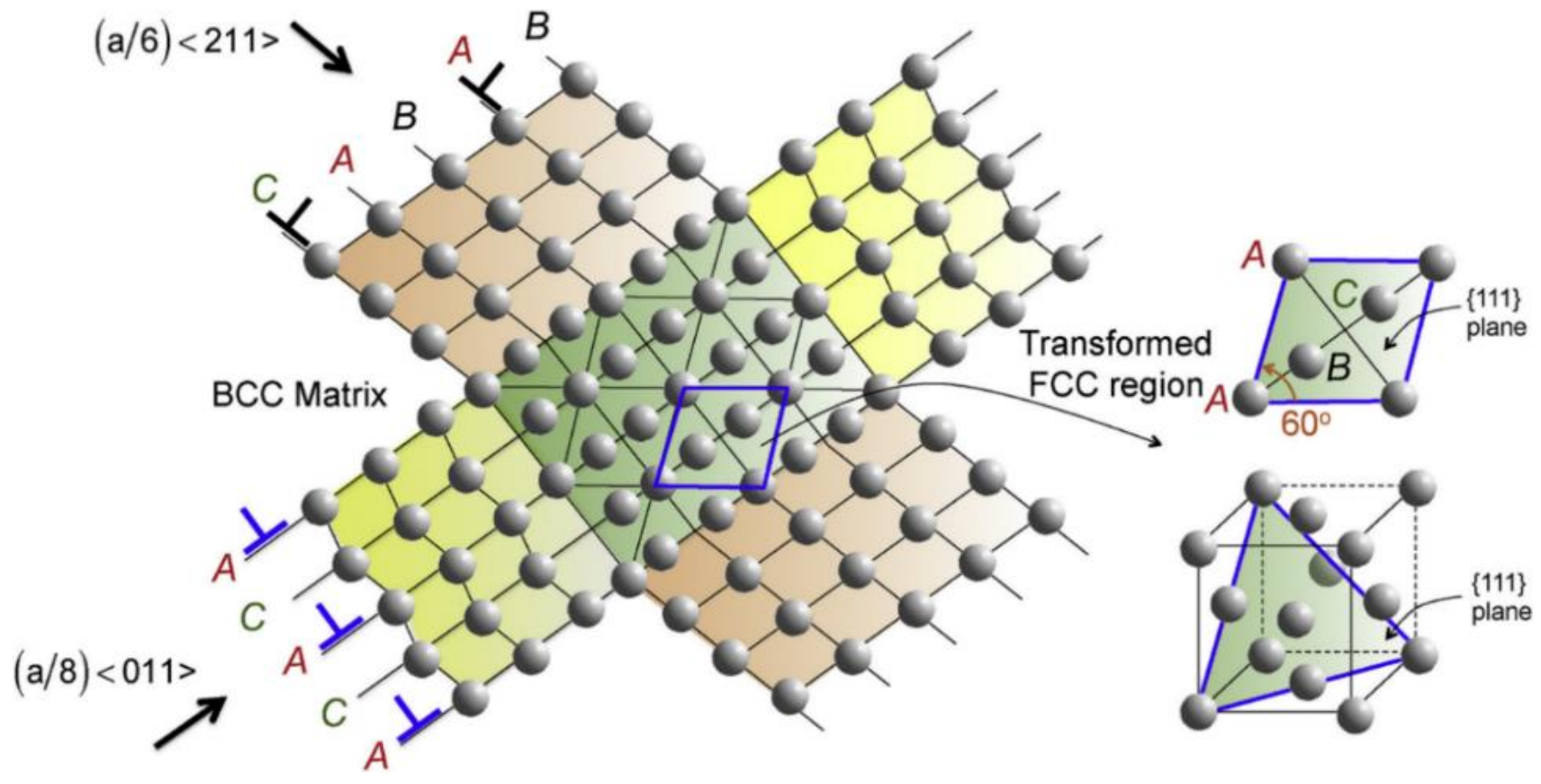

**Fig. 11. Dislocation-based transformation mechanism considered for the transformation stress modeling in FeMnAlNi SMA [87].**

## 4. Features and limitations of Fe-SMAs

### 4.1. Thermomechanical properties of Fe-SMAs

The novelty of Fe-SMAs and other SMAs lies in their capability to experience great deformations and restore their original shape by stress removal (SE) or heating (SME). Phase transformation is the key factor behind these smart properties. Stress-related transformation can result in SE, while temperature-related transformation can produce SME [57], [143], [224]. The SME was first observed by Chang and Read in an Au-Cd in 1951. Later, it was reported in numerous NiTi, Cu-based, and Fe-based alloys, which together make up the majority of SMAs. Several Fe-SMAs, including Fe-Ni, Fe-Mn, Fe-Mn-Si, Fe-Ni-Co-Ti, Fe-Pd, and Fe-Pt-based compositions, have been investigated [19], [143], [225]. For Fe-Ni-Co-Al-Ti SMA, precipitation heat treatments between 180 and 200 min at 600 °C were favorable for superelasticity with a high degree of recovery (>95%) and significant recoverable strains (7%), as illustrated in Fig. 12 (a). Low-temperature heat treatments changed the strength without producing superelasticity. The critical transition stress being larger than the slip resistance caused this phenomenon, which was attributed to plastic slip [117]. Fe–Mn–Si SMAs have been explored for a long time due to their prospective merits over other SMAs. These Fe-SMAs possess good weldability, workability, and corrosion resistance, as well as a low manufacturing cost [226]. By adding a particular quantity of Nb and C to traditional Fe-Mn-Si SMAs, the SME is significantly improved, with the generation of NbC precipitates during aging. The pre-rolling or a straightforward extension of austenite just before the aging can further improve SME [67], [70], [102], [227]. It was revealed that the shape recovery ratio of Fe–Mn–Si SMA reduces monotonically with increasing pre-strain as shown in Fig. 12 (a) [187]. This phenomenon is attributed to the buildup of irreversible plastic strain caused by dislocation slip under the influence of increasing pre-strain[196].

Fe-Mn-Ni-Al SMA has excellent thermomechanical properties and has shown a lot of promise recently [34], [112], [228]. It exhibits exceptional superelastic properties such as transformation strains and low thermal hysteresis. It has manifested among the most fascinating SMAs, with a large SE temperature range (> 400 °C). Fe-Mn-Al-Ni SMA displays a considerable work output because of high transformation strains (> 8%) and high transformation stress (500–700 MPa). The transformation stress displays relatively little temperature dependency throughout a wide range of (-196 °C to 240 °C) as compared to NiTi-SMA [229]. The Clausius–Clapeyron slope magnitude ($\partial\sigma/\partial T$) of this Fe-SMA was estimated to be less than 0.2 MPa/°C in compression, and 0.53 MPa/°C under tension. The $\partial\sigma/\partial T$ magnitude is much smaller than that shown in NiTi SMAs (6–8 MPa/°C) [229], [230]. The mean temperature dependency of the critical stress was estimated to be 0.514 MPa/K in Fe-Mn-Al-Ni while 2.87 MPa/K in Cu-Al-Mn [231], and 5.87 MPa/K in Ti-Ni [34]. During transformation, an exceedingly modest adiabatic temperature rise of less than 1 °C can be investigated. Fe-Mn-Al-Ni SMAs' SE is governed by a reversible transformation between the bcc and fcc lattices [89], [95]. Stress must be resistant to temperature changes for SE to be effective. Engineers and scientists can build components and structures with desirable qualities since the critical stress's temperature dependency is easily modifiable[91]. Fe-36Mn-11Al-7.5Cr-7.5Ni was studied and compared to other SMAs in terms of transformation stress and superelastic window, the findings showed that it has extremely advantageous properties [34], [89], [91]. The research on Fe-Mn-Ni-Al SMA is expected to expand due to these obvious outstanding functions. The size, composition, and volume fraction of precipitates were shown to strongly affect the superelastic strain, transformation temperature, stress hysteresis, and critical stress for stress-induced martensitic transformation of Fe-Mn-Al-Ni single crystals. The aging for 3 h at 200 °C produced 7.2% superelastic strain with approximately 6-10 nm precipitate as shown in Fig. 12 (b, c). Increasing the aging time or temperature led to a reduced superelastic recovery because of the growth of coarser precipitates [228]. Fe-SMAs consist of several elements, including Mn, Ni, Co, and Ti, which can be used to modulate flow stresses, TTs, and transformation strains, to enhance their functional qualities. The addition of aluminum and subsequent heat treatments produces cohesive precipitates, which are essential for SE. Although the precipitations do not change, they do produce internal stress fields that aid in transformation [21]. The nano-scale precipitates can also enhance slip resistance, which promotes martensitic transition and SE. The difficulty of performing reversible SE transformation, especially at ambient temperature, and the destruction of shape memory capabilities with cyclic loading are the main factors that limit SE in Fe-SMAs [34], [232], [233]. The pinning of the martensite interface boundary, which prevents perfect reversibility and causes the accumulation of residual strains with sustained loading, has been the main factor contributing to the deterioration of SE properties under functional fatigue [232]. To overcome such limitations, Fe-Ni-Co-Al-X (in which X is either Tantalum, Niobium, or Titanium) and Fe-Mn-Al-Ni alloys [230], [233] have received a lot of attention recently, and various investigations have found substantial superelastic and recoverable stresses at ambient temperature. Using alloying elements such as titanium, tantalum, and others in Fe-SMAs can improve their thermomechanical properties. They can also enhance other characteristics of Fe-SMAs such as surface roughness as they behave in other materials [21], [34], [36], [93], [112], [113]. The Recently investigated Fe-SMAs such as Fe-Mn-Ni-Al SMA can overcome most of the major limitations associated with Fe-SMAs as they offer remarkable properties such as a stable superelastic behavior over a large temperature range and less temperature dependency of the critical stress required for martensitic transformation [129], [130], [234]. Table 4 summarizes the functional properties of Fe-SMAs.

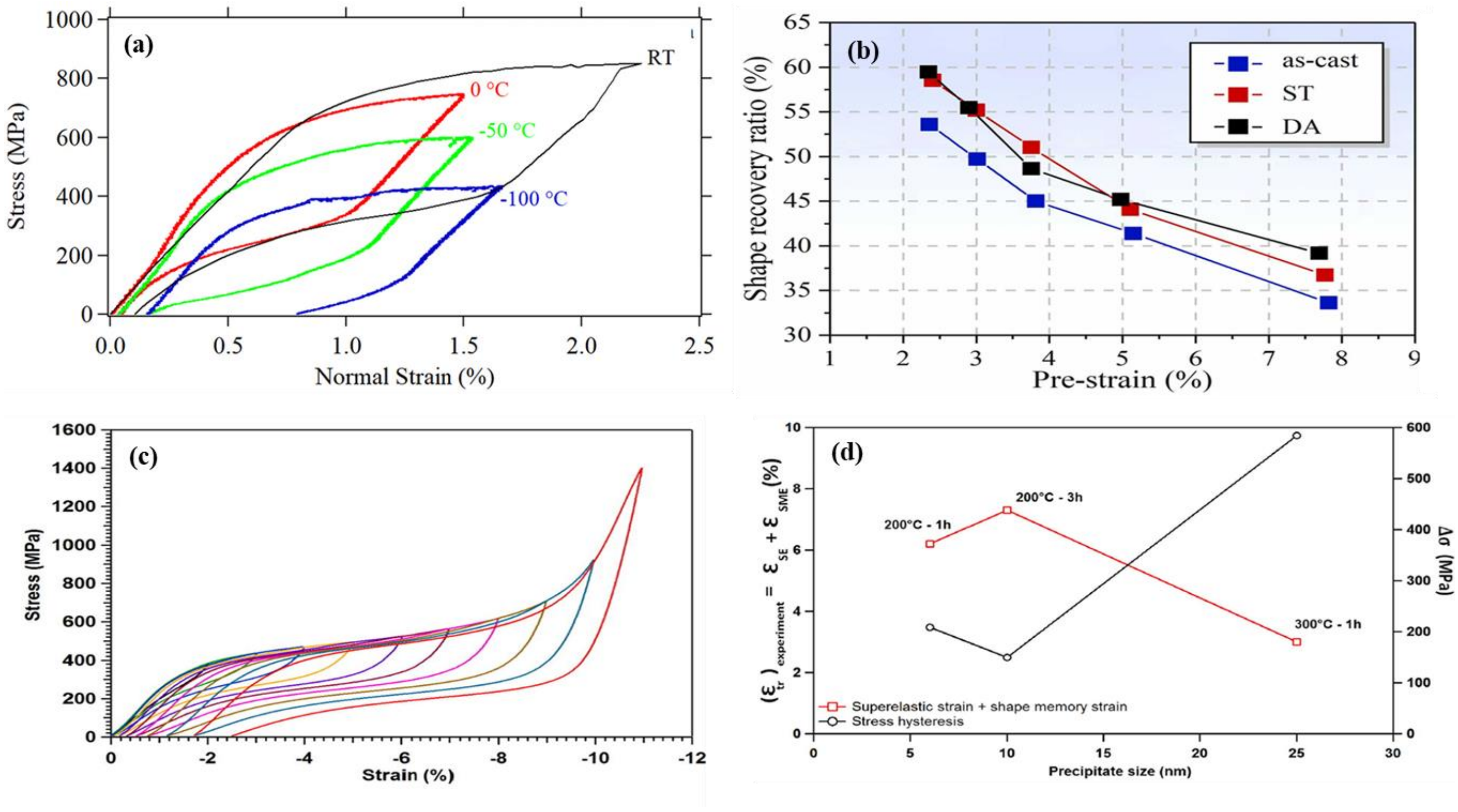

**Fig. 12. Factors influencing thermomechanical properties of Fe-SMAs. (a) Stress-strain behavior of Fe-Ni-Co-Al-Ti SMA aged for 180 min at 600 °C at various deformation temperatures** [117]. **(b) Pre-strain effect on shape recovery ratio of Fe-Mn-Si-Cr-Ni SMA at as-cast, solution-treated (ST), δ-annealed (DA) conditions**[187]. **The recovery annealing was applied for 15 min at 450 °C. (c) Superelastic behavior of Fe-Mn-Al-Ni SMA aged at 200 °C for 3 h. The compression test was performed at room temperature. (d) The effect of precipitate size on the stress hysteresis and superelastic + shape memory strains of Fe-Mn-AL-Ni SMA. (b and c are taken from** [228] **).**

**Table 4. Comparative study of different Fe-based SMAs and their functional properties**

| Fe-SMA | Maximum tensile elongation (%) | Superelastic strain (εSE) (%) | Shape memory effect (%) | γ′ solvus temperature (°C) | Thermal hysteresis, Th=TAf - TMs (K) | Ref. |
|---|---|---|---|---|---|---|
| Fe – 28Ni – 17Co – 11.5Al – 2.5Nb (single crystal) | 8.7 | 6.5 | 4.2 | 700 | 105 | [110] |
| Fe – 28Ni – 17Co – 11.5Al – 2.5Nb – 0.05B (single crystal) | 8.7 | 4.0 | 4.2 | 700 | 180 | |
| Fe–34Mn–15Al–7.5Ni | 6.1 | 5.5 | | N/A | 150 | [34] |
| Fe–28Ni–17Co–11.5Al–2.5Ta–0.05B | 15 | 13.5 | N/A | 932 | 24 | [21] |
| Fe–28Ni–17Co–11.5Al–2.5Ti–0.05B | 6.5 | 4.2 | | 888 | 31 | [108] |
| Fe–28Ni–17Co–10Al–2.5Nb–0.05B | 8.8 | 5 | N/A | 878 | 20 | [107] |
| Fe–34Mn–7.5Ni–13.5Al | 7.6 | 2.2 | 4.7 | N/A | N/A | [120] |
| Fe–28Ni–17Co–11.5Al–2.5Ti (single crystal) | 8.7 | 4.5-6 | 5.35 | 700 | 90-100 | [109] |

## 4.2. Mechanical properties of Fe-SMAs

### 4.2.1. Tensile characteristics of Fe-SMAs

The tensile stress-strain relationship was investigated for Fe-17Mn-5Si-4Ni-5Cr-1Ti-0.3C SMA to evaluate the tensile features [126]. The $\sigma - \varepsilon$ curves of its different tested samples are plotted in Fig.13 (a). The investigated Fe-SMA has a 0.2% proof yield value of 507 MPa and an ultimate extension of 38% in the as-rolled state. When the SMA was aged at temperatures below 700 °C, the yield strength improved, and its maximum stretch reduced with elevating the aging temperature. The textural variations of Fe-SMA during the aging process might be the reason for this phenomenon. Due to recrystallization, the yield strength significantly decreased when the aging temperature reached 800 °C. Additionally, Fe-17Mn-5Si-10Cr-4Ni-1(V, C) (mass%) SMA was examined, and its elastic modulus was discovered to be between 180 and 185 GPa, which is rather near to that of normal steel (i.e., 200 GPa) [235]. The high fracture strain of this Fe-SMA, which was shown to be over 30% in the absence of buckling while exceeding 8% in the presence of inelastic buckling, indicates that it is incredibly ductile. The yield and ultimate strengths of the unrestrained samples were approximately 400 MPa and 800 MPa, respectively. It was determined that Fe-17Mn-5Si-10Cr-4Ni-1(V, C) (mass%) SMA had a yield strength of 415 MPa when heated to 1100 °C with hot pressing, and 530 MPa when further cold rolled. This rise can be attributed to the last type's smaller average grain size (37 µm) compared to the first one [72]. Fig. 13 (b) displays the hot-rolled Fe-Mn-Al-Ni SMA's mechanical tensile characteristics. The stress-strain graphs show that there is no significant directional dependence. The yield and tensile strength results of the samples obtained parallel to the rolling direction (0° RD) were 585 MPa and 813 MPa, respectively, while they were 625 MPa and 855 MPa, respectively, for those samples produced perpendicular to the rolling direction (90° RD). This is ascribed to the elongated grains in the rolling direction, which finally explains the higher values for yield strength and tensile strength in accordance with the Hall-Petch relation by resulting in a slightly reduced effective grain size perpendicular to the rolling direction [236] The tensile response of Fe-34Mn-7.5Ni-13.5Al SMA under room temperature was tested as shown in Fig.13 (b) and a strain of more than 12% was indicated by the transformation front. At both room temperature and a wide range of temperatures, it exhibits outstanding superelastic performance under tension [89]. The findings of the Fe-17Mn-5Si-10Cr-5Ni (mass-%) SMA study revealed that it has good ductility with a maximum of 55% fracture strain under monotonic stress [237].

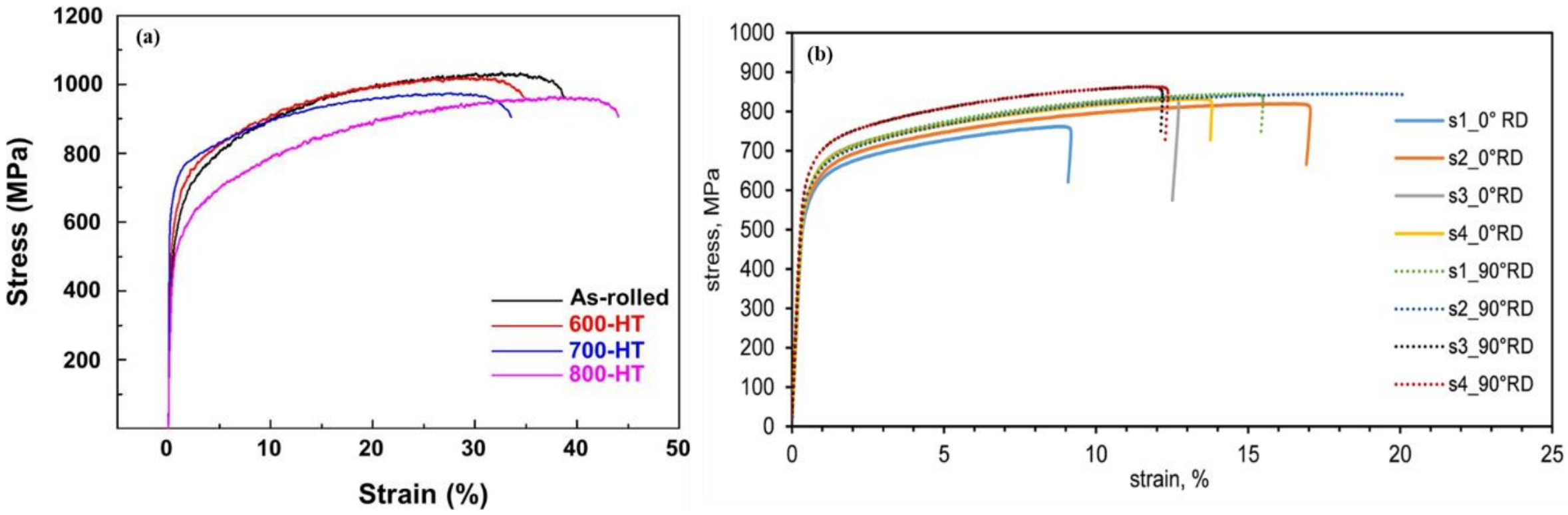

**Fig. 13. Tensile findings of different processing cases. (a) Stress-strain findings of aged heat-treated and rolled Fe-17Mn-5Si-4Ni-5Cr-1Ti-0.3C SMA** [126]**. (b) Mechanical properties of hot rolled Fe-Mn-Al-Ni SMA in tensile tests of 4 specimens parallel (0°) and perpendicular (90°) to the rolling direction** [236]**.**

### 4.2.2. Hardness of Fe-SMAs

Hardness is an important property of Fe-SMAs and needs to be discussed as suitable hardness demonstrates the capacity of Fe-SMA to withstand erosion, friction, and other types of wear. Fe-36.5Mn-10.7Al-6.2Ni-2.3Mo and Fe-35.2Mn-10.9Al-7.9Ni-2.4Mo SMAs' microhardness was investigated [238]. Vickers hardness was determined to be 414 and 423 in the solution-treated alloys of Fe-36.5Mn-10.7Al-6.2Ni-2.3Mo and Fe-35.2Mn-10.9Al-7.9Ni-2.4Mo, respectively; however, after 338 days of normal aging, as well as aging at 200 °C for various times, it rose to 428 and 429, as shown in Fig. 14 (a), and (b). The hardness was also evaluated for additively manufactured Fe-34Mn-14Al-7.5Ni (at. %) SMA [167]. The as-built samples displayed a notably increased hardness compared to those with 1 or 5 heat treatment cycles. In comparison to the earlier study of this work [180], the hardness in the as-built specimens rose to 420 HV. The Fe-Mn-Al-Ni SMA was subjected to LPBF processing with a preheating temperature of 200 °C to 500

°C, which induced the nanoscale β-phase precipitates and, as a result, generated a higher hardness than solution-treated samples [180]. The hardness of present phases in Fe–Mn–Si–Cr–Ni SMA was modified using samarium [239]. The hardness values of γ and ε phases were reported to be 254 and 286 VHN, respectively. Samarium was found to refine the grain size to 15%, which resulted in higher strength. As a result, the hardness has also improved. The development of fine precipitates in Fe-SMAs will reinforce their matrix and hence raise the hardness of the material [107], [228], [240], [241]. Using a 10 h aging at 200 °C, Fe-Mn-Al-Ni single crystals' hardness with the orientation 〈001〉 rose from roughly 370 HV to 430 HV. A precipitation size of ~ 10 nm after age hardening at 200 °C for 3 h showed the maxim hardness which confirms the substantial correlation between the precipitation size and the microhardness. The Fe-SMA also displayed great pseudoelastic ability under this peak aging environment [228]. Fig.14 (b) exhibits that the aging of Fe-Mn-Al-Ni-Ti SMA at 200 °C for 10 h very slightly increases hardness by about 20 HV1 [116]. Therefore, aging at 250 °C utilizing varied aging periods was carried out to account for the slower kinetics that was inferred from the suppression of the γ-phase growth in Fe-Mn-Al-Ni-Ti under the air cooling process. After solution treatment, the hardness rose from 440 HV1 to approximately 500 HV1 after an 18-hour aging heat treatment. After 15 minutes at 250 °C, a significant rise in hardness is already achieved, and the gradient becomes less steep as the aging time increases as illustrated in Fig. 14 (b) [116].

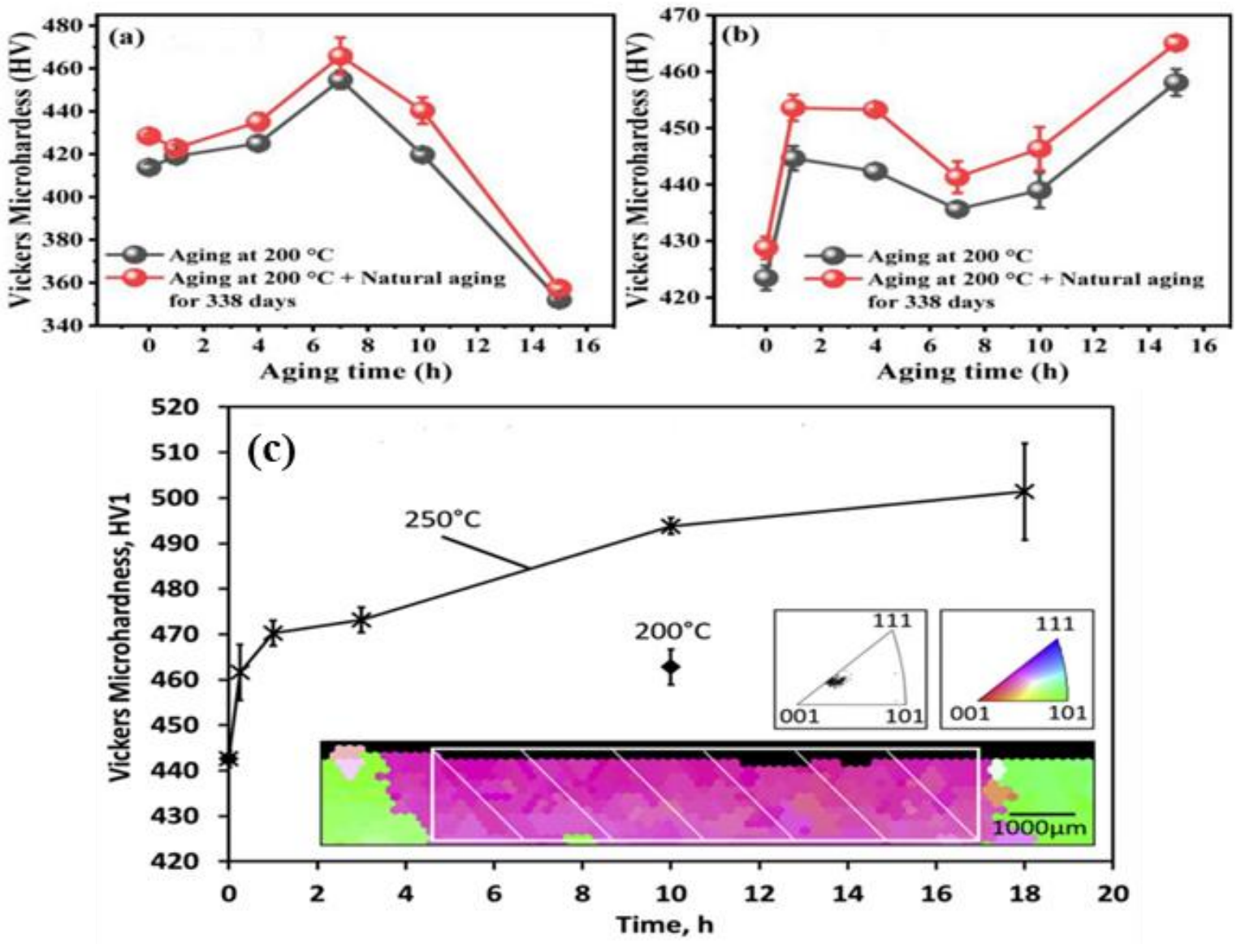

**Fig. 14. Microhardness findings of different Fe-SMAs. Microhardness of the 15 °C water-quenched Fe-SMAs. (a) Fe-36.5Mn-10.7Al-6.2Ni-2.3Mo. (b) Fe-35.2Mn-10.9Al-7.9Ni-2.4Mo. Underwent aging at 200 °C for varying amounts of time before undergoing 338 days of natural aging.** [238]**. (c) Ambient temperature Vickers microhardness readings in Fe-34Mn-15Al-7.5Ni-1.5Ti (at. %) after different aging periods at 250 °C. A benchmark hardness value after aging at 200 °C is supplied for comparison** [116]**.**

### 4.2.3. Fatigue behavior of Fe-SMAs

The fatigue resistance of Fe-SMA demonstrates its ability to absorb energy under cyclic loads. Fatigue evaluation of Fe-SMA is vital to properly match these materials with suitable applications. The cyclic behavior of the Fe-17Mn-5Si-10Cr-4Ni-1(V, C) was examined. Fe-SMA's elastic limit ($\sigma_{y,0.2\%}$=371 MPa) was substantially exceeded by the high cycle fatigue and endurance limits. Around 500 MPa marked the shift from high to low cycle fatigue, and 450 MPa was the value when the endurance limit was established to resist more than 2×106 cycles under fatigue loading. Tested Fe-SMA maintained high fatigue lifetimes even when subjected to maximum stresses that were considerably more than Fe-SMA's elastic limit [145]. The Fe-17Mn-5Si-10Cr-4Ni-1(V, C) SMA fatigue behavior was also evaluated under strain-controlled conditions following pre-straining and thermal activation[185]. The recovery stress was found to decrease under high-cycle fatigue loading, as illustrated in Fig. 15 (a), and this should be included in design evaluations even if the alloy's stiffness remained essentially the same. The reduction in recovery stress was supposed to be primarily the product of relaxation brought on by transformations during cyclic loading. The applicability of using a constant life diagram model to measure the fatigue limit of the Fe-17Mn-5Si-10Cr-4Ni-1(V, C) SMA for various stress ratios was investigated [185]. The current fatigue evaluation findings demonstrated complete agreement with the suggested fatigue design criterion and other reported work[145] as illustrated in Fig. 15 (b). This formulation was presented for a safe design of the Fe-SMA under a high-cycle fatigue loading. The structural fatigue of Fe-34Mn-15Al-7.5Ni-1.5Ti (at%) SMA was conducted at 1% constant strain amplitude. The recovered strains were then saturated at about 1.15% in local residual martensite regions. Global residual stresses increase due to the production of residual martensite in the newly stimulated region. Due to transition occurring in formerly untransformed regions, intermittent augmentation of recoverable strains was investigated. After 2046 cycles, fatigue failure eventually occurred, and the austenite/martensite interface's microcrack initiation and coalescence were found to be the primary cause of this failure. The interfacial dislocations, which are important for Fe-SMAs' SE functioning, always have an impact on the structural fatigue performance since they are the microstructure's weakest link [242]. Fe-17Mn-5Si-10Cr-5Ni (mass%) SMA results showed that the SMA has excellent low-cycle fatigue resistance for damping applications, meeting the demand for more resilient, long-lasting, and possibly fatigue-free applications in seismically active regions. When the strain amplitude rose from 1% to 9%, the fatigue life was demonstrated to be between 4007 and 83, and the measurements might be 10 times that of typical structural steel. The subsequent damper testing revealed that the fatigue life of the SMA under a rotational angle of 4% was observed to be 173 cycles, as opposed to 16 cycles for its standard steel equivalent, which further confirms the Fe-SMA priority [237].

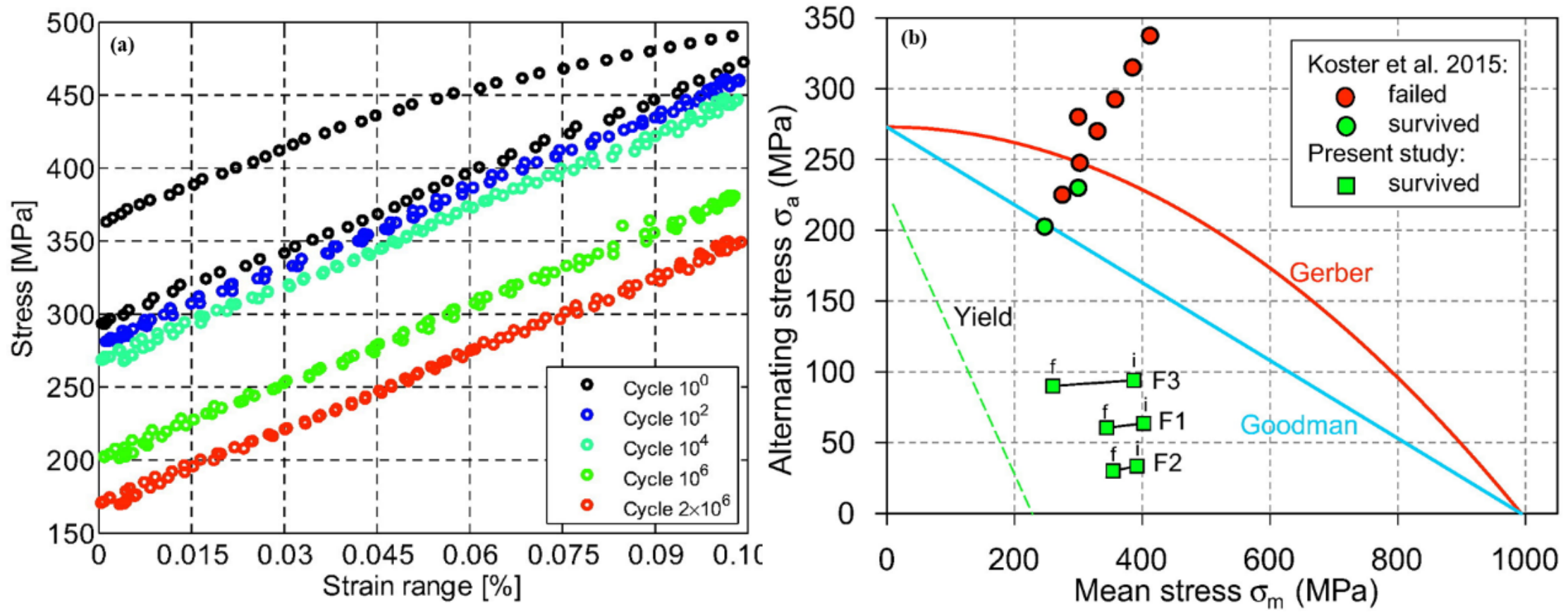

**Fig. 15. Fatigue behavior of Fe-SMA. (a) The stress-strain behavior of the fatigue-loaded activated Fe-17Mn-5Si-10Cr-4Ni SMA with a strain range of $\Delta\varepsilon_0$ = 0.105%. (b) Findings of the fatigue tests on the Fe-17Mn-5Si-10Cr-4Ni SMA were evaluated using the constant life diagram methodology** [185]**.**

### 4.2.4. Corrosion resistance of Fe-SMAs

The correlation between the chemical constituents and oxidation resistance of Fe-Mn-Si-Cr-Ni SMA, while exposed to air at 800 °C, was examined [243]. The more rapid diffusion of Mn, which controls the rate of oxidation, was shown to be possible by an increase in ferrite stabilizer components like Cr and Si in the Fe-SMA at the start of oxidation exposure. The austenite stability must be preserved during the oxidation process to enable the Fe-SMA to function properly [243]. Fe-17Mn-5Si-10Cr-4Ni-VC SMA underwent cyclic oxidation tests at 800, 900, and 1000 °C [244]. An unusual mass variation was discovered when the mass variation was assessed, and oxide layers were investigated using several characterization methods. After initial spallation, the material continued mass gain, as shown in Fig.16 (b) [244]. On the metal/oxide interface, the created Mn-depleted zone converted the austenite structure into ferrite, and the resultant roughness led to enhanced oxide anchoring. As a result, this alloy is a remarkable alternative for uses where cyclic oxidation is a significant concern [244]. The corrosion performance of this Fe-SMA was compared to that of S500 structural steel, a more common type of steel (EN 10149 PT2 standard) [182]. Fe-SMA was employed as a reinforcement in concrete without experiencing any significant corrosion which confirms that it has higher corrosion resistance than steel S500 [182]. The effect of cerium on the corrosion characteristics of Fe-14Mn-4Si-9Cr-4Ni was examined using anodic potentiodynamic polarization (Fig.16 (a)) and electrochemical impedance spectroscopy in 0.6 M NaCl solution [183]. The findings confirmed that cerium is crucial for improving the corrosion characteristics, while there is a limit to where it starts to be damaging [183]. The investigation of Fe-Mn-Si-Cr-Ni-(Co) SMA further demonstrated that the exceptional protection of passive coatings produced anodically on the SMA in 0.5 M H2SO4 solution is due to a protective layer made of a (Fe, Cr)-mixed silicate [245]. Fe-SMA's high Si content was confirmed to significantly improve Fe-SMA's resistance to intergranular attack in high oxidizing conditions [245]. At 800 °C under air for up to 120 h, the oxidation of a Fe-8.26Mn-5.25Si-12.80Cr-5.81Ni-11.84Co SMA was investigated [182]. The results demonstrated that exposure to oxidation favors the development of the sigma (σ), chi (χ), and ferrite phases in the support layer. The oxidation pattern followed a parabolic trend, with Mn2O3 oxide formation in the early hours and MnCr2O4 spinel and Mn3O4 growth in the last 24 hours of exposure controlling the oxidation's kinetics [182]. The corrosion of Fe–Mn–Al–Ni SMA was studied in a 5.0 wt.% NaCl solution [246]. Corrosion characteristics were found to be comparable to those of pure iron. The polarization curves of individual crystals, however, revealed the existence of an unsteady passive system [246]. The Pure Fe and Fe-34Mn-17Al-5Ni SMA's polarization behavior were both examined in NaCl-contaminated Ca(OH)2 and NaCl-free solutions [247]. FeMnAlNi and Fe behaved similarly in the NaCl-free solution and are characterized by strong $O_2$ development at high overpotentials and pronounced passivity at low overpotentials, while the FeMnAlNi SMA exhibited higher pitting corrosion sensitivity than Fe in the NaCl-contaminated solution [247].

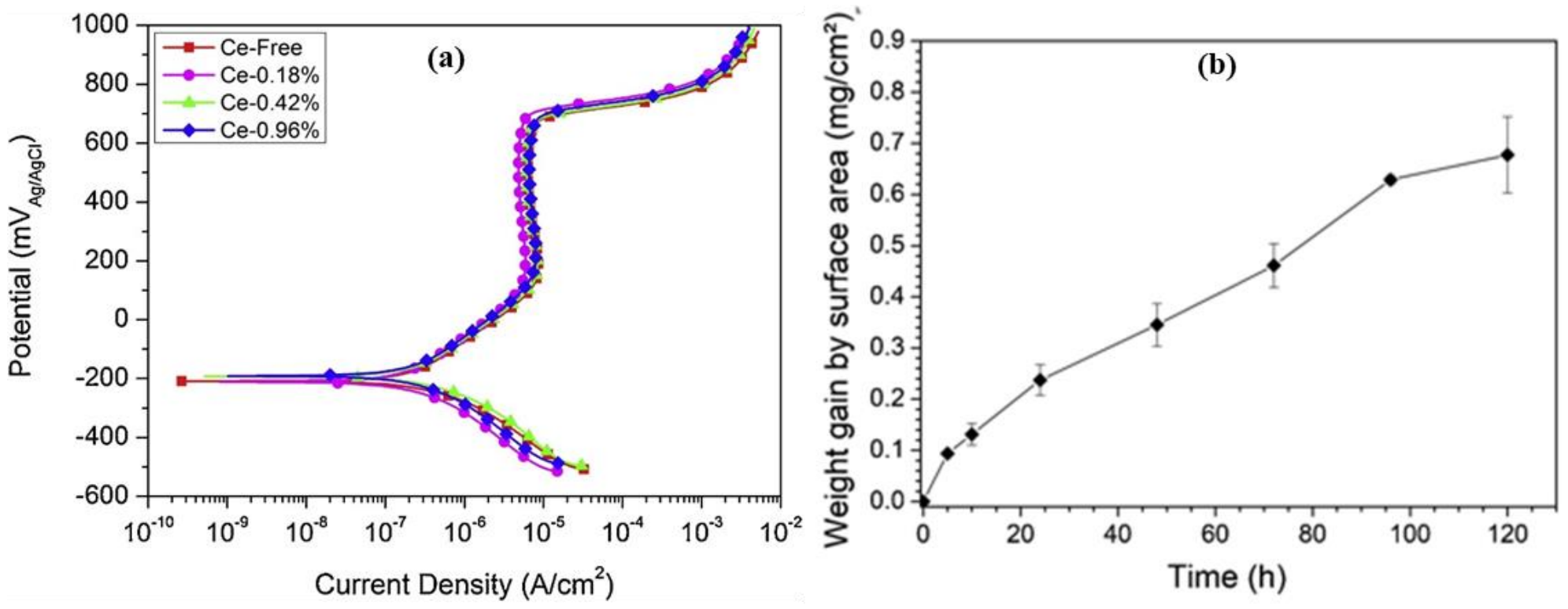

**Fig. 16. Corrosion resistance of Fe-SMA. (a) The curves of potentiodynamic polarization for Fe-14Mn-4Si-9Cr-4Ni SMA with different Ce contents in the simulated pore solution** [183]. **(b) The average weight gain of air-oxidized Fe-8.26Mn-5.25Si-12.8Cr-5.81Ni-11.84Co SMA at 800 °C** [182].

### 4.3. Manufacturing cost of Fe-SMAs

The demand for a cost-effective SMA with more economically appealing production pathways prompted the creation of iron-based, or ferric SMAs. This intriguing feature sets them apart from other SMAs. The main reason behind that can be the low price of their constitutive chemical elements including Fe, Mn, Si, Al, and Cr [59], [248], [249]. Their cost is reported to be a very small fraction compared to that of NiTi SMAs [23]. Therefore, they have attracted significant interest in the fields of engineering and metallurgy, and they are being thoroughly researched for use in construction where a huge amount of material is needed. The Fe-Mn-Si-SMAs are much cheaper than NiTi SMAs because of the reduced cost of their raw components and their processability in an atmospheric setting. Future costs are anticipated to be like those of highly alloyed stainless steel (about 8–10 €/kg) [250]. A cost analysis was conducted to evaluate the two strengthening options for 6.4 m girders using Fe-SMA and carbon fiber-reinforced polymer [251]. The cost comparison exhibited that both strengthening solutions were observed to be nearly comparable from an economic standpoint when the attainable mechanical clamping systems cost and prestressing force are considered, even though the actual cost of the Fe-SMA strips exceeds that of the regular carbon fiber-reinforced polymer plates [251].

## 5. Processing variables of Fe-SMA performance

### 5.1. Alloying elements' role in Fe-SMAs

Fe-SMAs' properties are induced using the reversible transformation between austenite (γ) and martensite (ε) thus, any factor that promotes the austenite-martensite transformation or their reversal can highly boost those properties. On the other side, the variables that hinder such a transition would be detrimental to Fe-SMAs' properties. The investigation of the various factors impacting the different characteristics of Fe-SMAs is a key factor in easing the choice of the optimum method for enhancing Fe-SMAs' properties. Alloying elements, precipitate, grain size, texture [252], grain orientation [253], [254], as well as the type of loading (creep, cyclic or constant loading), have all different effects on the MT [145].

Nickel (Ni), manganese (Mn), chromium (Cr), niobium (Nb), titanium (Ti), tantalum (Ta), cobalt (Co), and boron (B) are incorporated into Fe-SMAs to modify TTs, flow stresses, and transformation strains. Their inclusion into Fe-SMAs with further heat treatments can also produce coherent precipitates that induce enhanced SE. The formed precipitates do produce internal stress levels that aid in the process of transformation [89], [255]. The incorporation of aluminum (Al) into the FeNiCoAlX system resulted in the development of nanoscale precipitates favorable for SE [89]. SE in Nb-containing compositions was restricted to low temperatures (less than 0 °C), but in the case of FeNiCoAlTi SMA [117] and FeNiCoAlTa SMA [256], an SE of 7% was displayed at ambient temperature and higher but with quick cyclic deterioration.

Using 10 wt.% of cobalt in Fe–28Ni–6Si SMA with heat treatment at 400 oC for three days, enhanced Fe-SMA's shape recovery by about 49%. The samples were distorted at −196 °C with a surface strain of 2% and then heated to 1100 °C. The production of martensitic thin plates and the rise in austenite hardness caused by the incorporation of cobalt are the key factors that enhanced the SME [104]. The embedding of Ta into Fe–Ni–Co–Al stabilized the γ′ phase and improved the α′ martensite's strength and tetragonality. However, after aging to produce the γ′ phase, the brittle phase precipitated at grain boundaries, making the Fe–Ni–Co–Al–Ta SMA very brittle. Therefore, a modest quantity of B was incorporated into Fe-SMA to reduce the undesired precipitation at grain boundaries [21]. In addition, β-phase precipitation at grain boundaries with concurrent γ'-phase precipitation inside the grains of systems like Fe–Ni–Co–Al–X (X = Ta, Nb, and Ti) results in a rapid decline in the specimen's ductility and brittle fracture. This phenomenon makes it impossible to examine the SE and SME in these SMAs [21]. Low-boron content of ~ 0.05 at% was embedded in those Fe-SMAs, this modification reduced β-phase precipitation and elevated the SE from 5 to 13% at ambient temperature [21], [34], [108]. In Fe-SMAs containing boron, the thermal and mechanical hysteresis was observed larger than in other Fe-SMAs without boron because boron can reduce the Ms temperature and slow down the aging process [110].

The inclusion of the Ti element was also shown to be beneficial in stabilizing the γ′ phase, with the temperature of γ′ solvus and hardness increased to 888 °C and 420 HV, respectively. Therefore, the Fe–30Ni–15Co–10Al–2.5Ti–0.05B SMA experienced a thermoelastic martensitic transition with a temperature hysteresis of 31 °C [108]. The use of Nb in the Fe–Ni–Co–Al–Nb–B SMA increased the hardness to around 450 HV due to the stability of the γ′ phase and induced a thermoelastic martensitic transition, but grain boundary precipitation was retained. The grain boundary precipitation can be dramatically suppressed by a small quantity of B [107]. The addition of a modest amount of Nb and C to traditional Fe–Mn–Si SMA significantly improved the SME by causing small NbC precipitates to form during aging. SME can be improved further by pre-rolling austenite before aging treatments [67], [257]. The inclusion of samarium to Fe–14Mn–3Si–10Cr–5Ni (wt.%) SMA was also observed to enhance the SME by 27% [258].

### 5.2. Heat treatment of Fe-SMAs

In Fe-SMAs, heat treatment is frequently employed as a processing technique to homogenize the alloy, change the microstructure, and improve the desired properties. Aging was employed in FeNiCoAlNb SMA to generate nano-sized precipitates to tailor Fe-SMA properties. Fe-SMA displayed strong tension-compression asymmetry and shape recovery of roughly 8.8% in compression and 4.5% in tension after 3 h aging at 700 °C [113]. The critical stress of transformation showed a minimal temperature dependence as shown by the Clausius-Clapeyron curves in Fig.17 (a) [113]. The temperature and time of aging had an impact on the recovery stress, yield stress, and SE of Fe-17Mn-10Cr -5Si-4Ni-1(V, C) (wt.%) SMA [259]. The Fe-SMA displayed increased yield and recoverable stress at relatively moderate aging temperatures (e.g., 600 and 660 °C), as illustrated in Fig. 17 (b). On the other hand, lower yield and recovery stress were observed at higher aging temperatures (e.g., 774 °C). The SE of all aging situations improved after the aging time increased [259]. Fe-34Mn-14Al-7.5Ni SMA was quenched to mitigate intergranular cracking after solution treatment at 1200 °C [181]. Based on changing the quenching procedures, it was demonstrated that regulated γ-phase precipitate at the grain boundary can minimize intergranular cracking without impacting the SE [181]. The impact of heat treatment on the features of additively manufactured Fe-17Mn-5Si-10Cr-4Ni SMA was analyzed [260]. Heat treatment at temperatures below 800 °C was inadequate to completely accomplish the phase transition from bcc-δ to fcc-γ, whereas heat treatment at temperatures over 800 °C resulted in the thickening and production of new grains in the hcp-ε phase. The latter significantly hindered shape memory [260]. Additionally, σ-phase production using around 3 h heat treatment at 800 °C had a detrimental effect on the SME by altering the chemical makeup of the fcc-γ phase. 800 °C for 0.5 h was shown to be the ideal procedure for achieving a maximum recovery strain as shown in Fig.16 (c) [260]. The influence of aging time on transformation stress and hysteresis was investigated in Fe-34Mn-15Al-7.5Ni (at%) SMA. The tensile strength at ambient temperature and critical stress for inducing the MT rose without loss of SE with aging at 200 °C [261]. As shown in Fig.16 (d) and (e), the transformation stress rises as the aging time rises, while the hysteresis fluctuates with the aging time. Transformation stress of ~ 600 MPa with a recovered strain of ~ 6.7% was produced after 24 hours of aging at 200 °C [261]. Fe-34Mn-15Al-7.5Ni (at%) SMA was aged at ambient temperature. After 30 days of aging, the MT began at a stress of about 156 MPa, and the deformation was restored after unloading [262]. The superelasticity was ~ 5% which was lower than those aged at 200 °C. The transformation stress increased to 216 MPa after a total of 60 days of ambient-temperature aging. This transformation stress is quite close to those aged at 200 °C for 1 h [262].

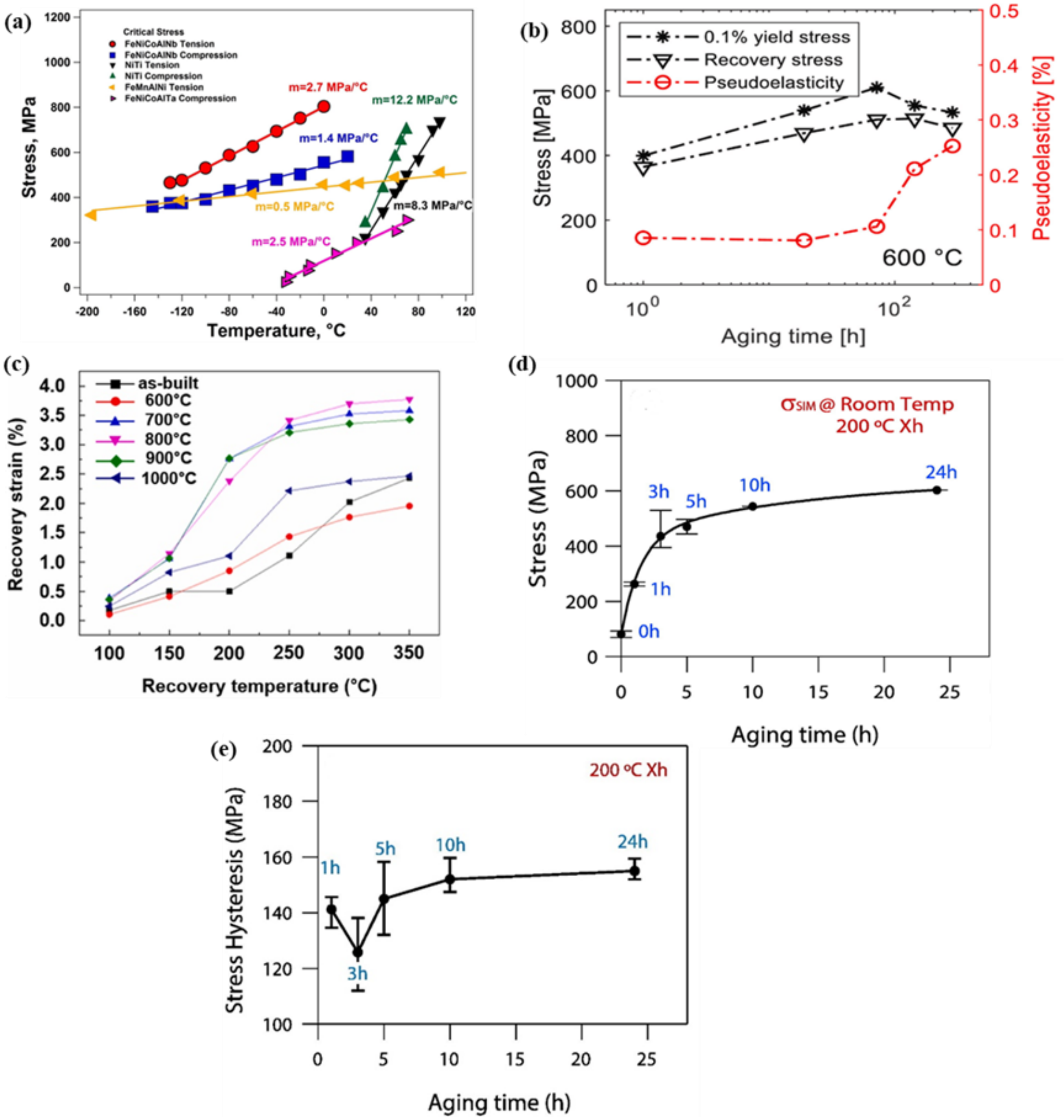

**Fig. 17. Heat treatment processing findings. (a) FeNiCoAlNb critical stress for transformation in compression and tension vs. temperature** [113]**. The Clausius-Clapeyron curves of FeNiCoAlTa, NiTi, and FeMnAlNi are included for comparison** [34], [36], [263]**. (b) Impact of aging time on the thermomechanical characteristics (yield and recovery stress, as well as SE) of FeMnSi-based SMA at 600 °C** [259]**. (c) Recovery strain vs. recovery temperature of heat-treated additively manufactured Fe–17Mn–5Si–10Cr–4Ni SMA with the same heat-treatment time at varying temperatures** [260]**, and the effect of aging time at 200 °C on (d) The critical stress for inducing the MT and (e) Stress hysteresis of the Fe-34Mn-15Al-7.5Ni SMA. Stress hysteresis was evaluated at 3% applied strain in every aging cycle** [261]**.**

### 5.3. Grain and grain boundary effect in Fe-SMAs

Grain size control is an important strategy for modifying and promoting the functionality of numerous materials. Grain boundaries must be minimized to achieve decreased creep rates in Ni superalloys or to generate wafers in semiconductors [121]. The reduction of grain size was widely employed in Fe-SMAs to provide exceptional mechanical characteristics, such as increased formability, higher resistance to intergranular cracking, and increased fatigue strength [264], [265], [266], [267]. The grain size control was shown as a key factor to enhance the reversed transformation [268]. The bamboo-like microstructure was shown to be particularly effective in finding good SE and shape-memory abilities [269], [270]. Grain size has been shown to influence the stability of the constituent phases of martensitic transformations. The decline in grain size exhibited a significant enhancement in the austenite phase stability. This results in a rise in the thermal hysteresis of tiny grains and a drop in MT temperatures [270], [271].

The bamboo-like microstructure substantially supports SE in the Fe–Mn–Al–Ni system, demonstrating the importance of grain size in improving desirable characteristics. The SE of the Fe-34Mn-15Al-7.5Ni SMA was found to be significantly influenced by increasing the average grain diameter, width, and thickness of the sheet sample of the relative grain size [34], [154], [272]. Large grains with a diameter of several millimeters were shown to be generated using a cyclic heat treatment procedure, and their average size can be regulated by the cycle quantity of heat treatment. The SE of large grain size samples was enhanced with improved elongation to fracture, lower threshold stress for transformation, and better reversibility [272]. Due to more substantial grain boundary limitations, small grain size results in increased resistance to the superelastic response and reduced reversibility. As grain size increases, grain restrictions lessen, and each grain can change to martensite almost autonomously [34], [154], [272].

Titanium incorporation into Fe-Mn-Al-Ni SMA was confirmed to overcome its limited recoverability. Introducing modest amounts of titanium into the Fe-SMA significantly boosted abnormal grain growth, because of the high subgrain refinement, as shown in Fig.18 [121]. It is feasible to fabricate bars having large single crystals by tailoring and inducing abnormal grain development. Other systems with comparable microstructural properties can be tailored using this process [121]. Twin boundaries and their effect on strain recovery were investigated in Fe-18.8Mn-5.0Si-8.5Cr-5.0Ni and Fe-20.2Mn-5.6Si-8.9Cr-5.0Ni SMAs [198]. The presence of Twin boundaries and their interactions can considerably inhibit the gamma (γ) to martensite (ε) transformation. Therefore, after suppressing their formation, an SR of 8.4% and tensile SR of about 7.6% in a cast-annealed sample with coarse grains of around 1.10 mm was achieved.

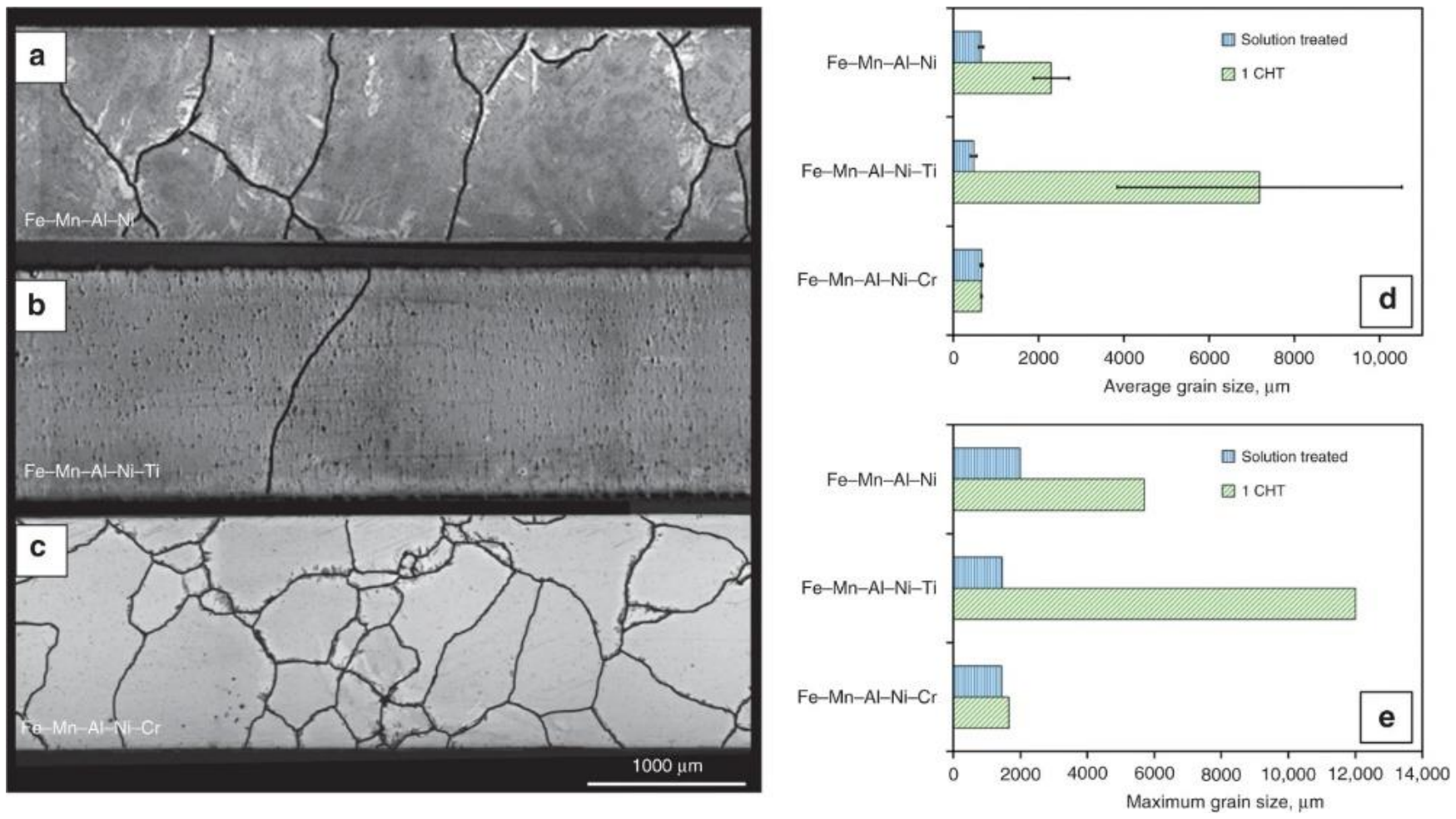

**Fig. 18. Fe-Mn-Al-Ni grain growth behavior assessment. OM images of the typical structure with grain boundaries under one heat treatment cycle. (a) Fe-Mn-Al-Ni. (b) Fe-Mn-Al-Ni-Ti. (c) Fe-Mn-Al-Ni-Cr. (d) Average grain size. (e) Maximum grain size. d and e findings are after solution treatment for 1 h at 1225 °C and one heat treatment cycle [121].**

### 5.4. Effect of texture in Fe-SMAs

The texture is reported to be an important element in evaluating the SME of polycrystalline Fe-SMAs. The texture is considered the basis for the difference observed in the recovered strain of Fe-SMAs. Modulation of the texture significantly influences both SME and SE features because the transition strain in the crystal is highly reliant on the deformation orientation [273], [274]. The presence of coincidence site lattice boundaries and low-angle boundaries increases when recrystallization texture is generated by appropriate thermomechanical treatment. Consequently, in polycrystalline conditions, the heavily textured Fe-SMAs including FeNiCoAlTaB and FeNiCoAlNbB had a remarkable ductility exceeding 8% and 20%, respectively. Precipitates' boundary control was shown to be effective in reducing the grain boundary energy, but no clear connection between grain boundary and precipitation feature has been demonstrated [108]. In FeNiCoAlNbB SMA, the development of cold-rolling and recrystallization textures was illustrated [275]. With the rolling decrease of 98.5% (Fig.19), a rolling texture with a robust brass orientation was improved dramatically. After 1 hour of solution treatment at 1220 °C with a strong {hk0} 〈001 texture, the 98.5% cold rolled FeNiCoAlNbB SMA displayed a good SE of 3.2% with a tensile value of around 960 MPa and residual strain of stability of ~ 0.7%. The significantly improved SE in this Fe-SMA, as compared to the non-SE in the as-forged condition, is primarily due to the development of strong textures and the inhibition of the precipitation in the grain boundaries [275].

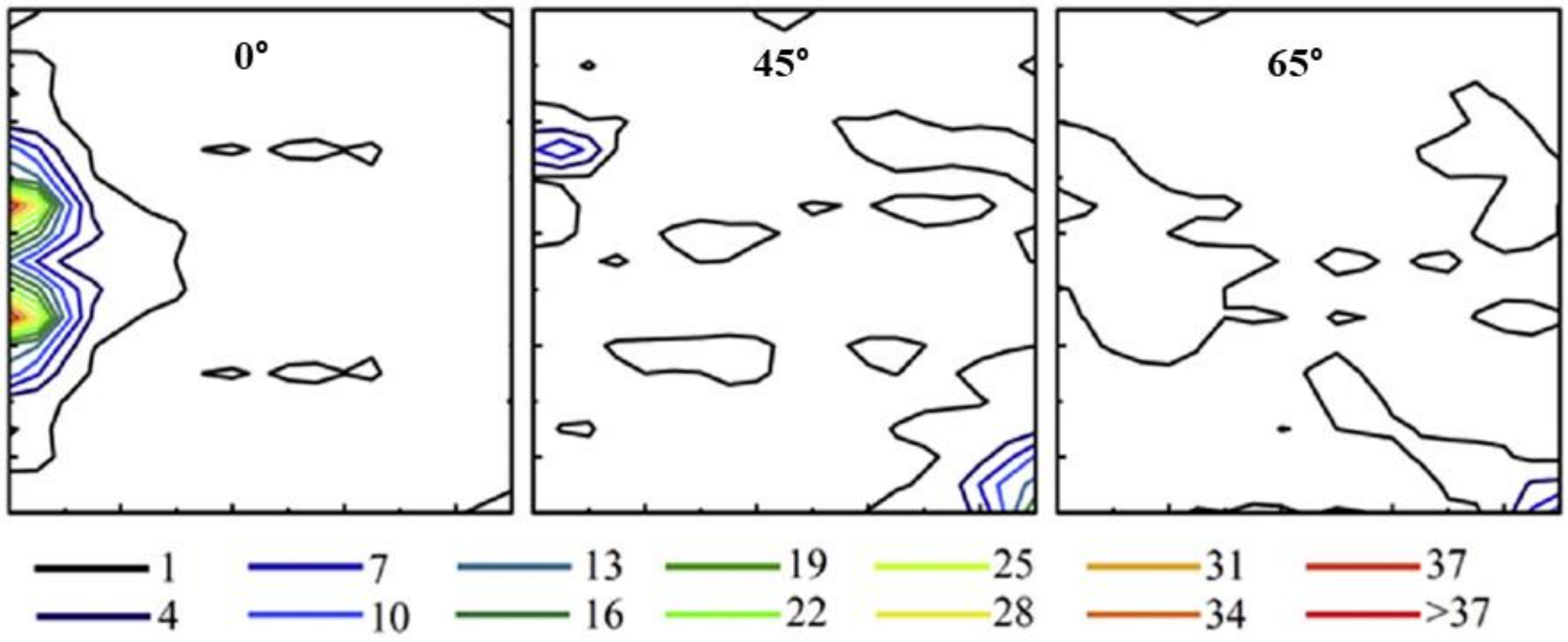

**Fig. 19. Texture investigation of Fe-SMA. Orientation distribution functions for recrystallized textures of FeNiCoAlNbB alloy with 98.5% cold-rolling reduction (φ2= 0°, 45°, and 65° sections)** [275]**.**

## 6. Additive manufacturing of Fe-SMAs: Recent investigations and future trends

AM technologies create three-dimensional parts from a digital model by assembling thin layers of materials using a layer-by-layer technique under computer control. This distinctive feature enables the manufacturing of intricate or customized components, eliminating the need for costly tooling, punches die, or casting molds used in traditional processes [276], [277]. Significant advancements in the constitutive innovations of AM metal processing, such as lower-cost, more reliable lasers, cheaper high-performance computing software and hardware, and metallic powder feedstock tech, have empowered it to be the ultramodern processing method over the last two decades [278], [279]. Laser powder bed fusion (LPBF) is a powder bed-based technology that melts and fuses the powdered material utilizing a high-power-density laser as an energy source as shown in Fig. 20 (a). The parameters that should be considered in LPBF processing are the laser power, scanning speed, powder layer thickness, hatching type, and scanning speed as illustrated in Fig. 20 (b). LPBF has been shown to generate near-net-shape objects with a relative density of 99.9% [280], [281]. It is a more common and most effective printing technique among the various types of AM techniques, such as laser-engineered net shaping, electron beam melting, and wire-arc AM (WAAM), and is typically used for processing most SMAs including Fe-SMAs [170], [280], [281], [282], [283], [284], [285].
Fe-SMAs fabricated through melting and casting under high vacuum or high purity shielding gas have been the subject of numerous investigations. The final shape is typically obtained through additional machining, such as hot forging and cold rolling, which restricts production to parts with straightforward geometries, such as strips or bars [72]. AM technologies like LPBF and WAAM can offer new ways to get beyond the complexity constraints of conventional manufacture and to fully utilize the appealing functions of these Fe-SMAs. They enable the creation of complicated parts with high densification in a single manufacturing step as the raw metal powder is entirely melted throughout the process [286].

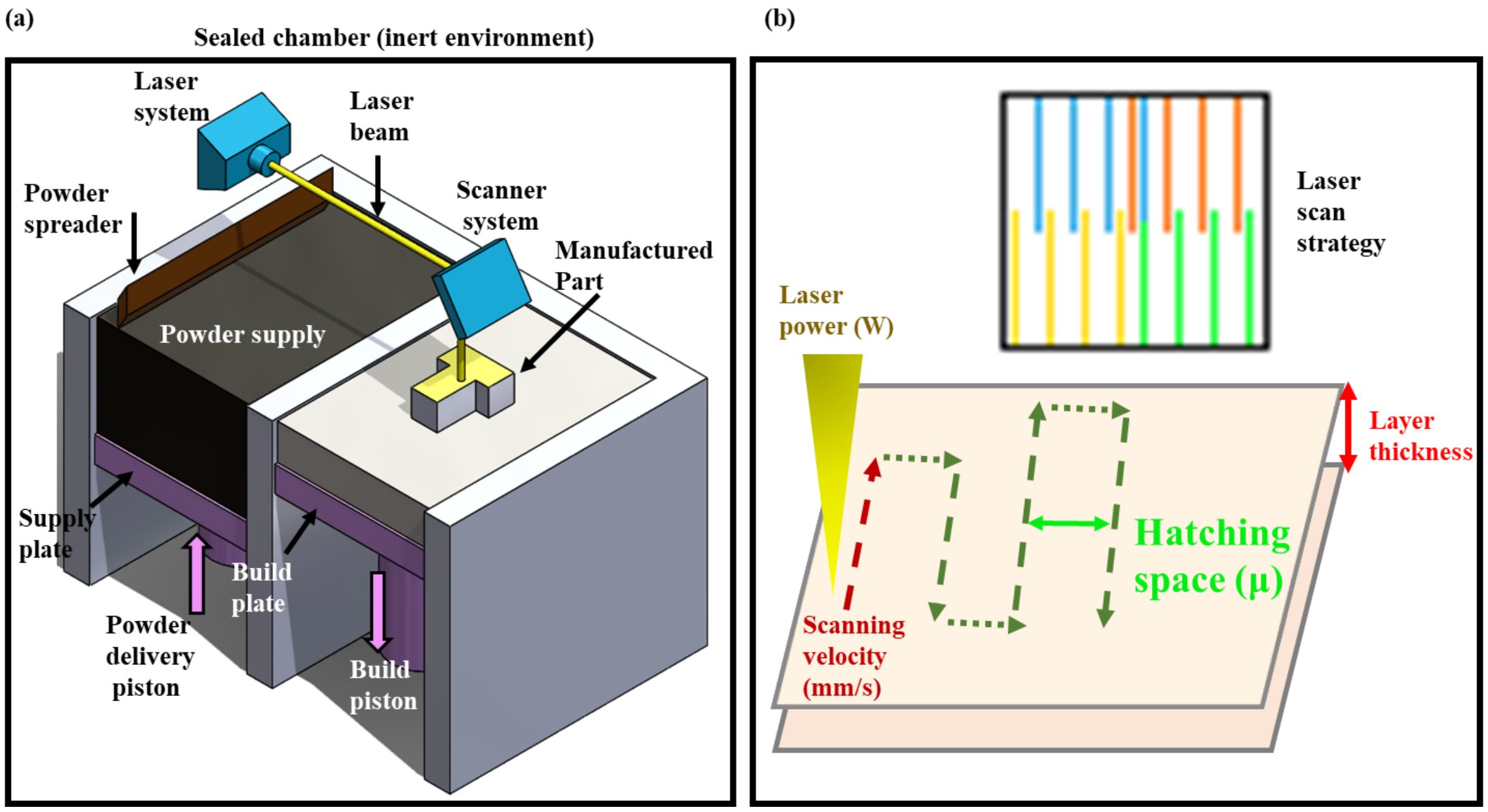

**Fig. 20. A schematic illustrating the LPBF technique utilized in the processing of Fe-SMAs. (a) The LPBF machine's mechanism and parts. (b) The main parameters are optimized to produce high-performance Fe-SMA parts.**

### 6.1. Additively manufactured Fe-SMAs

AM of Fe-SMAs has not received much attention until now and appears to be extremely rare, with only a few Fe-SMA systems investigated. Fe–Mn–Si SMAs, among other Fe-SMAs, were shown to be the focus of AM-based research. More details about the recent work related to AM-researched Fe-SMAs can be found in [130], [132], [133], [140], [171], [260], [287], [288], [289], [290], [291], [292], [293], [294], [295], [296], [297], [298], [299], [300], [301], [302], [303], [304]. AM-based studies focused on the LPBF as an AM-based potential process for manufacturing Fe-Mn-Si-based SMAs with promising properties.

In recent studies, Fe-Mn-Ni-Si-Cr SMA was first manufactured using LPBF. The aim of the research was to use a processing window of 115 W – 175 W combined with a scanning speed of 300 mm/s – 600 mm/s to achieve a high-densified Fe-Mn-Si-based SMA [171]. Fe-Mn-Si-based cubic samples of 10x10x10 $mm^3$ were fabricated to investigate the densification and microstructural behavior of the SMA. The highest-density material was then used for mechanical and thermomechanical investigation. The study showed that the used powder had a chemical makeup of Fe-17Mn-5Si-10Cr-4Ni (wt.%). The results of the powder investigation indicated that the particles were mostly spherical, with a mean particle diameter of 29.7 μm and a size distribution range of 10–50 μm [130], [171], [174]. To identify the optimal conditions necessary for printing Fe-Mn-Si components with enhanced properties, tensile and intricate shape specimens were printed at varying scanning speeds, laser powers, build orientation, as well as constant hatching space, and layer thickness. The specimens printed with high volumetric energy density (VED) showed minimal defects including lack of fusion, contour, and cracks defects compared to other parts with lower VEDs. Higher VEDs also produced the best densities and porosities [130], [171], [174]. Fe-Mn-Si SMA was also used to fabricate intricate 3D structures utilizing LPBF [174]. Successful fabrication of a completely working Fe-SMA with remarkable mechanical qualities was accomplished. Yield stress of ~ 230 MPa with an elongation of ~ 50% was attained by a 30-minute straightforward heat treatment at 800 °C following LPBF processing. This strength was at least three times greater than the greatest value for shape memory polymers that have ever been recorded (70 MPa). Heating the deformed samples at 200 °C achieved a shape recovery of ~ 36% and ~ 47% in the parallel and perpendicular directions to the building direction, respectively [174].

The impact of post-heat treatment on the characteristics and microstructure of additively produced Fe–17Mn–5Si–10Cr–4Ni using LPBF was examined [260]. It was found that using a temperature less than 800 °C was insufficient to fully perform the bcc-δ to fcc-γ transformation, but temperatures over 800 °C resulted in grain expansion and hcp-ε phase thickening, which had a significant impact on SME. σ-phase production via ~ 3 h heat treatment at 800 °C had a detrimental effect on the SME by altering the chemical constituents of the fcc-γ phase. The SME parallel to the fabrication direction was higher than that parallel to the laser scanning direction and 45° to the scanning direction. In contrast to the random but weaker {111} and {100} textures that were seen in the scanning direction and 45 with the scanning direction, the higher SME parallel to the fabrication direction was attributable to the texture {110} in the plane perpendicular to the fabrication direction [260]. Additionally, Fe-Mn-Al-Ni SMA was processed using LPBF with a high-temperature build plate of 500 °C to overcome constraints associated with the crack formation during AM [167]. Increased diameter in the as-built samples led to a strong texture in the 001> direction and a columnar-grained microstructure. An increased hardness was observed in as-built samples as well. The treated alloy's applied cyclic heat treatment induced abnormal grain development as well. Last but not least, tensile load testing showed an identified stress plateau and reversible strains of approximately 4% [167]. The Fe-Mn-Si-Cr-Ni SMA was also fabricated using LPBF under a high energy density of 222–250 J/mm$^3$ [133]. The Fe-SMA achieved a combination of a high strength of over 480 MP, good ductility of about 30%, superior ultimate tensile strength of higher than 1 GPa, and a significant recovery strain of roughly 6%. The strain and shape recovery findings are summarized in Fig. 21. Fe-Mn-Al-Ni SMA was another system processed using LPBF [180]. Temperature variations and solidification speed, and therefore processing variables including the actual specimen design, had a significant impact on the microstructural development during processing. A noticeable grain growth was started by a single-step heat treatment, which produced microstructures with good reversibility. The greatest reversible strain, as determined by the stress-strain response, was 7.5% [180]. The summary of AM-based processing parameters, powder characteristics, and sample specifications is provided in Table 5.

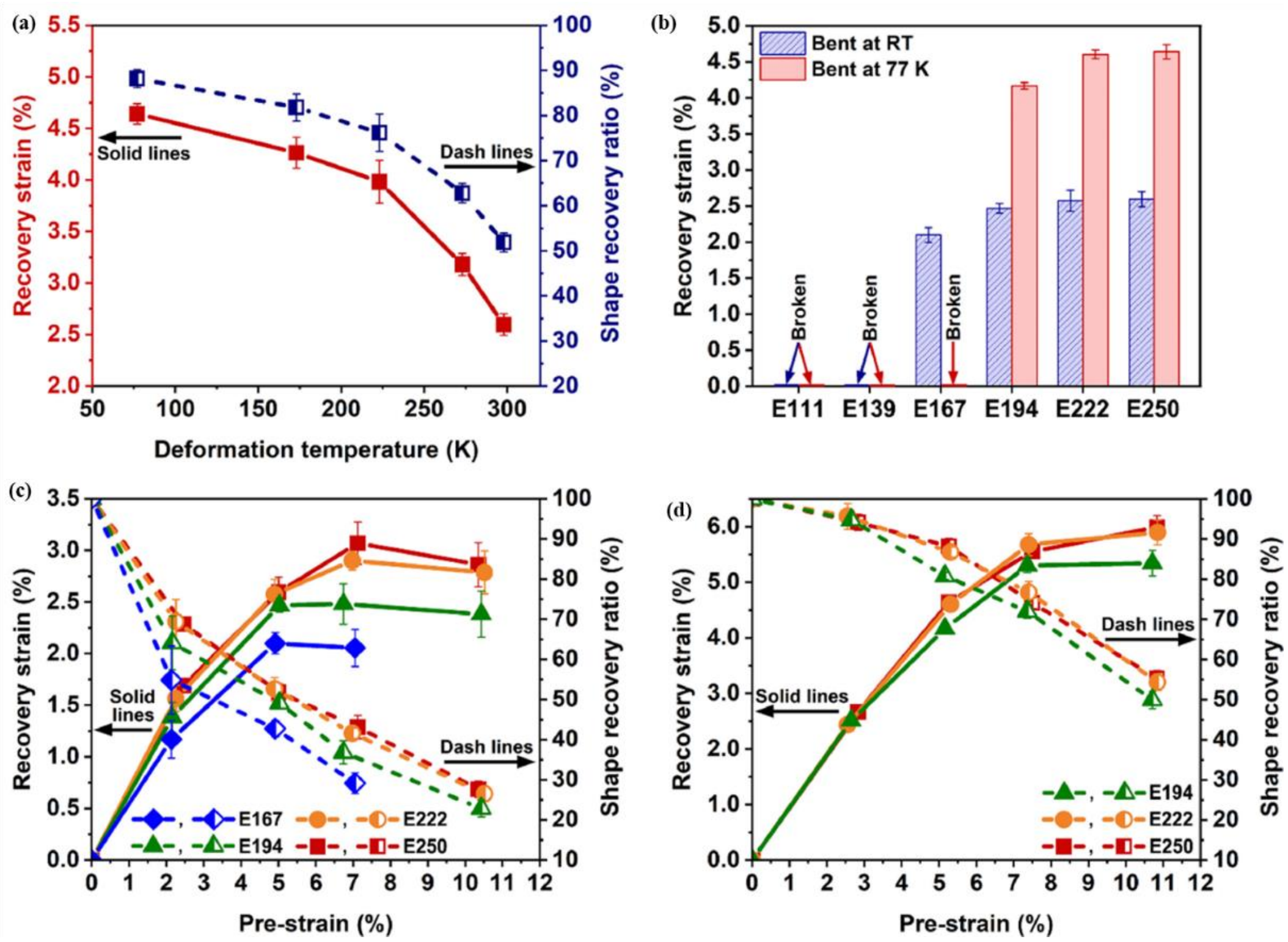

**Fig. 21. Shape memory characteristics of the LPBF-fabricated Fe-Mn-Si-Cr-Ni SMA. (a) The impact of deformation temperature on shape recovery ratio and recovery strain in the specimen manufactured with an E of 250 J/mm$^3$ after 5%**

**bending. (b) Recovery strain after 5% bending at 77 K and room temperature in the specimen produced with varying energy input. (c) Shape recovery ratio and recovery strain at room temperature as a function of the fa**[171]**brication's varying energy input. (d) Shape recovery ratio and recovery strain at 77 K as a function of the fabrication's varying energy input** [133]**.**

**Table 5 Processing parameters and specifications of AM-produced Fe-SMAs**

| Alloying system | Raw material specifications | AM-based process | Process parameters | | | | Samples | Reference |
|---|---|---|---|---|---|---|---|---|
| | | | P (W) | V (mm/s) | H(mm), L (mm) | VED ($J/mm^3$) | | |
| Fe–17.25Mn–5.10Si–9.90Cr–4.04Ni (wt.%) | Gas atomization, spherical particles, size distribution of 10-50 μm, and $d_{50}$ of 29.7 μm. | LPBF | 115 – 175 | 300 – 600 | 0.1, 0.03 | 63.89 – 194.44 J | Cubic samples of10 $mm^3$ and dog-bone tensile samples | [171] |
| Fe-17.70Mn-4.50Si-10.10Cr-4.30Ni-0.70V-0.20C (wt%) | Gas atomization, spherical morphology, size distribution ranging from 22 to 48 μm, and a $d_{50}$ of 33 μm. | LPBF | 420 | 800 | 0.1, 0.05 | 105 | Rectangle blocks | [291] |
| Fe-21.48Mn-5.36Si-9.42Cr-5.29Ni-0.02C-0.05O (wt%) | Gas-atomized powder using electrode induction melting, spherical morphology, particle size of 8.97-81.4 μm, and $d_{50}$ of 30.9 μm. | LPBF | 100-360 | 500-1000 | 0.08, 0.03 | 42-300 | Cubes of 8 $mm^3$, cuboid blocks of 46 mm x 10 mm x 8 mm, and dog-bone samples | [133] |
| Fe-17.70Mn-4.50Si-10.10Cr-4.30Ni-0.70V-0.20C (wt%) | Gas atomization under argon gas, $d_{50}$ of 33 μm, and a size distribution between 22 and 48 μm. | LPBF | 175 | 225 | 0.1, 0.03 | 178-389 | Cubes of 10 $mm^3$, and dog-bone samples | [130] |
| Fe–17.0Mn–4.6Si–9.9Cr–4.6Ni (wt.%) & Fe-17.70Mn-4.50Si-10.10Cr-4.30Ni-0.70V-0.20C (wt%) | SMA I: Size distribution of 10-50 μm with a $d_{50}$ of 29.7 μm.<br>SMAII: Size distribution of 22-48 μm with a $d_{50}$ of 33 μm. | LPBF | 130-175 | 100-600 | 0.1, 0.03 | 72-583 | Cubes of 10 $mm^3$ | [290] |
| Fe 48.1-Mn 36.1. Al 7.3 Ni 8.5 (wt.%) | Vacuum induction melting produced Fe-Mn-Al-Ni was first cast and then gas-atomized with $d_{50}$ 40 μm. | LPBF | 200 | 680 | 0.12 | 0.03 | Cubes of 5x5x7$mm^3$, cylinders of 5 mm in diameter and 40 mm in height, rods/pillars of 0.5-3 mm in diameter and 7 mm in height. | [167] |
| Fe-20.0Mn-5.5Si-8.8Cr4.9Ni (wt.%) | The powder was prepared from pure elements (99.9%) under gas atomization using vacuum induction melting. | DED | 1000 | 13.33 | 0.001, 0.3 | N/A | 50 × 50 × 30 $mm^3$ samples | [298] |
| Fe-17Mn-5Si-10Cr-4Ni-1(V, C) (wt. %) | 1 mm diameter wire of Fe-17Mn-5Si-10Cr-4Ni-1(V, C) (wt. %) | WAAM | Wire feed speed of 3 m/min, travel speed of 480 mm/min, voltage of 18V, contact tip to work distance of 10 mm, shielding gas of 82% Argon and 18 $CO_2$, and gas flow rate of 15L/min. | | | | 45 mm length structure | [294] |

DED, directed energy deposition; $d_{50}$; mean diameter of powder particles, H, hatching space; L, layer thickness; LPBF, laser powder bed fusion; P, laser power; VED, volumetric energy density, V, laser scanning speed; WAAM, wire arc additive manufacturing.

## 6.2. Features and limitations of additively manufactured Fe-SMAs

The research on Fe-SMAs such as Fe–Mn–Si SMAs has mainly centered on the microstructural and thermomechanical characteristics of conventionally manufactured parts. Various microstructures and textures, as well as different shape memory properties, are believed to be achieved using the LPBF technique [171]. Since the SME is achieved using fcc-γ → hcp-ε transformation and its reversal [196], [305], the as-built additively manufactured specimens with optimum LPBF parameters were heat treated for 30 min at 800 °C to achieve an austenitic phase and melt any ferritic phase produced in the rapid consolidation [171]. After annealing, high elongation, strength, and ductility were exhibited. The SME and SE were obtained, which were higher than those observed in the conventionally produced Fe–17Mn–5Si–10Cr–4Ni-1(V, C) [254]. SME and SE were significantly dependent on the build direction compared to the loading direction. Strain recovery was enhanced when both loading and building were in a parallel direction, this is due to the increase of (101) orientated grains [171]. After being heat-heated at 200 °C and then cooled, the restored strain of the tensile specimens including horizontal and vertical samples elongated to about 4 % strain as illustrated in Fig. 22 (b) [171]. The mechanical and thermomechanical findings of the LPBF-generated FeMnNiSiCr SMA are summarized in Fig. 22. Successfully created complex structures that had dramatic shape recovery, good dimensional precision, and complex geometry. After deformation and subsequent heating, as seen in Fig. 22 (e), shape changes take place. Every object tested exhibits a clear restoration of the original printed shapes after heating, which is attributable to the back MT. Additionally, the Fe-Mn-Al-Ni process employing the selective laser melting technique demonstrated outstanding features. Temperature changes and solidification speed, and therefore processing variables including the actual sample shape, had a significant impact on the microstructural evolution during processing. Heat treatment caused strong grain development through single step solutionizing, which produced microstructures with good reversibility. The maximum reversible pseudo-elastic strain was found to be around 7.5 % based on the alloy's compressive stress-strain response [180], [304]. In-situ digital image correlation combined with EBSD was performed to investigate the phase distribution within differently oriented grains of Fe–17Mn–5Si–10Cr–4Ni SMA [295]. It was applied after deformation at 2% Fig. 23 (b) and 4% Fig. 23 (c) while the EBSD (Fig.23 (a) was applied prior to deformation. The digital image correlation analysis demonstrated the localization of the strain, which indicates that the strain was accommodated by the emergence of slip, martensitic transformation, and stacking faults [306]. Different grains were analyzed to have different strain distributions and concentrations of stacking faults (Fig. 23 (d)). It was concluded that the grains with 〈1 1 0〉 orientation along the loading direction were preferred for the martensitic transformation and resulted in more pronounced work hardening as well as improved shape memory properties.

A common challenge in LPBF technology is that the processing parameters must be optimized to minimize defects and produce specimens with higher quality. Several defects including keyholes, lack of fusion, cracks, irregular voids, and spherical gaseous pores are likely to arise in additively manufactured specimens, resulting in low-density parts with degraded mechanical and functional properties. The reason for the development of such defects is the fabrication of the samples at processing parameters that are far from the optimal window [171], [280], [281], [307], [308]. The concentrated heating and rapid cooling along with using powdered material as the feed establishes a favorable environment for porosity formation by leaving cracks, partially melted powder, lack of fusion zones, or generating gas bubbles [309]. A lack of bond formation between the scanning tracks and layers will result in the creation of voids if the powder does not melt completely and the melt pool fails to penetrate far enough into the previously solidified layers [278]. Cracking has been investigated to occur when parts are manufactured with a low specific energy density. The fundamental reason for crack generation is excessive levels of the bcc- δ phase, which is exceedingly brittle comparable to fcc-γ, as well as the residual stress and surface roughness caused by a significant temperature differential around the laser spot [310], [311], [312].

Cracks typically form and propagate near the sample's edge, in which the surface roughness functions as a stress concentration source known as the residual stresses, and the notch effect is relatively high. For high-energy-density samples, on the other hand, the high ratio of power to scanning speed results in increased penetration depth and melt pool size, as well as a drop in molten material viscosity. Throughout the deposition, great track bonding, as well as suitable wetting, distributing, and flattering features of the melted tracks, are promoted, ensuring the production of dense bulk elements. Because the fcc-γ phase has higher ductility than bcc-δ, the predominantly austenitic microstructure resists cracking [171], [310]. Preheating at a suitable temperature and post-heat treatment processing are two common strategies used to minimize cracking in additively manufactured Fe-SMAs [167], [180].

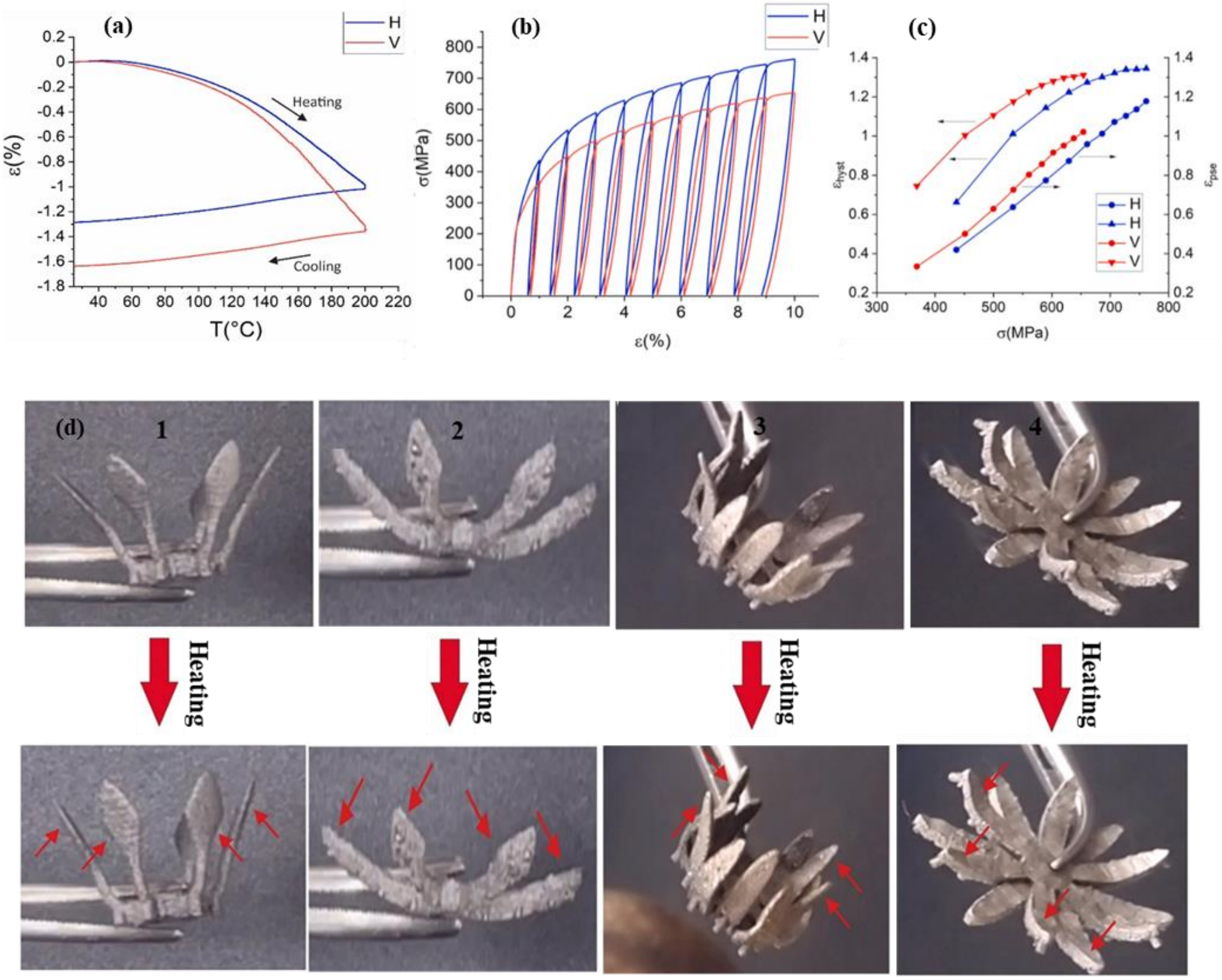

**Fig. 22. LPBF findings of Fe–17Mn–5Si–10Cr–4Ni SMA. (a) Recovery strain as a function of temperature after 4% pre-straining for LPBF FeMnNiSiCr SMA in both horizontal and vertical orientations. (b) Cyclic loading-unloading curves. (c) Superelastic strain and hysteresis width vs. the applied stress. (d) Testing the SME of Fe-based 3D-printed complex shapes at 300 °C after deformation, (1,2) is defined as type I, while (3,4) is known as type II** [171].

Recently, the effect of in-situ preheating, in-situ heat treatment, or post-heat treatment on AM-produced Fe-SMA has been investigated[167], [291], [292]. The impact of two different post-heat treatment settings on the mechanical, microstructural, and shape memory characteristics of a Fe-17Mn-5Si-10Cr-4Ni-(V, C) (wt%) SMA produced using LPBF was explored [291]. Specimens aged following solution treatment at 1050 °C for 2 hours revealed fully crystallized equiaxed grains with annealed twins, while specimens aged immediately at 750 °C for 6 hours from the as-built condition had elongated fcc-γ grains with a substantial quantity of low angle boundaries, as illustrated in Fig. 24. The as-built and direct aged samples had average fcc-γ grain sizes of around 10 and 30 μm, respectively. Thus, substantial grain growth occurred throughout the direct aging process.VC precipitates were finer and more evenly distributed in the direct aging than in the aging after the solution heat treatment. Moreover, the solidification cell structure was preserved in the direct aging since the VC precipitates were primarily generated along the margins of the solidification cells. Similar recovery strains and superelasticity were displayed by both conditions. Conversely, the direct energy exhibited significantly higher yield strength and recovery stress in comparison to the aging after the solution heat treatment. The fine consolidation of structures and fine distribution of VC precipitates in the direct aging followed by quick solidification of the LPBF technique could be attributed to the higher yield and recovery strains in the direct aging compared to the aging after the solution treatment.

The effect of in-situ treatment on the bcc- δ formation in LPBF-produced Fe-Mn-Si SMA and its transformation into the fcc- γ phase was explored [292]. To apply the in-situ treatment, the alloy was manufactured utilizing a laser power range of 380–420 W. Throughout the manufacturing process, each layer was subjected to additional laser scanning with the same laser power. The δ phase transforms into the γ phase by increasing the laser power from 380 W to 420 W, with the maximal transformation occurring at 400 W. The sample produced at 420 W has an almost entirely fcc-γ microstructure (γ of ~ 93%). Investigations were conducted on the microstructure of the laser-rescanned areas. Compared to other areas, the rescanned area had a much lower bcc-δ phase fraction. On considering this, the additional bcc-δ to fcc-γ phase transformation that the laser remelting effect causes in this area can account for the comparatively low bcc-δ fraction in the rescanned area.

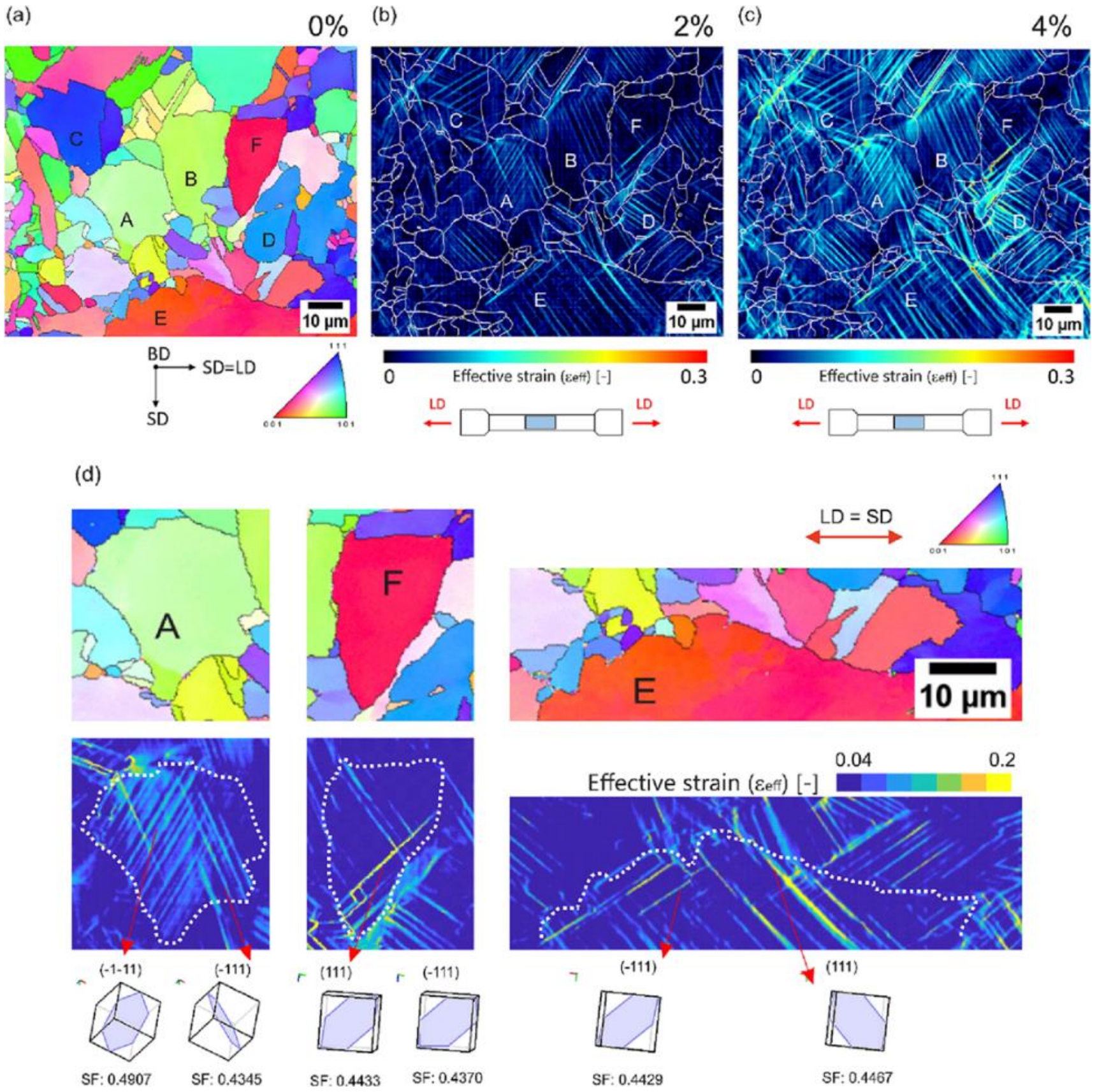

**Fig. 23. Digital image correlation of LPBF-fabricated Fe-Mn-Si SMA. (a) EBSD map with IPF coloring associated with the loading direction. (b) Strain maps were obtained at a deformation of 2 %. (c) Strain maps were obtained at a deformation of 4 %. d | strain and EBSD maps for the three chosen grains. The red arrows represent the hcp-ε lamellae nucleation along these planes** [295].

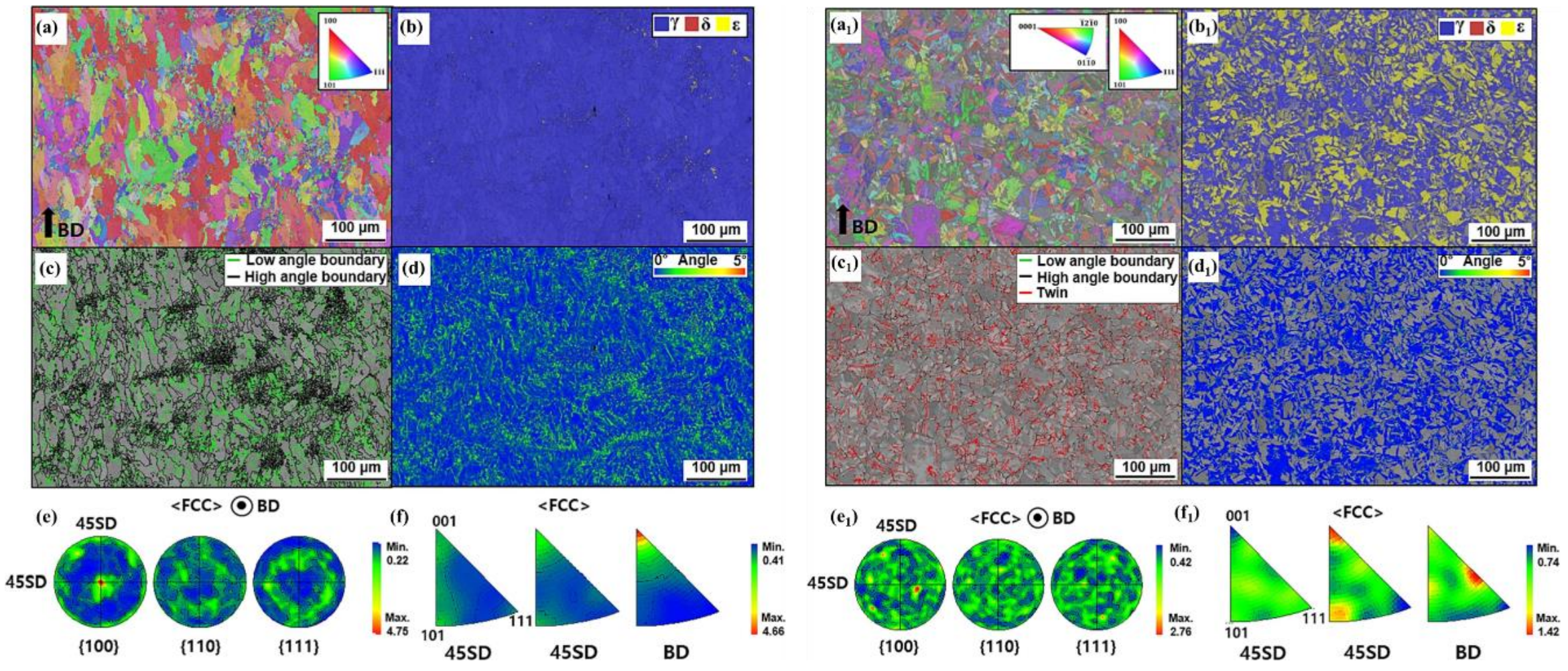

**Fig. 24. (a, $a_1$) IPF map, (b,$b_1$) phase map, (c, $c_1$) grain boundary map, (d, $d_1$) KAM map, (e,$e_1$) PF direction, and (f, $f_1$) IPF of LPBF-produced Fe-Mn-Si-Cr-Ni-V-C SMA in both conditions direct aging (a-f) and aging after solution heat treatment ($a_1$-$f_1$). Where IPF, KAM, and PF stand for inverse pole figure, Kernel average misorientation, and pole figs, respectively** [291]**.**

Using AM methods to produce functionally graded microstructures of Fe-SMA is another innovative approach that has been developed recently [290]. This novel approach was implemented to tailor the microstructure of two Fe-SMAs (Fe-17Mn-9.9Cr-4.6S-4.6Ni (wt.%) and Fe-17.8Mn-10.6Cr-4.8Si-4.2Ni-0.7V-0.2C (wt.%)) using LPBF. Two distinct methodologies have been employed to generate functionally graded specimens featuring microstructures that vary spatially. The first included locally altering the scanning speed at various sample locations which led to variable VEDs on the different locations, as represented in Fig. 25. The second one concentrated on the rescanning plan. A second laser scan of a few selected areas of the powder layer was conducted after the first scan was completed. Using different laser scanning speeds (VED1, 100 mm/s) and (VED2, 600 mm/s during the manufacturing process resulted in two different phases with the sample δ and γ phases as shown in Fig.25. The regions where the reduced scan speed was implemented exhibit refined austenitic material grains with a texture of <101 >. Approximately 7.5 ± 3.1 μm is the typical grain size (Fig. 25-b,d). Reduced VED yields areas of coarse elongated grains of bcc-δ ferrite that are preferentially <001 > oriented and alternating with austenitic regions (Fig. 25-b,c,d). The process of rescanning or remelting was also utilized to create samples with a mixed bcc-δ/fcc-γ microstructure. EBSD was used to examine the sample as it was manufactured. Under these processing conditions, the rescanned area could completely change from the fine-grained austenitic structure (average grain size: 11.0 ± 4.4 μm) produced during the first scan at 400 mm/s to a coarse-grained ferritic microstructure (average grain size: 49.3 ± 11.7 μm). The development of graded specimens with zones containing finer-grained austenite and ferrite (rescanned regions) and regions of coarser austenite grains (single-scanned areas) is made possible by the local microstructure alteration brought about by remelting.

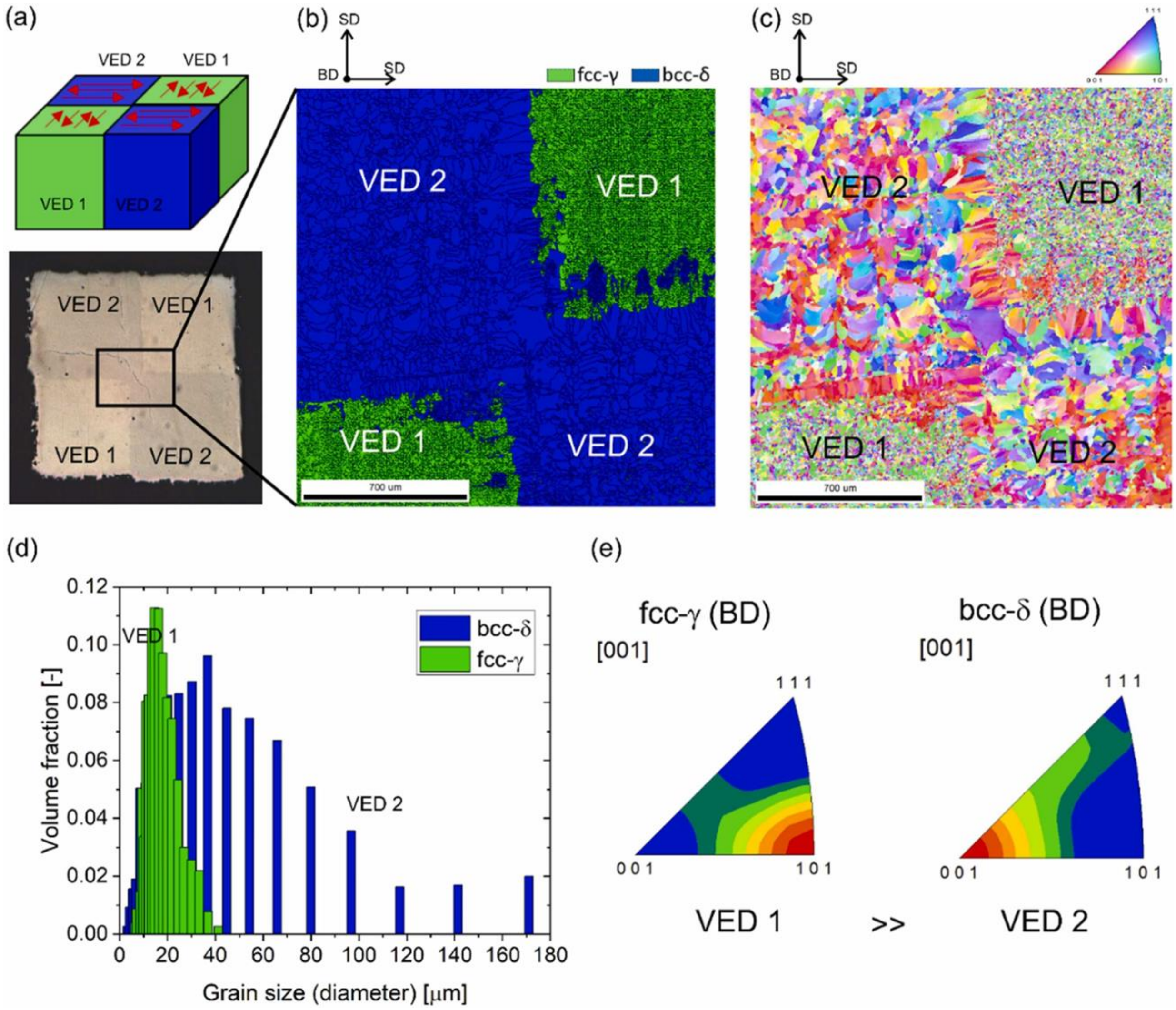

**Fig. 25. The development of variable microstructure in LPBF-produced Fe-SMA using variable processing parameters. (a) An OM image demonstrating light contrast as a result of distinct phases existing in the various sample locations, (b) EBSD maps, (c) IPF coloring, (d) grain size distribution, and (e) IPFs of Fe-17Mn-9.9Cr-4.6S-4.6Ni (wt.%) SMA. The laser scanning speed was modified on the different regions of the samples combined with a variable VED over the different areas of the sample** [290].

Another novel method that has been validated as having the potential to produce Fe-SMA with improved mechanical properties is the WAAM technique [294]. Fe–Mn–Si–Cr–Ni–V-C SMA was manufactured via arc-based directed energy deposition AM or WAAM as shown in Fig. 26 (a). The mechanical/functional behavior and structural evolution of the SMA were investigated. The WAAM-produced Fe-SMA had a low porosity and excellent deposition performance. The microstructure of the deposited material was shown to be consistent and homogenous as exhibited in Fig. 26 (b). The as-deposited material's microstructure analysis showed that it is mostly made up of the γ-FCC phase, with trace amounts of the VC, ε, and σ phases. The exceptional mechanical and functional response of the Fe-SMA was demonstrated by tensile and cyclic testing. According to tensile tests, the material had a fracture strain of 26%, yield strength, and fracture stress of 472 and 821 MPa, respectively. Post-mortem synchrotron X-ray diffraction investigation provided unambiguous evidence of the γ to ε phase transition following uniaxial tensile loading to fracture. As demonstrated in Fig. 26 (c), the hardness assessment showed a fairly steady and flat pattern between 250 and 300 HV0.5, with a slight rise on the upper area of the wall and tiny oscillations. This phenomenon can be explained by the existence of finer grains in the border region, which amplifies the mechanical characteristics as a result of the Hall-Petch effect. The potential applicability of the manufactured material for structural applications was demonstrated by the evaluation of its cyclic stability over 100 load/unloading cycles.

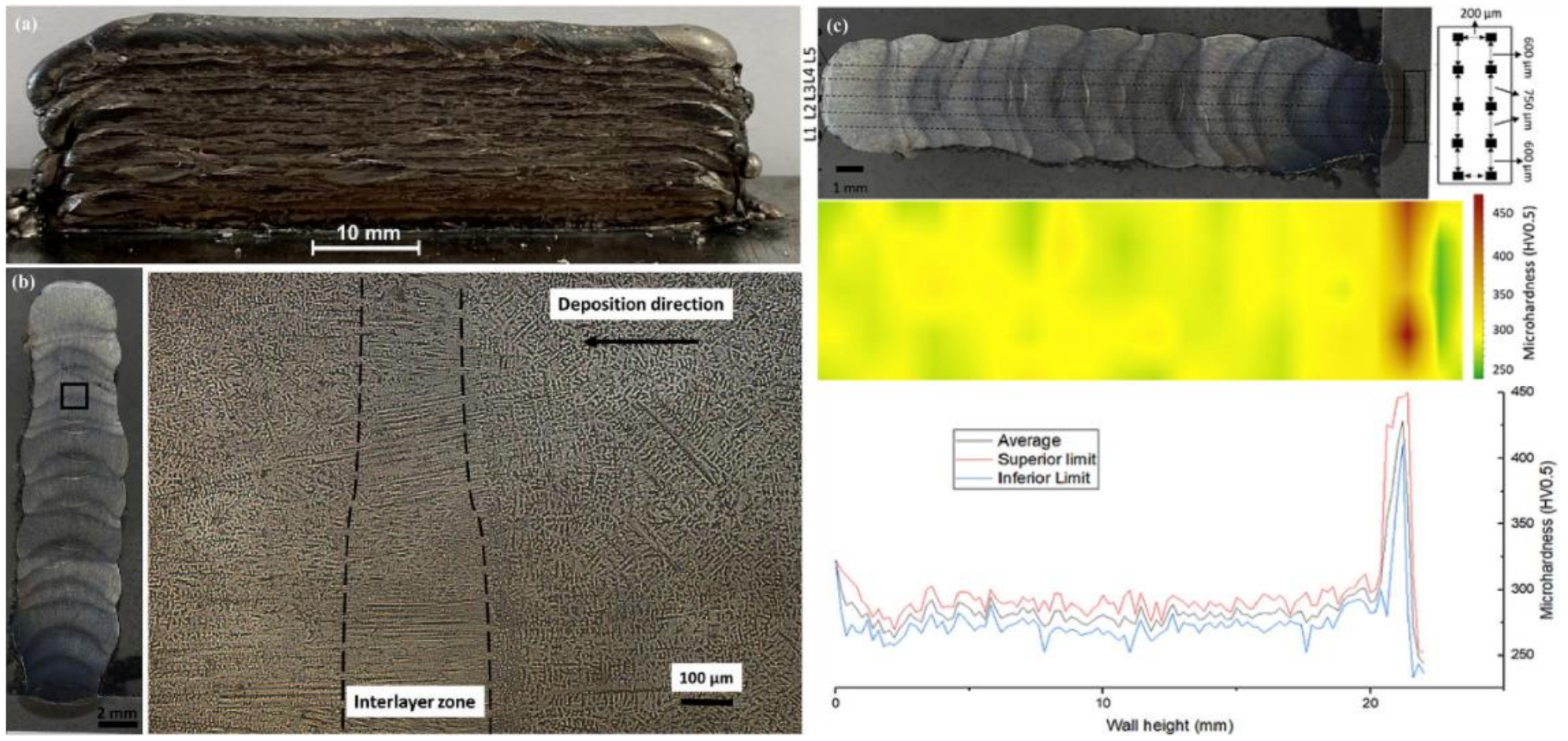

**Fig. 26. (a) As-built WAAM-manufactured Fe–Mn–Si–Cr–Ni–V-C SMA after full deposition, (b) optical image showing the microstructure of the as-built material, and (c) microhardness evaluation showing the measured lines on the material (upper), intensity map (middle), and average hardness** [294]**.**

Ultimately, it was demonstrated that Fe-SMA developed with the directed energy deposition (DED) AM approach had superior shape memory capabilities than Fe-SMA produced with traditional manufacturing techniques [298]. Fe-20Mn-5.5Si-4.9Ni-8.8Cr SMA was processed using laser-based DED. The optical analysis of the produced Fe-SMA (Fig. 27(a-f)) revealed the existence of spherical pores that can return to the gas produced during the manufacturing process from the volatilization of Mn and water vapor. The absence of visible cracks in the manufactured material can be attributed to the martensitic phase transformation being driven by residual stress from the forming process, which efficiently releases it and suppresses the creation and growth of cracks. The preferable growth orientations of <001> and <101> are visible in the columnar grains on the vertical surface of the EBSD analysis (Fig. 27($a_1$)), with <001> being the more prominent orientation. As such, a <001> preferred growth direction is seen in the columnar crystals that form on the build surface throughout the deposition process. The equiaxed grains in the horizontal plane (fig. 27($d_1$)) don't possess a preferable direction. This is explained by the fact that equiaxed grains develop without a particular preferred orientation. The equiaxed crystal area revealed a more consistent distribution of grain size than the columnar crystal region. As observed in Fig. 27 ($b_1$) and ($e_1$). On the horizontal surface, the grain size is more uniformly distributed and significantly smaller, usually less than 60 μm, as seen in Figs. 27. (c1) and (f1). According to the results of the tensile testing, the shape recovery was 68.1%, 44.2%, 31.7%, and 17.6% for deformation levels of 3%, 7%, 11%, and 15%, respectively. Interestingly, the DED-manufactured Fe-Mn-Si SMA's maximum recoverable deformation achieved 3.49%, higher than the conventional Fe-SMA (<3%).

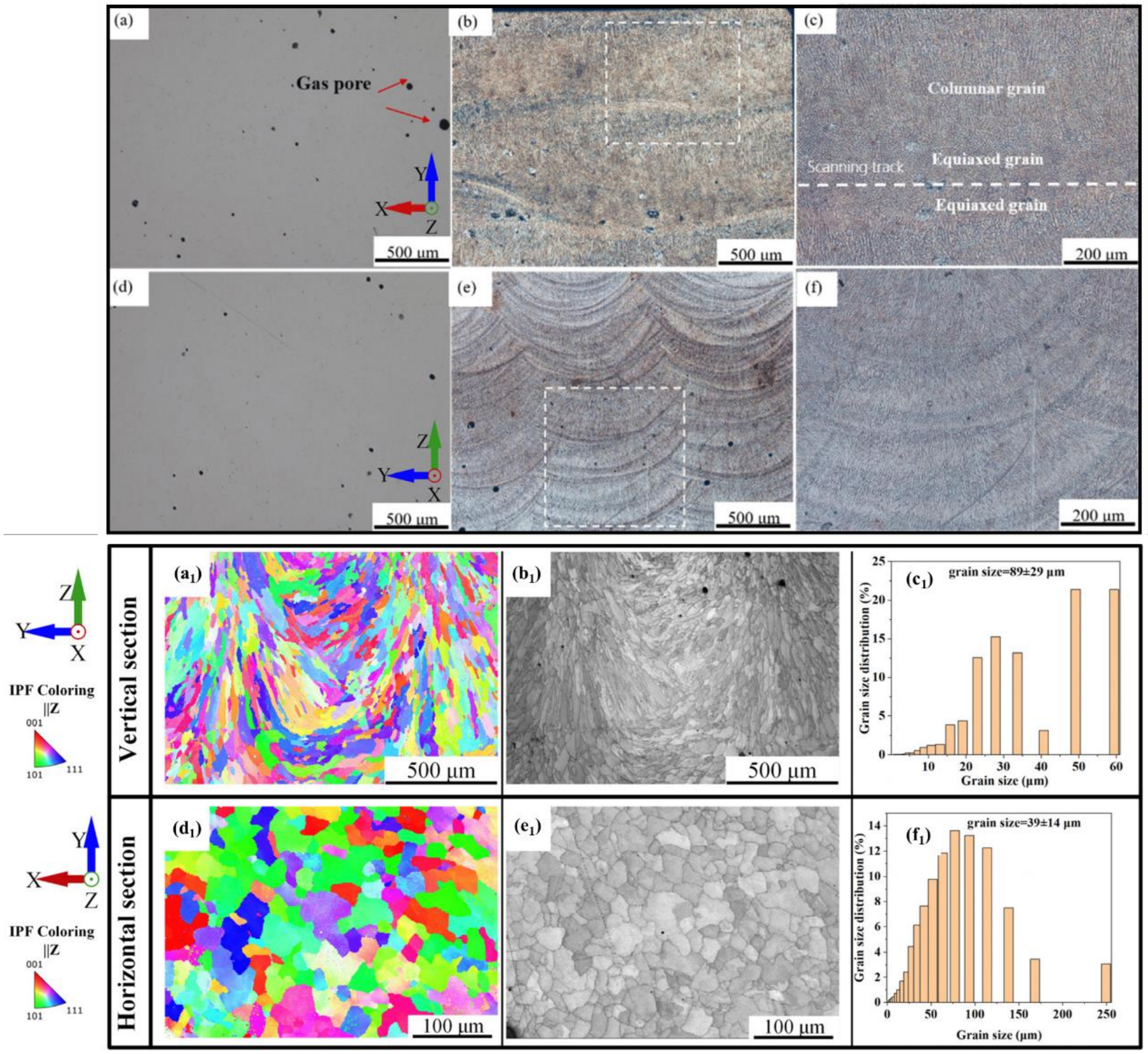

**Fig. 27. (a-f) Optical images showing the morphology and the microstructure of the as-deposited Fe-SMA. (a) Horizontal surface, (b) laser scanning track, (c) the morphology of grains within the scanning track, (d) vertical surface, (e) molten pool morphology, and (f) the morphology of the grains within the molten pool. ($a_1$-$f_1$) EBSD characterization of the vertical ($a_1$-$c_1$) and horizontal ($d_1$-$f_1$) surfaces of the as-deposited Fe-SMA. ($a_1$, $d_1$) inverse pole maps, ($b_1$,$e_1$)grain quality map, and ($c_1$,$f_1$) grain size distribution** [298]**.**

**Table 6. Summary of recent work on Fe-SMAs**

| Fe-SMA type | Investigated parameters | Treatment methods | Findings summary | Reference |
|---|---|---|---|---|
| Fe-20Mn-5.5Si-9Cr-5Ni (wt.%) | • The density of grain boundaries with low angles.<br>• The size of γ grain.<br>• The precipitation of χ phase. | • Lowering annealing twin boundary density.<br>• Characterization techniques: OM, SEM, XRD, EBSD, and TEM.<br>• Pre-strains were applied to the samples and then annealed at 450 °C for 15 min to restore their shapes.<br>• Bending test coupled with recovery annealing for studying the shape recovery ratio. | • Increased densities of low-angle grain boundaries.<br>• Larger γ grain sizes.<br>• χ precipitates with a volume fraction of less than 1.6%.<br>• The growth of γ grain size improves the SME. | [187] |
| Fe–34Mn–7.5Ni–13.5Al (at. %) | • The effect of aging treatments.<br>• Crystal orientation.<br>• SME or SE. | • Cyclic heat treatment was used.<br>• EDM was used for machining dog bone samples.<br>• The SMA was encapsulated in quartz tubes before applying the heat treatment. | • Achieved superelastic strain of 4.5 %.<br>• Aging treatments, crystal orientation, and bending temperature all had a substantial effect on SME and the magnitude of SE. | [120] |
| Fe–30Ni–15Co–10Al–2.5Ti–0.05B (at. %) | • The microstructure, texture, and low energy boundaries.<br>• Ductility.<br>• MT and SE. | • Induction melting in an argon environment was used to produce the alloy.<br>• Cold-rolling with a reduction ratio of 90% or 98.5% was utilized after hot-rolling at 1200°C to extract samples from the ingot.<br>• The samples were heated in a solution for three hours at 1200 °C, then quenched in water and aged for 12 to 48 hours at 550 °C.<br>• OM, EBSD, electrical resistivity measurement, DSC, Vickers hardness test, and cyclic tensile test were used to evaluate the microstructure, texture, TTs, hardness, and SE. | • The thermoelastic transformation was observed at an aging of 550 °C, having a structure of γ (FCC) + γ′ (L12).<br>• {0 1 2} 〈1 0 0〉 recrystallization texture was generated.<br>• The high presence of low energy boundaries in 98.5% cold-rolling.<br>• Aging reduced the brittle precipitation and led to an enhancement of ductility.<br>• SE of higher than 4% was investigated under a tensile test at ambient temperature. | [108] |
| Fe–30Ni–15Co–10Al–2.5Ti (at. %) (single crystal) | • SE and functional fatigue. | • The Bridgman procedure was used to make single-crystalline ingots of the alloy under a helium atmosphere.<br>• An Instron servo-hydraulic load frame was utilized to perform deformation.<br>• Heat treatment was applied to improve precipitation and induce superelasticity.<br>• Cyclic tests combined with heat treatment were carried out to investigate the functional fatigue behavior. | • The heat treatments at 600 °C achieved SE with a high strain recovery of ~7%.<br>• No SE is shown at ambient temperature.<br>• Treatment time >200 min attributed to a brittle response before the transformation.<br>• With continuous loading, superelastic strains were reduced and functionality was compromised. | [117] |
| Fe-17Mn-5Si-5Cr-4Ni-0.3C-1Ti (wt.%) | • The relationship between shape memory behavior, microstructure, and mechanical properties. | • Homogenization and post-aging heat treatment were carried out.<br>• XRD, OM, SEM, EBSD, EDX, and TEM were applied for characterizing phases, microstructure, and texture.<br>• Thermo-Calc software was utilized to simulate the phase diagram equilibrium.<br>• Recovery and bending tests were followed to evaluate the shape memory performance. | • TiC precipitates appeared in a few micrometers size ∼ 5 nm.<br>• Both the nano-sized and coarse TiC particles showed good stability during the aging treatment.<br>• The as-rolled SMA displayed a good recovery response.<br>• The aging heat treatment did not significantly improve shape recovery performance. | [126] |
| Fe-13.51Mn-4.82Si-8.32Cr-3.49Ni-0.15C (wt.%) | • The feasibility of utilizing rolling technology to create ultrafine grains in the SMA.<br>• Investigate the features of shape memory. | • Homogenization, hot forging and rolling, annealing, and quenching were used for processing the samples.<br>• Recovery of stress and strain was evaluated by loading/unloading and heating the samples.<br>• TEM and EBSD were applied to analyze the microstructure. | • Ultrafine grains were produced and improved the shape recovery stress.<br>• Ultrafine grains samples have higher strength than coarse grains samples. | [115] |

## 7. Applications of Fe-SMAs

Scientists have performed a substantial study into many structural applications and developed novel technologies that take advantage of their unique features. They've been investigating how smart materials can be used to improve infrastructure performance by constructing adaptive and intelligent structures. SE and SME are two noteworthy properties of SMAs that have been demonstrated to significantly enhance the overall performance of civil structures such as bridges and others. Because of these exceptional qualities, SMAs can endure massive nonlinear deformations and fully recover after the load is removed [313]. Moreover, a superelastic pattern in the austenite phase can be employed to reduce the amount of residual deformation created by the excitations, like those resulting from earthquakes. Additionally, in situations where the recovery strain is constrained due to external restrictions including end anchorages and clamps, recovery stress is produced in the SMAs, which can be exploited for prestressing [314]. Investigation has been done on the development of recovery stresses in a constrained Fe–Mn–Si–Cr–Ni–VC SMA that is utilized for prestressing and mechanical coupling, with a focus on non-ideal constraints that may lower the recovery stress [315]. An initial gap has formed between a Fe-SMA element and the structure that is to be prestressed or linked to examine the stress recovery following partial free shape recovery. For instance, as schematically illustrated in Fig. 28 (a), when an SMA ring is employed for pipe joining, the inner diameter of the pre-extended SMA ring must be greater than the outer diameter of the pipes to be connected. When heating in this instance, the SMA initially exhibits free shape recovery until the initial gap closes. Stress-related to recovering only then begins to accumulate. Uniaxial tensile tests with a closing gap were conducted, caused by pipe joining and confinement. After pre-straining the samples to 4%, they were heated to 160 °C and subsequently cooled to room temperature. Two phases were separated in the heating and cooling process to replicate the gap effect. The samples were initially given unrestricted shrinkage through shape recovery. During this phase, the alloy recovers to its original shape until the strain approaches the contact strain, which signifies the gap closure. The sample was maintained at a constant strain equal to the contact strain in the second stage. It is anticipated that the sample would experience less stress recovery in the second stage than in the case of stiff constraints with no gaps since the sample has already experienced some strain recovery in the first step. Fe-SMA was employed in concrete construction as a prestressing element to identify its stress recovery. In this instance, the recovery stress's development is dependent not only on the ultimate strain but also on the temperature route and strain. Uniaxial tensile tests have been conducted with elastic constraints, motivated by the prestressing of concrete. The associated stress-strain pathways are schematically displayed in Fig. 28. (b). The stress-strain routes have a negative slope because tensile stress causes the restraining supports to shrink, which releases tension. The samples were heated to 160 °C, cooled to room temperature, and then pre-strained to 4% for the testing. The testing machine's integrated controller was used to implement compliance during heating and cooling. This controller accepts several inputs and is additionally employed to adjust strain caused by the extensometer's thermal expansion. The investigation concluded that testing was done on the evolution of recovery stresses in the presence of various starting gaps and elastic restrictions. The findings demonstrate that, even in less-than-ideal restricting circumstances, the Fe-SMA is still capable of producing significant recovery stresses.

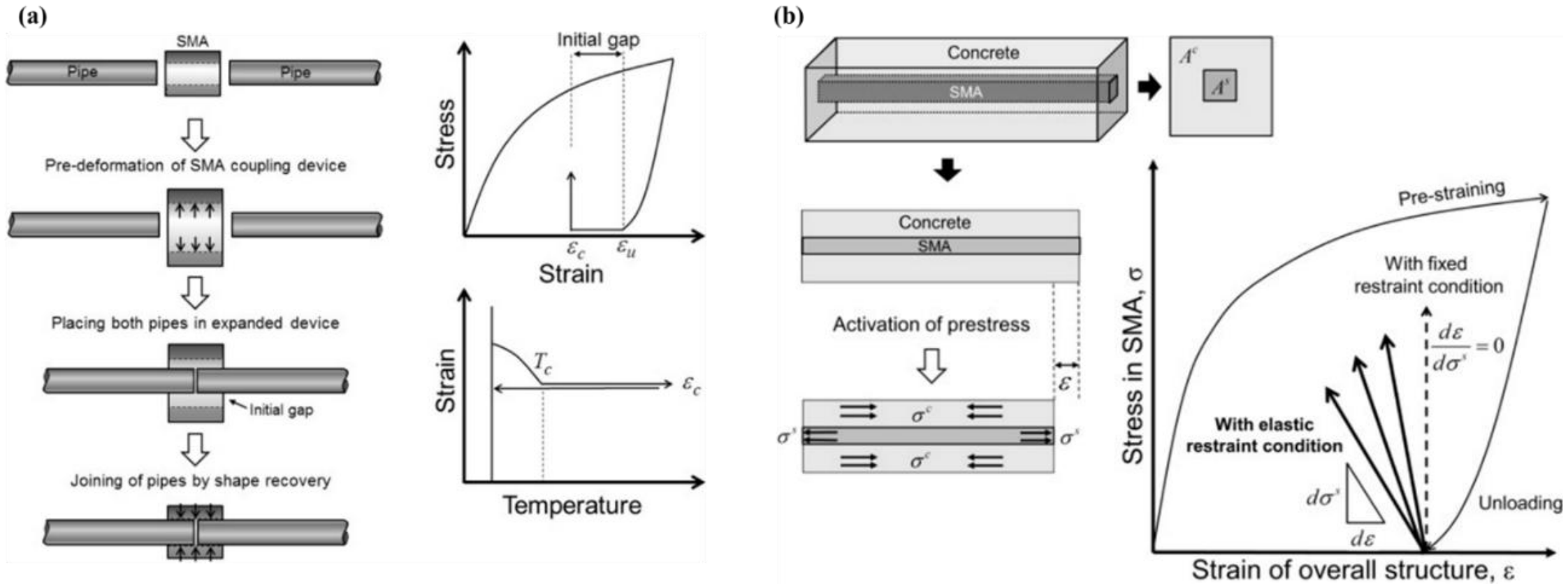

**Fig. 28. (a) The evaluation of stress recovery in a partially unrestricted shape recovery. The diagram shows the joining process with an initial gap between the pre-strained Fe-SMA and the pipes. (b) the development of stress recovery under elastic restrictions** [315]. **Where $\varepsilon_c$, $T_c$, $\sigma^s$, and $\sigma^c$ are the contact strain, contact temperature, tensile stress in the SMA, and compressive stress in the concrete, respectively.**

Fe-SMAs are predicted to be employed for superelastic materials, as well as dampening and sensing materials [24], [250]. They feature exceptional properties like high strength, high stiffness, and low manufacturing costs. Furthermore, their manufacturing costs are expected to be very small compared to those of Ni-Ti SMAs, making them the ideal alternative in structural engineering. Fe–Mn–Si–X alloys are well-known Fe-SMAs and have good mechanical properties and machinability, making them ideal for large-scale civil applications [23], [316], [317]. A new Fe-Mn–Si–Cr–Ni– (V, C) with relatively high recovered stress, high strength and stiffness, and excellent corrosion resistance has been produced and implemented in pre-stress concrete [72], [73]. In addition, a FeNiCuAlTaB SMA with a strain recovery of 13.5 % and austenite finish temperature of − 62 °C, was found to be effective in natural rubber bearings used under large shear strain amplitude and they are employed as a supplemental element to enhance their energy dissipation capability and residual distortion [318]. Moreover, Fe-SMAs have attracted a lot of interest in retrofitting older bridges to increase their load-carrying capability [35], [319]. Steel members were also reinforced using Fe-SMA strips for strengthening purposes, as illustrated in Fig. 29. To attach the Fe-SMAs to the steel structures, a direct fastening mechanism using a nail-anchor system was developed [320]. Two Fe-SMA strips, each measuring 50 mm in width and 1.5 mm in thickness, were mechanically fastened to the girders as part of the retrofit system. The strips have been activated/prestressed by heating them to a certain temperature using the SME of the Fe-SMA. The primary benefit of the suggested Fe-SMA-retrofit system is that, in contrast to traditional systems, it can prestress itself without the need for bulky hydraulic jacks. This leads to a major decrease in the amount of time, labor, and cost needed for the prestressing procedure. To assess the effectiveness of the created system, a series of static and fatigue four-point bending tests were carried out on a 6.4-m Fe-SMA retrofitted beam. According to the test findings, the Fe-SMAs' attained prestressing values, or recovery stresses, at activation temperatures of 100, 160, and 260 °C were around 160, 330, and 430 MPa, respectively. The bottom flange experienced induced compressive loads between 10 and 30 MPa. It was shown that the Fe-SMAs could be repeatedly reactivated, even up to temperatures higher than the temperature of initial activation. This would lead to increased degrees of prestressing. The findings of high-cycle fatigue testing revealed no slippage in the anchoring system and a steady prestressing in the Fe-SMA components during the tests, which confirms the developed system's reliability under a high-cycle fatigue loading regime.

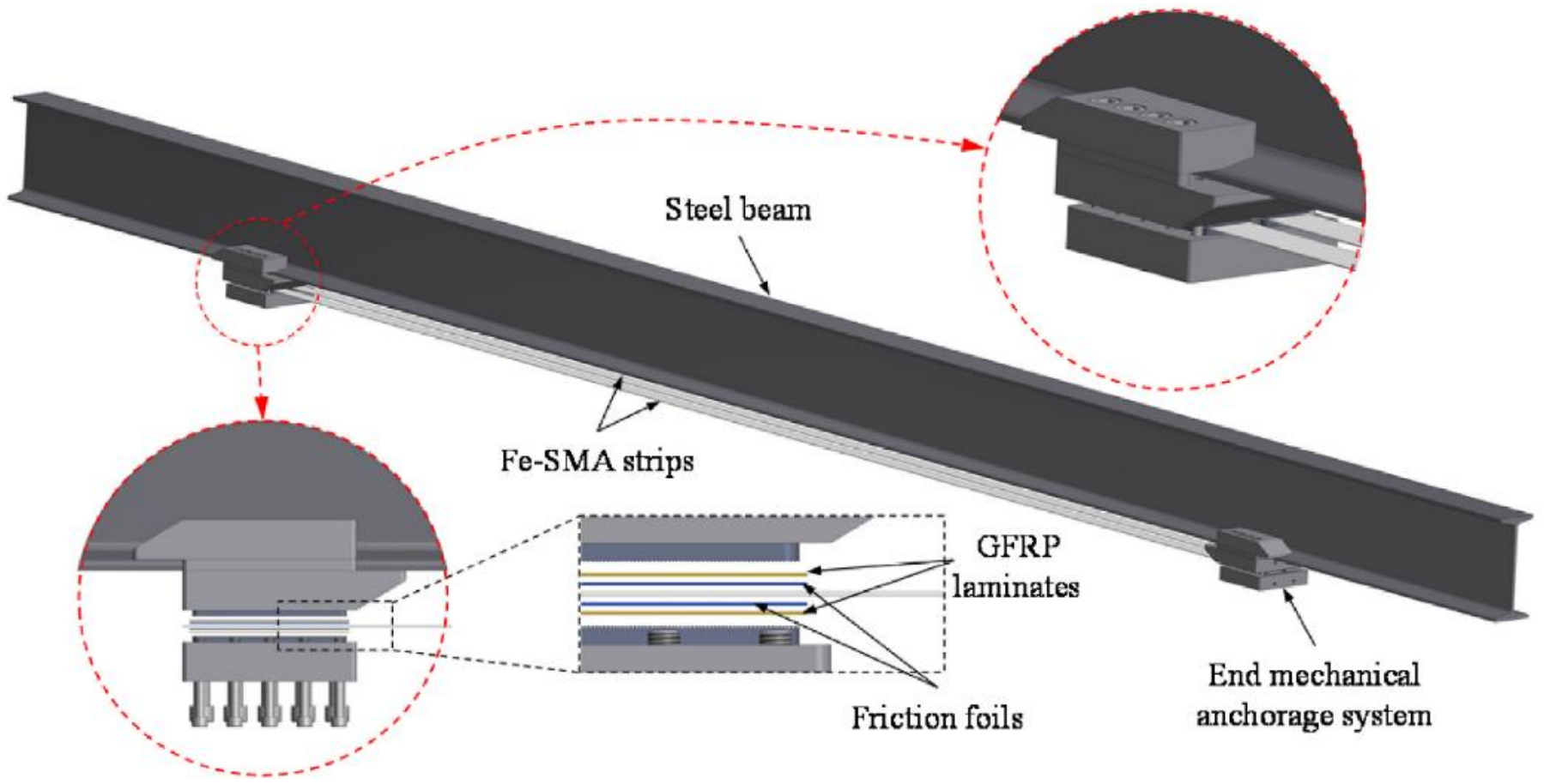

**Fig. 29. Strengthening the steel members using Fe-SMA strips. A direct fastening mechanism using a nail-anchor system is to attach the Fe-SMAs to the steel structures** [320]**.**

Fe-SMAs can be integrated into concrete without damaging its surrounding concrete due to the high temperatures generated during actuation [71]. Additionally, Fe-SMAs were employed as rebars in concrete bridge piers to enhance the seismic function of the bridges during a major earthquake [321]. Fe-SMA strips were used to reinforce the concrete constructions. As part of a well-known procedure for strengthening concrete known as near-surface mounted reinforcement, the strips were bonded into grooves in the concrete's top layer. Since prestressing the Fe-SMAs strips was found to be simpler than that of the fiber-reinforced polymer strips, because there are no manual jacks or anchoring heads used in this prestressing, there are fewer additional apertures on the concrete surface along the grooves required to clamp the NSMR. As a result, their usage as a means of strengthening concrete structures was recommended [24].

While some Fe-SMAs, such as Fe29Ni18Co5Al8Ta0.01B and Fe36Mn8Al8.6Ni, have reasonable mechanical properties at ambient temperature, high recovery strain, and adequate tensile strength, they could be highly useful in situations involving SE and damping capacities. However, these alloys are still not extensively employed in civil applications due to a lack of mass manufacturing improvements and high processing costs [34], [143], [241]. The seismic performance of the Fe-SMA prestressed columns using 3D-printed concrete was recently investigated [322]. The system's development and assembly as well as the Fe-SMA activation process are illustrated in Fig. 30. (a-h). The incorporated Fe-SMA rebars were heated by electric resistance to generate recovery stress and prestress the column. Prestressing had two purposes for the column: it added self-centering properties and held the sections connected. Before the concrete was formed, every pair of Fe-SMA rebars was joined into a single continuous conductor to facilitate simple access for attaching the heating power supply as seen in Fig. 30. (i-k). Large-scale tests were carried out on two columns with combined lateral and gravity stress in order to assess the viability and seismic performance of the suggested system. The variable taken into account in this study was the steel-to-Fe-SMA reinforcement ratio in the column design. The testing findings demonstrated that the permanent 3D-printed concrete formwork exhibited neither early failure nor delamination and that the columns could tolerate lateral drifts of up to 5% without collapsing. Additionally, when the reinforcing ratio of steel to Fe-SMA rebars was 0.3, the columns demonstrated self-centering capabilities, retaining a residual drift of 1% up to a target drift of 4%.

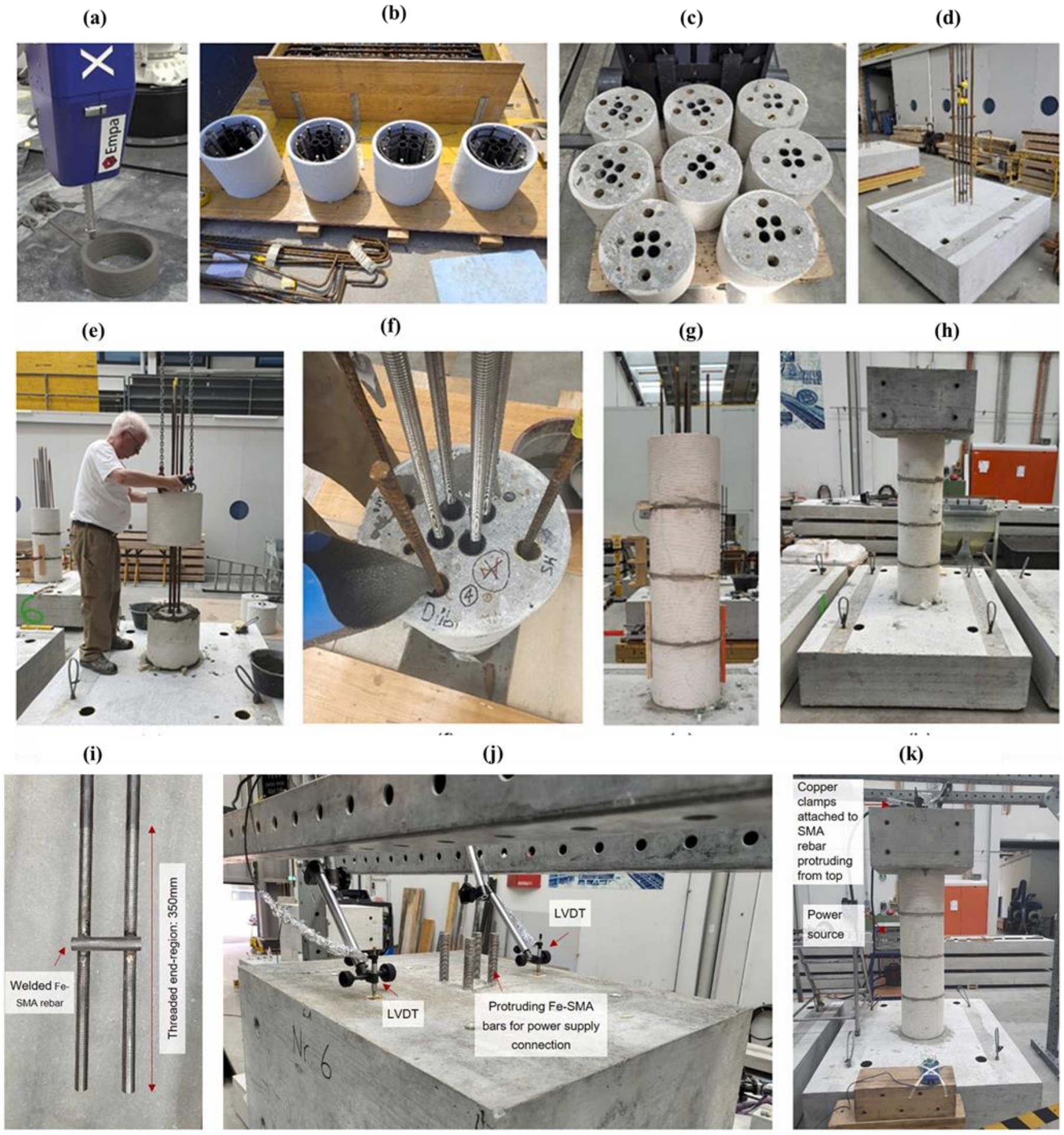

**Fig. 30. (a) Fabrication of shells of 3D-printed concrete, (b) installation of reinforcing cage and ducts, (c) adding the cast concrete, (d) ED steel and Fe-SMA rebars over footing, (e) segments stacking over footing ( f) ED steel and Fe-SMA rebars grouting, (g) segments assembly, ( h) assembly with top block, (i) joining Fe-SMAs within column footing for electrical continuity, (j) Fe-SMA rebars protruding to serve as axial displacement measurement instruments and electrical connections, (k) overall setup for the activation process** [322]**.**

Reinforced concrete was strengthened and pre-stressed using Fe-SMA bars [323]. The load-bearing ability of the bridge girder was substantially improved by this technology. The use of sprayed mortar and innovative U-shaped iron-SMA ribbed bars for shear reinforcement of reinforced concrete structures was examined, as illustrated in Fig. 31. (a) [35]. By using electric resistive heating, the Fe-SMA ribbed bars with a nominal diameter of 12 mm were activated. The activation caused the memory-steel reinforcement to prestress by around 300 $N/mm^2$, which in turn caused vertical compressive stresses in the RC beams' web. Extensive trials were conducted on T-beams measuring 0.75 m in height and 5.2 m in total length to demonstrate the feasibility and effectiveness of memory-steel shear strengthening. Positive outcomes demonstrate that the novel strengthening method is effective in real-world settings. The level of stress in the inner steel stirrups was lowered when the SMA stirrups were prestressed. Different slabs reinforced using memory steel bars were prestressed and tested under quasi-static loading. The activation memory steel bars improved the deflection of the SMA-reinforced slabs [35], [323], [324]. Fe-Mn-Si-SMA was employed by Zhang et al. to develop U-shaped dampers with exceptional low cycle fatigue (LCF) resistance which can be used for energy dissipation in structures [172]. Various applications for U-shaped dampers are depicted in Fig. 31. (b). Fe-Mn-Si characteristics were evaluated and compared with samples of standard carbon steel. The material tests revealed that under monotonic loading, the Fe-Mn-Si displayed a non-obvious yielding and modest strain hardening. The Fe-Mn-Si dampers displayed slightly narrower hysteresis responses than standard steel dampers, according to the damper tests. Depending on the factors considered, the total energy dissipation and LCF life of the Fe-Mn-Si dampers are about 7 times higher than those of their steel counterparts.

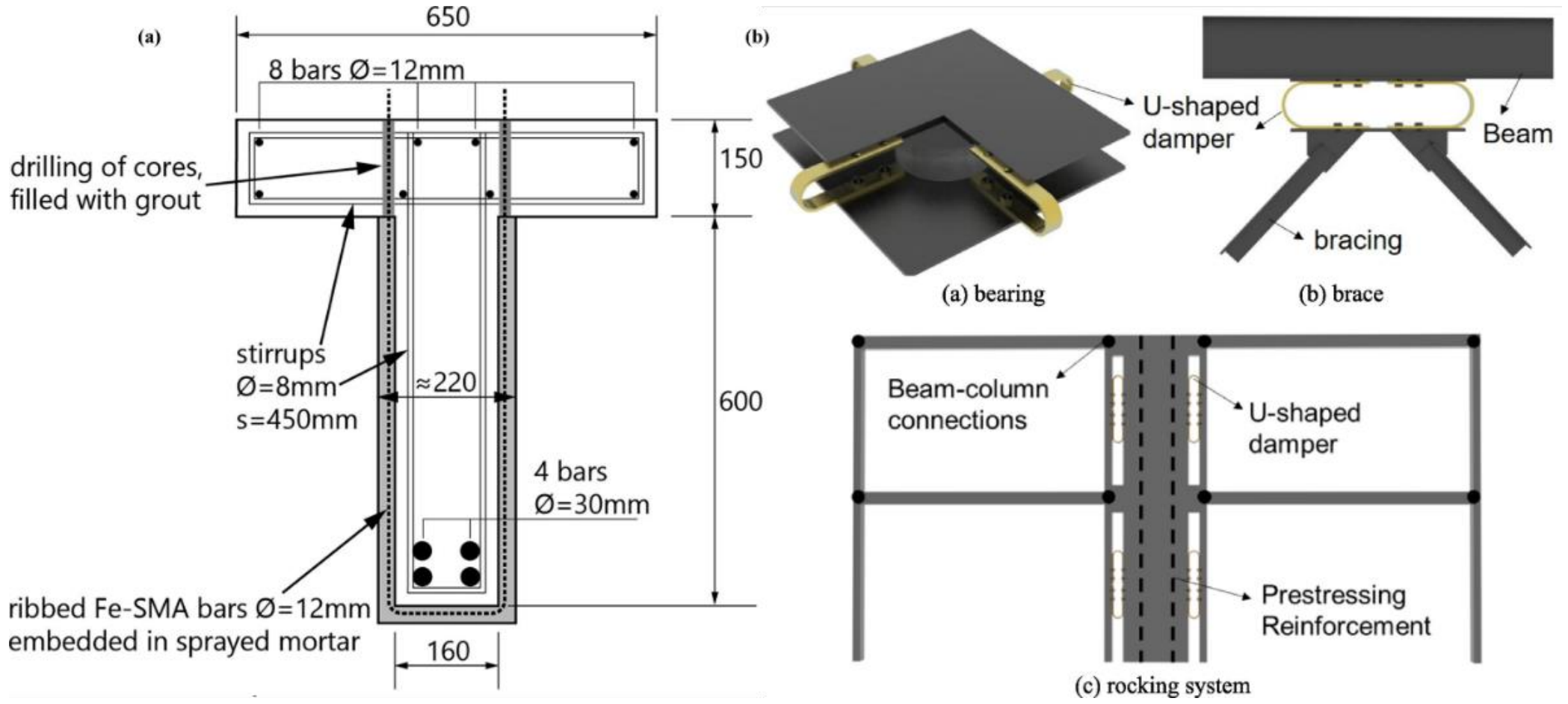

**Fig. 31. (a) A cross-section of RC beam reinforced by U-shaped ribbed Fe-SMA bars** [35]**. (b) Fe-SMA is used to manufacture U-shaped dampers characterized by their outstanding low cycle fatigue** [172]**.**

Shahverdi and his coauthors created ribbed bars with promising qualities for prospective applications in civil engineering using a Fe-17Mn-5Si-10Cr-4Ni-1VC (mass-%) SMA. To enhance the RC structure, they embedded the designed ribbed Fe-SMA bars into a shotcrete layer. The prestressed Fe-SMA bars considerably improved the performance at the serviceability phase. Using this method, the cracking load also rose in comparison to the reference sample. The results demonstrated that the strengthening strategy worked as intended and that the installation of Fe-SMA bars inserted in a freshly performed coating of shotcrete on the RC beams bottom was effective [325]. The magnitude of recovery stresses produced during the martensitic transformation is a key factor in the effectiveness of prestressed Fe-Mn-Si SMA in repairing and strengthening the RC structures. SMA's recovery stresses are a function of their thermal hysteresis, hence in prestressed reinforcements, Fe-Mn-Si SMAs are favored over NiTi SMAs because they have a higher thermal hysteresis and, consequently, higher stable recovered stresses. To improve their application as intelligent reinforcements in civil constructions, many attempts have been made to enhance the recovery stress in Fe-Mn-Si SMAs [24], [73], [106], [315].

Furthermore, based on Fe-SMA, a novel technique for reinforcing steel plates was developed by Izadi et al. [326]. The steel plates were prestressed using the SME of the Fe-SMA. To attach the prestressed Fe-SMA strips to the steel substrate, a mechanical anchorage method was established as shown in Fig. 32. (a). The SME Fe-SMA sample was then induced by heating with electrical resistive heating to a temperature of roughly 260 °C. The recovered stress was approximately between 350 and 400 MPa, which caused the maximum compressive stress in the steel plate to be around 74 MPa. The fatigue and yield strength of the steel specimens can be greatly improved by this compressive stress, which can also be extremely advantageous. Finally, static loading was applied to the strengthened specimens under displacement control until failure. It was established that the suggested strengthening method overcomes the challenges of traditional prestressing and provides a quick installation process because it doesn't need surface modification or curing for bond implementation. In contrast to mechanical prestressing, the use of SMAs during the prestressing of concrete structures eliminates the need for jacking devices and holding anchors. To exclusively apply prestressing force to the middle of a prestressed concrete component, SMA tendons can be put there. To prevent crack ignition on the outermost concrete fibers, the SMA tendons are activated sequentially using electronic power. Alternatively, prestressing might be added to the center of the beam by utilizing hot guns to heat the SMA tendons put from end to end only in the middle. This would assist in preventing the prestressed concrete from cracking [327], [328], [329]. A summary of the Fe-based SMA application is represented in Fig. 32. (b).

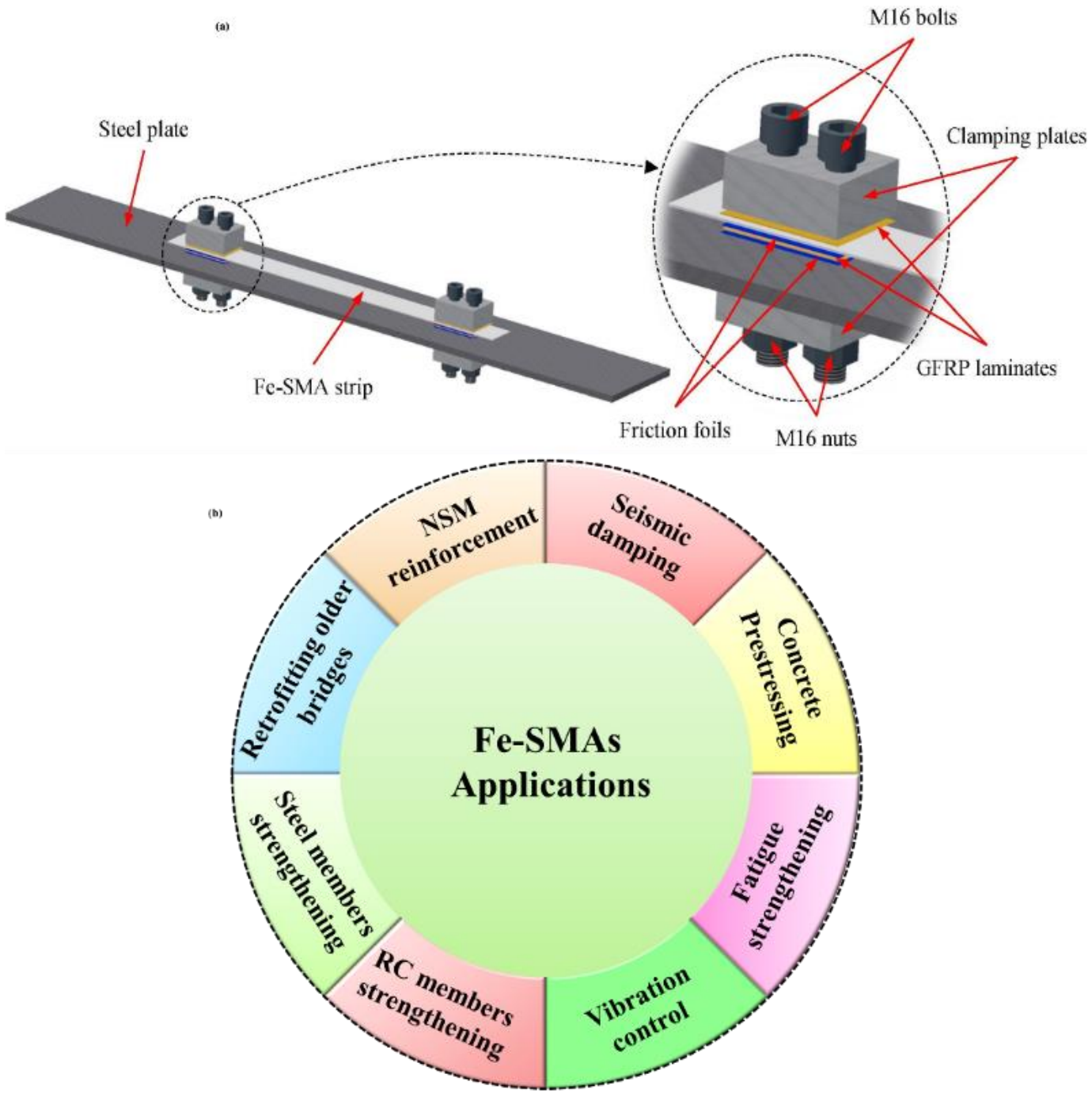

**Fig. 32. (a) SMA-to-steel joints' mechanical anchorage system's parts** [326]. **(b) Summary of common applications of Fe-SMAs.**

**Table 7. Fe-SMAs' applications with a detailed description of their implementation**

| Author | Application | Implementation procedure |
|---|---|---|
| [24], [317], [319], [320], [324], [326], [328], [330], [331], [332] | Structures strengthening | Fe-SMA strips or bars are incorporated into the specified components using anchorage systems, simply by gluing them into grooves in the outer layer of the concrete or making ducts through the structures, due to the high strength, outstanding recovery stress, and excellent stiffness that Fe-SMAs have. They are used as a reinforcement for metal I-girders. Fe-SMAs could be repeatedly triggered even at temperatures greater than their initial activation temperature, leading to larger prestressing levels. Additionally, the Fe-SMA-enhanced components displayed an exceptional high cycle fatigue resistance. |
| [185], [327], [333], [334] | Prestressing of concrete component | The smart prestressing technique uses the SME in prestrained SMA parts, such as cables, strands, or wires, to induce recovery stress in the concrete. Compared to traditional approaches, this one seems more practical and requires fewer devices. In addition, only external heating sources are necessary to produce the SME. |
| [80], [237], [335], [336], [337], [338], [339] | Seismic damping | Owing to their advantageous features, Fe-SMAs have a great deal of potential in seismic systems. With changing strain amplitudes and various loading procedures, the cyclic properties of Fe-SMAs, including recovery capacity, hysteretic response, as well as fracture and fatigue behaviors were assessed. According to the test results, under cyclic tension-compression loadings, adequate hysteretic loops with |
| | | outstanding deformation capacity can be produced. In addition to their other distinctive intrinsic qualities, Fe-SMA bars demonstrated reasonably stable performance under multistage loadings, offering a potential method for creating high-performance seismic devices. |
| [316], [325], [329], [340], [341] | Near-surface mounted reinforcement | By inserting and fastening pre-strained SMAs such as ribbed bars or strips into grooves through the beam tension side, the intended component (for example, an RC beam) is strengthened. The element is then heated to produce pre-stressing force. However, in other situations, a cementitious composite is used to cover the SMA element once it has been bonded to the concrete surface. |
| [342], [343], [344] | Fatigue strengthening | In numerous constructions, including steel bridges, fatigue-cracked riveted connections are retrofitted using Fe-SMAs. The application of the prestressed strengthening technique was discovered to be a successful strategy for overcoming fatigue issues in riveted connections. Fe-SMAs have exceptional characteristics that make prestressing them simple. On any direction of the connection, the activated Fe-SMA strips are secured to the steel I-beam's flanges. The results showed that the inserted Fe-SMA strips significantly increased the fatigue life and prevented the fatigue crack. |

## 8. Key challenges and future directions of Fe-SMA

Fe-SMAs have not yet gained widespread market acceptance, due to their higher price than other reinforcement materials like carbon fiber reinforced polymer and their slower production process. Therefore, there is an essential need to advance Fe-SMA production technology to lower the price and expand its application in the construction industry. Fe-SMAs have superior weldability than other SMAs which can significantly decrease the cost of materials joining during the applications. The weldability of Fe-SMAs has not been thoroughly investigated in earlier studies. Therefore, it is crucial to develop ideas for improving the weldability of Fe-SMAs with other dissimilar alloys. Fe-SMAs are used in humid, unexpected, and even corrosive conditions, therefore improving their corrosion resistance should be considered. The lack of theoretical research and the absence of engineering-focused design approaches are two key challenges to the actual implementation of Fe-SMAs. Microscale low-cycle fatigue mechanisms and welding methods associated with the SME of Fe-SMA are also potential areas for research. The Fe-SMAs are mostly manufactured using traditional methods, such as melting and casting as the first steps in the preparation, followed by heat treatment for homogenization, and finally hot working and other treatments. The current trend in Fe-SMA production has been attributed to AM techniques, especially LPBF ones, however, few studies have reported the use of additive manufacturing to manufacture Fe-SMAs. Further research is needed to explore the potential of AM techniques in a wide range of Fe-SMA production and how to overcome the related challenges. There are several challenges and opportunities to explore in the AM of Fe-SMAs. The first step must be the investigation of the printability of Fe-SMAs, followed by optimizing and determining a good window for manufacturing each Fe-SMA. The effect of processing parameters including laser power, scanning speed, layer thickness, hatching space, scanning strategy, spot size, rotation angle, chamber temperature, preheating temperature, type of inert gas, chamber oxygen concentration, and others on their mechanical and thermomechanical properties should then be investigated. Finally, the type and properties of the used powder, as well as post-heat treatments' effect on Fe-SMAs' properties can be studied.

## 9. Conclusion

Fe-SMAs are promising candidates for many practical applications such as large-scale civil and structural engineering applications, sensing and damping systems, tube coupling, reinforced concrete, etc. due to their potential merits such as inexpensive alloying elements, remarkable fatigue and corrosion resistance, notable SE, and good SME. The evolution of Fe-SMAs has shown an increase in the number of Fe-SMAs in recent years, indicating that the new Fe-SMAs have fascinating properties when compared to present SMAs. Thermomechanical characteristics of Fe-SMAs were investigated and compared, confirming that these SMAs can provide excellent SME and SE. The maximum SE of Fe-SMAs was achieved using FeNiCuAlTaB alloy, and it was estimated to be around 13.5 %. Many factors influence the manufacture of new Fe-SMAs and the development of their properties, including alloying components, texture and structure modification, deformation temperature, grain size, as well as heat treatment procedures. Theoretical work on the SMAs has been motivated by a desire to understand physics and develop a direction of improvement. Significant effort was put into developing computational models that would make it easier to develop alloys, design geometry, and fabrication. The Fe-SMAs are mostly manufactured using traditional methods, such as melting and casting as the first steps in the preparation, followed by heat treatment for homogenization, and finally hot working and other treatments. The current trend in Fe-SMA production has been attributed to AM techniques, especially LPBF ones, however, few studies have reported the use of additive manufacturing to manufacture Fe-SMAs. Up to now, only Fe-Mn-Si and Fe-Mn-Al-Ni SMAs have been additively manufactured with enhanced properties. Further research is needed to explore the feasibility of additive manufacturing for a wide range of Fe-SMA production and to overcome the related challenges in expanding Fe-SMA's application.

**Declaration of Competing Interest**

The authors declare that they have no known competing financial interests or personal relationships that could have appeared to influence the work reported in this paper.

**Funding**

This work was funded by the National Science Foundation. Grant No. 2301766 and the Toyota

**Data availability**

No data was used for the research described in the article.